\documentclass[final,3p,times,fleqn]{elsarticle}
\usepackage{mathtools,booktabs}
\usepackage{bm}
\usepackage{array}
\usepackage{multirow}
\usepackage{autobreak}
\usepackage{amsmath}
\usepackage{amsthm}
\usepackage{gensymb}
\usepackage{subfigure}
\usepackage{epstopdf}
\usepackage{color}
\usepackage{hyperref}

\biboptions{sort&compress}
\begin{document}
	\begin{frontmatter}
		\title{Multiple-distribution-function finite-difference lattice Boltzmann method for incompressible Navier-Stokes equation}
		\author[a]{Xinmeng Chen}
		\author[a,b,c]{Zhenhua Chai }
		\author[d]{Yong Zhao}
		\author[a,b,c]{Baochang Shi \corref{cor1}}
		\ead{shibc@hust.edu.cn}
		\address[a]{School of Mathematics and Statistics, Huazhong University of Science and Technology, Wuhan 430074, China}
		\address[b]{Institute of Interdisciplinary Research for Mathematics and Applied Science, Huazhong University of Science and Technology, Wuhan 430074, China}
		\address[c]{Hubei Key Laboratory of Engineering Modeling and Scientific Computing, Huazhong University of Science and Technology, Wuhan 430074, China}
		\address[d]{School of Mathematics and Statistics, Changsha University of Science and Technology, Changsha 410114, Hunan, China}
		\cortext[cor1]{Corresponding author.}
		\begin{abstract}
			In this paper, a multiple-distribution-function finite-difference lattice Boltzmann method (MDF-FDLBM) is
  proposed for the convection-diffusion system based incompressible Navier-Stokes equations (NSEs). By Chapman Enskog analysis, the
  convection-diffusion system based incompressible NSEs can be recovered from MDF-FDLBM.
  Some quantities, including the velocity gradient, velocity divergence, strain rate tensor, shear stress and vorticity, can be computed
  locally by the first-order moment of the non-equilibrium distribution function. Through the von Neumann analysis, we conduct the stability analysis for the MDF-FDLBM and incompressible finite-difference lattice Boltzmann method (IFDLBM).
  It is found that the IFDLBM will be more stable than that of MDF-FDLBM with small kinematic viscosity, and the MDF-FDLBM will be more stable than that of
    IFDLBM with large Courant-Friedrichs-Lewy condition number.
  Finally, some simulations are conducted to validate the MDF-FDLBM.
  The results agree well with the analytical solutions and previous results. Through the numerical testing, we find that the MDF-FDLBM has a second-order convergence rate in space and time. The MDF-FDLBMcombined with non-uniform grid also works well. Meanwhile, compared with IFDLBM, it can be found that MDF-FDLBM offers higher accuracy and computational efficiency, reducing computation time by more than $36\%$.
		\end{abstract}
		\begin{keyword}
			Finite difference lattice Boltzmann method \sep Coordinate transformation \sep Power-law fluid \sep Trapezoidal cavity
		\end{keyword}		
	\end{frontmatter}
	\section{Introduction}
	
	As a mesoscopic numerical method, the lattice Boltzmann method (LBM) has received great attention in the past 30 years. It has a great performance in simulating complex fluid flows, including the thermal flows \cite{2001Succi,SM2015}, microscale flows \cite{CY2002application,raabe2004overview,Guo2015}, porous media flows \cite{Higuera1989Lattice,Guo2002Lattice,AM2015porous}, turbulent flows \cite{strumolo1997new,chen2003extended}, multiphase flows \cite{shan1993Lattice,swift1995Lattice,Yeomans2004mulitiphase,Falcucci2010multiphase,Sb2012multphase,chai2019two}, multicomponent flow and other fluid flows. And the finite difference lattice Boltzmann method (FDLBM), which combining LBM with finite-difference scheme, also has attracted much attention. Compared with LBM, FDLBM can promote the geometrical flexibility, due to the discrete-velocity being decoupled with lattice and time steps. This feature makes non-uniform meshes can be used to FDLBM to improve computational efficiency.

In 1995, a finite-difference lattice Boltzmann equation for the simulation of the incompressible Navier-Stokes equations was proposed by Reider and Sterling \cite{reider1995accuracy}. Subsequently, by using body-fitted coordinates with non-uniform grids, Mei and Shyy \cite{mei1998finite} explored a FDLBM in curvilinear coordinates. Based on above mentioned works, Guo et al.~\cite{guo2003explicit} proposed a FDLBM with a mixed difference scheme dealing with the advection term. Then, Wang et.al. \cite{Wang2017FDLBM} extend the FDLBM to solve nonlinear convection-diffusion equations (NCDEs).
Guo et al.~\cite{guo2005finite} developed a new FDLBM by using the second-order Lax Wendroff scheme and first-order Euler$'$s formula to discrete space and time derivatives, respectively. Wang et al.~\cite{wang2009implicit} explored a high-order FDLBM with a third-order implicit-explicit Runge-Kutta scheme and a fifth-order weighted essentially non-oscillatory scheme for time and space discretization. Chen et.al. \cite{ChenFDLBM2020} develop second-order FDLBM for the NSEs with a mixed difference scheme for space discretization. Then the second-order FDLBM has been combined with the multiple-relaxation-time model to solve the NCDEs \cite{Chen2021}.
In addition, other high-order accuracy FDLBMs are also developed \cite{hejranfar2014high,hejranfar2018preconditioned}, and these models have been used to simulate complex flow problems, including three-dimensional
incompressible flows \cite{ezzatneshan2019simulation}, two-phase liquid-vapor flows \cite{hejranfar2015simulation}, natural convection in some special geometries \cite{sai2019natural,Khakrah2019Numerical}, and blood flow \cite{Sakthivel2019blood}.
The immersed boundary method was also incorporated into the FDLBM to simulate the incompressible flows \cite{kim2016immersed}.
Besides, there are many scholars studied the stability of FDLBMs \cite{GV2014Stability,el2013stability,GV2015Stability,Chen2021}.
El-Amin et.al. used the Fourier expansion to analyze a new explicit FDLBM \cite{el2013stability}.
As a common stability analysis method, the Von Neumann method has been used to study the stability of three-level finite-difference-based lattice Boltzmann schemes \cite{GV2014Stability},
the FDLBM with upwind differences schemes \cite{GV2015Stability} and MRT-FDLBM \cite{Chen2021}.

For incompressible NSEs, there are two kinds of FDLBM. One is the standard FDLBMs \cite{guo2003explicit,wang2009implicit}, where the equilibrium distribution function is defined by density and velocity. The other is incompressible FDLBM \cite{IFDLBM}, where the equilibrium distribution function is defined by pressure and velocity.
Recently, Chai et el. \cite{Chai2022} developed a multiple-distribution-function LBM (MDF-LBM) for convection-diffusion-system based on incompressible NSEs.
The authors converted incompressible NSE into a convection-diffusion-system. Instead of solving incompressible NSEs directly, MDF-LBM is used to solve convection-diffusion system.
Based on this work, we can also introduce this idea to the FDLBM, and develop a MDF-FDLBM with a second-order convergence rate in space and time for the incompressible NSEs.
The MDF-FDLBM has some advantages as follows.

$\bullet$ Compared with the MDF-LBM, MDF-FDLBM can adopt non-uniform grid simulation to improve the computational efficiency of the model.

$\bullet$ Compared to IFDLBM and standard FDLBM, one can use fewer discrete velocities to construct the MDF-FDLBM. Take the two-dimensional problems as an example, the D2Q4 or D2Q5 lattice model is enough for the MDF-FDLBM, while the D2Q9 model needs to be used for the IFDLBM and standard FDLBM.

$\bullet$ Some physical variables, including the velocity gradient, velocity divergence, strain rate tensor, shear stress and vorticity, can be computed locally by the first-order moment of the non-equilibrium distribution function. But, for the IFDLBM and standard FDLBM, it is difficult to develop a local computation scheme for the velocity gradient and vorticity, and the computation of strain rate tensor need to use the second-order moments of the nonequilibrium distribution function.

The rest of the the paper is organized as follows. In Section 2, the incompressible NSEs is transformed into the convection-diffusion-system, and a MDF-FDLBM for the convection-diffusion-system based incompressible NSEs is developed. And then, the stability of IFDLBM and MDF-FDLBM are studied in Section 3. In Section 4, some numerical simulations are conducted. Finally, we summarizes the results and concludes in Section 5.
	
	\section{Physical model and governing equation}\label{GoverEqs}

    \subsection{The incompressible Navier-Stokes equations}
	
	The incompressible fluid flows can be described by the incompressible Navier-Stokes equations, which the NSEs can be written as
\begin{equation}
\begin{split}
  \nabla\cdot \bm u&=R,\\
  \frac{\partial \bm u}{\partial t}+\nabla\cdot(\bm u\bm u)&=-\nabla P+\nabla\cdot(\nu\nabla \bm u)+\bm F,
\label{2.1.1}
\end{split}
\end{equation}
where $R$ is the source term, $\bm u$ is a $d$-dimensional velocity vector, $P$ is the pressure,
$\nu$ is the kinematic viscosity and $\bm F$ is the $d$-dimensional external force term.
As noted in Ref \cite{Chai2022}, the above NSEs can also be rewritten as a coupled convection-diffusion system,
\begin{equation}\label{2.1.2}
  \frac{\partial \bar{u}_\alpha}{\partial t}+\nabla\cdot(\bar{u}_\alpha\bm u+P\bm E_\alpha)=\nabla\cdot(\nu\nabla\bar{u}_\alpha)+\bar{F}_\alpha,\quad \alpha=0,1,...,d
\end{equation}
where $\bar{ u}_0=\rho_0$ and $\bar{u}_\alpha=u_\alpha(\alpha=1-d)$, $\bar{F}_0=R$ and $\bar{ F}_\alpha=F_\alpha(\alpha=1-d)$, $\bm E_0=\bm 0$ and $\bm E_\alpha(\alpha=1-d)$ is the unit
vector in $d$-dimmensional space.

It can be noticed that the NSEs can be reformulated as $(d+1)$ CDEs. And the NSEs \eqref{2.1.1} are equivalent to the NCDEs \eqref{2.1.2}.
However, the FDLBM is more effective to solve the NCDEs, due to fewer discrete velocity is needed for the NCDEs.

    \subsection{The MDF-FDLBM for the convection-diffusion-system based NSEs}
    In the FDLBM, there are two kinds of model, including the single-relaxation-time FDLBM and multiple-relaxation-time FDLBM. Here, we use the MRT-FDLBM to ensure the higher stability
and accuracy. Based on this work \cite{Chen2021}, the evolution equation of the MDF-FDLBM for the convection-diffusion-system based NSEs \eqref{2.1.2} can be decuded
\begin{equation}\label{2.2.1}
\begin{split}
  \hat{f}_{j,\alpha}(\bm x,t+\Delta t)=& \hat{f}^+_{j,\alpha}(\bm x,t)-\Delta t\bm c_j\cdot\nabla f_{j,\alpha}(\bm x,t+\frac{1}{2}\Delta t)\\
  & +\Delta t(F_{j,\alpha}+G_{j,\alpha})(\bm x,t)+\frac{\Delta t^2}{2}(\partial_tF_{j,\alpha}+\partial_tG_{j,\alpha})(\bm x,t),
  \end{split}
\end{equation}
where $f_{j,\alpha}(\bm x, t)$ is the distribution function at position $\bm x$ and time $t$ along the discrete velocity $\bm c_j$. $\Delta t$ is the time step and $\nabla$ is the spatial gradient operator.
The $F_{j,\alpha}$ and $G_{j,\alpha}$ are the source terms. The $\hat{f}$ and the $\hat{f}^+$ are the combination of distribution functions and equilibrium distribution functions. They can be defined as
\begin{equation}\label{2.2.2}
  \hat{f}=f+\frac{\Delta t}{2}\tilde{\Lambda}(f-f^{eq}),
\end{equation}
and
\begin{equation}\label{2.2.3}
  \hat{f}^+=f-\frac{\Delta t}{2}\tilde{\Lambda}(f-f^{eq}).
\end{equation}
In Eqs. \eqref{2.2.2} and \eqref{2.2.3}, $\tilde{\Lambda}=(\tilde{\Lambda}_{jk})$ is a $q\times q$ invertible collision matrix, and $q$ is the number of the discrete velocities.
According to the MRT-FDLBM \cite{Chen2021}, for the collision matrix $\tilde{\Lambda}$, some appropriate requirement need to be satisfied:
\begin{equation}\label{2.2.4}
  \sum_j\bm e_j\tilde{\Lambda}_{jk}=s_0\bm e_k,\quad \sum_j\bm c_j\tilde{\Lambda}_{jk}=\tilde{\bm S}\bm c_k,\quad \forall k=1,2,...,q,
\end{equation}
where $\bm e=(1,1,...,1)\in R^q$, $\tilde{\bm S}$ is an invertible $d\times
d$ relaxation matrix corresponding to the kinematic viscosity $\nu$.
The equilibrium distribution function should be defined as
\begin{equation}\label{2.2.5}
  f_{j,\alpha}^{eq}=\omega_j\left[\bar{u}_\alpha+\frac{\bm c_j\cdot(\bar{u}_\alpha\bm u+P\bm E_\alpha)}{c_s^2}\right].
\end{equation}
And the source terms are defined as
\begin{equation}\label{2.2.6}
  F_{j,\alpha}=\omega_jF_\alpha,
\end{equation}
\begin{equation}\label{2.2.7}
  G_{j,\alpha}=\omega_j\frac{\bm c_j\cdot\partial_t(\bar{u}_\alpha\bm u+P\bm E_\alpha)}{c_s^2}.
\end{equation}

\begin{figure}[ht]
\centering
\includegraphics[width=10cm]{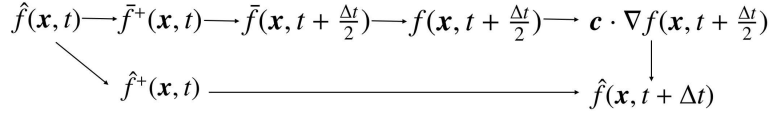}
\caption{The implementation process of the MDF-FDLBM.} \label{MRT}
\end{figure}

In addition, the second evolution equation used to evaluate distribution
function $f_{j,\alpha}(\bm x,t+\frac{1}{2}\Delta t)$ can be written as
\begin{equation}
 \bar{f}_{j,\alpha}(\bm x,t+h)=\bar{f}^+_{j,\alpha}(\bm x-\bm c_jh,t)+h\left[F_{j,\alpha}+\bar{G}_{j,\alpha}+\frac{h}{2}\partial_tF_{j,\alpha}+\frac{h}{2}\partial_t\bar{G}_{j,\alpha}\right](\bm x-\bm c_jh,t),
 \label{2.2.8}
\end{equation}
where $h=\Delta t/2$, and
\begin{subequations}
\begin{equation}
 \bar{f}_{j,\alpha}  =  f_{j,\alpha}-\frac{h }{2}(-\tilde{\Lambda}_{jk}\left(f_{k,\alpha}-f^{eq}_{k,\alpha})\right),\label{2.2.9a}
 \end{equation}
 \begin{equation}
 \bar{f}^+_{j,\alpha}  =  f_{j,\alpha}+\frac{h }{2}(-\tilde{\Lambda}_{jk}\left(f_{k,\alpha}-f^{eq}_{k,\alpha})\right).
\label{2.2.9b}
\end{equation}
\end{subequations}
The discrete source terms $\bar{G}_{j,\alpha}$ is
given by
\begin{equation}
 \bar{G}_{j,\alpha}(\bm x,t)=\omega_j\frac{\bm c_j\cdot (\bm I+\frac{h}{2}\bm \tilde{S})^{-1}\partial_t(\bar{u}_\alpha\bm u+P\bm E_\alpha)}{c_s^2}
 \label{2.2.10}
\end{equation}
Applying Taylor expansion to Eq.~\eqref{2.2.8} and ignoring the term
$O(h^2)$, we have
\begin{equation}
\begin{split}
 \bar{f}_{j,\alpha}(\bm x,t+h)=& \bar{f}^+_{j,\alpha}(\bm x,t)-h \bm c_j\cdot \nabla \bar{f}^+_{j,\alpha}(\bm x,t)+h[F_{j,\alpha}(\bm x,t)+\bar{G}_{j,\alpha}(\bm x,t)\\
 & +\frac{h}{2}\partial_tF_{j,\alpha}(\bm x,t)+\frac{h}{2}\partial_t\bar{G}_{j,\alpha}(\bm x,t)].
\label{2.2.11}
\end{split}
\end{equation}
The gradient terms $\nabla f_j$ in Eq. \eqref{2.2.1} and
$\nabla\bar{f}_j^+$ in Eq. \eqref{2.2.11} can be discretized by a mixed
difference scheme \cite{guo2003explicit},
\begin{equation}
 \nabla \Pi_j^*=\frac{\partial \Pi_j^*}{\partial \chi_\alpha} \Bigg{|}_m=\eta \frac{\partial \Pi_j^*}{\partial \chi_\alpha} \Bigg{|}_c+(1-\eta)\frac{\partial \Pi_j^*}{\partial \chi_\alpha} \Bigg{|}_u,
\label{2.2.12}
\end{equation}
where $\Pi_j^*$ represents $f_j$ or $\bar{f}_j^+$, and the parameter $\eta
\in [0,1]$. The terms $\dfrac{\partial \Pi_j^*}{\partial \chi_\alpha}
\Bigg{|}_u$ and $\dfrac{\partial \Pi_j^*}{\partial \chi_\alpha} \Bigg{|}_c$
represent second up-wind difference and central-difference schemes, and are
given by
\begin{subequations}
\begin{equation}
 \frac{\partial \Pi_j^*}{\partial \chi_\alpha} \Bigg{|}_c=\frac{\Pi_j^*(\chi_\alpha+\Delta \chi_\alpha,t)-\Pi_j^*(\chi_\alpha-\Delta \chi_\alpha,t)}{2\Delta \chi_\alpha},
\label{2.2.13a}
\end{equation}
\begin{equation}
 \frac{\partial \Pi_j^*}{\partial \chi_\alpha} \Bigg{|}_u=
 \begin{cases}
 \dfrac{3\Pi_j^*(\chi_\alpha,t)-4\Pi_j^*(\chi_\alpha-\Delta \chi_\alpha,t)+\Pi_j^*(\chi_\alpha-2\Delta \chi_\alpha,t)}{2\Delta \chi_\alpha}, \quad &if \quad {c_{i\alpha} \geq 0},\\
 -\dfrac{3\Pi_j^*(\chi_\alpha,t)-4\Pi_j^*(\chi_\alpha+\Delta \chi_\alpha,t)+\Pi_j^*(\chi_\alpha+\Delta \chi_\alpha,t)}{2\Delta \chi_\alpha}, \quad &if \quad {c_{i\alpha}< 0}.
 \end{cases}
 \label{2.2.13b}
\end{equation}
\end{subequations}
In the calculation, the time
step $\triangle t$ is given by the $CFL$ condition number,
\begin{equation}
 \triangle t=CFL\frac{\triangle x}{c},
\label{eq:3.1}
\end{equation}
where $\triangle x$ is the minimum grid scale, and $c=|\bm c_j|$.
Courant-Friedrichs-Lewy ($CFL$) condition number is an important parameter to evaluate the stability
and convergence of method.
The evolution process of the MDF-FDLBM is shown in Fig.~\ref{MRT}, and can
be listed as follows.\\
Step (1): Calculate $\hat{\bm f}^+(\bm x,t)$ by Eq.\eqref{2.2.14}, and calculate $\bar{\bm f}^+(\bm x,t)$ by Eq.\eqref{2.2.15}.
\begin{equation}
 \hat{\bm f}^+  = (\bm I-\bm \Lambda)\hat{\bm f}+\bm\Lambda\bm f^{eq},
\label{2.2.14}
\end{equation}
where $\bm\Lambda =\bm I-\left(\bm I-\frac{\Delta t}{2}\tilde{\bm \Lambda}\right)\left(\bm I+\frac{\Delta t}{2}\tilde{\bm \Lambda}\right)^{-1}.$
\begin{equation}
 \bar{\bm f}^+  =  ( \bm I - \frac{3 \bm \Lambda}{4}) \hat{\bm f} + \frac{3 \bm \Lambda}{4}\bm f^{eq}=\frac{3}{4} \hat{\bm f}^++\frac{1}{4}\bm \hat{f}.
\label{2.2.15}
\end{equation}\\
Step (2): Conducting the second evaluation function Eq. \eqref{2.2.11}.\\
Step (3): Evaluate spatial gradient term $\bm c\cdot \nabla \bm f(\bm
x,t+h)$ by Eqs. \eqref{2.2.16} and \eqref{2.2.12}.
\begin{displaymath}
 \bar{\bm f}(\bm x,t+h)\stackrel{\eqref{2.2.16}}{\longrightarrow}\bm f(\bm x,t+h)\stackrel{\eqref{2.2.12}}{\longrightarrow}\nabla \bm f(\bm x,t+h),
\end{displaymath}
\begin{equation}
\bm f=( \bm I + \frac{h }{2}\tilde{\bm \Lambda})^{-1}
      ( \bar{\bm f}+\frac{h }{2}\tilde{\bm \Lambda}\bm f^{eq} ),
\label{2.2.16}
\end{equation}\\
Step (4): Compute $\hat{\bm f}(\bm x,t+\Delta t)$ by Eq. \eqref{2.2.1}.

\subsection{The Chapman Enskog analysis of the MDF-FDLBM}

First, the CE analysis is conducted to the main evolution
equation \eqref{2.2.1} to recover the NCDE. The distribution function
$f_{j,\alpha}$, the source terms $F_{j,\alpha}, G_{j,\alpha}$, the time and space derivatives can be expanded as,
\begin{eqnarray}
f_{j,\alpha}=f_{j,\alpha}^{(0)}+\varepsilon f_{j,\alpha}^{(1)}+\varepsilon^2f_{j,\alpha}^{(2)},\quad \partial_t=\varepsilon\partial_{t_1}+\varepsilon^2\partial_{t_2},\quad\nabla=\varepsilon\nabla_1,\nonumber\\
\tilde{G}_{j,\alpha}=\varepsilon\tilde{G}_{j,\alpha}^{(1)}+\varepsilon^2\tilde{G}_{j,\alpha}^{(2)},\quad F_{j,\alpha}=\varepsilon F_{j,\alpha}^{(1)}+\varepsilon^2 F_{j,\alpha}^{(2)}.
\label{2.3.1}
\end{eqnarray}
\begin{eqnarray}
\partial_tf_{j,\alpha}+\frac{\Delta t}{2}\partial_t^2f_{j,\alpha}+\bm c_j\cdot \nabla f_{j,\alpha}+\bm c_j\cdot \nabla \frac{\Delta t}{2} \partial_t f_{j,\alpha}  =  \left[-\tilde{\Lambda}_{jk}(f_{k,\alpha}-f_{k,\alpha}^{eq})\right] & &\nonumber\\
+\frac{\Delta t}{2}\partial_t\left[ -\tilde{\Lambda}_{jk}(f_{k,\alpha}-f_{k,\alpha}^{eq}) \right] +F_{j,\alpha}+G_{j,\alpha}+\frac{\Delta t}{2}(\partial_tF_{j,\alpha}+\partial_tG_{j,\alpha})& &.
\label{2.3.2}
\end{eqnarray}
Substituting  Eq. \eqref{2.3.1} into Eq.~\eqref{2.3.2} yields
\begin{subequations}
\begin{equation}
O(\varepsilon^0): -\tilde{\Lambda}_{jk}(f_k^{(0)}-f_k^{eq})=0\Leftrightarrow f_j^{(0)}=f_j^{eq}
\label{2.3.3a}
\end{equation}
\begin{equation}
O(\varepsilon^1): \partial_{t_1}f_j^{(0)}+\bm c_j\cdot \nabla_1f_j^{(0)}=-\tilde{\Lambda}_{jk}f_k^{(1)}+G_j^{(1)}+F_j^{(1)},
\label{2.3.3b}
\end{equation}
\begin{eqnarray}
&   &O(\varepsilon^2): \partial_{t_2}f_j^{(0)}+\partial_{t_1}f_j^{(1)}+\frac{\Delta t}{2}\partial_{t_1}^2f_j^{(0)}+\bm c_j\cdot \nabla_1f_j^{(1)}+\bm c_j\cdot \nabla_1\frac{\Delta t}{2}\partial_{t_1}f_j^{(0)}\nonumber\\
& = & -\tilde{\Lambda}_{jk}f_k^{(2)}+\frac{\Delta t}{2}\partial_{t_1}(-\tilde{\Lambda}_{jk}f_k^{(1)})+G_j^{(2)}+F_j^{(2)}+\frac{\Delta t}{2}(\partial_{t_1}F_j^{(1)}+\partial_{t_1}G_j^{(1)}).
\label{2.3.3c}
\end{eqnarray}
\label{eq:34}
\end{subequations}

According to Eqs. \eqref{2.2.5}, \eqref{2.2.6} and \eqref{2.2.7}, one can determine the moment conditions of $f_{i,\alpha}^{eq}$, $G_{i,\alpha}$ and $F_{i,\alpha}$,
\begin{equation}\label{2.3.4}
  \sum_if_{i,\alpha}^{eq}=\bar{u}_\alpha,\quad \sum_i\bm c_if_{i,\alpha}^{eq}=\bar{u}_\alpha\bm u+P\bm E_\alpha, \quad \sum_i\bm c_i\bm c_if_{i,\alpha}^{eq}=\bar{u}_\alpha c_s^2\bm I
\end{equation}
\begin{equation}\label{2.3.5}
  \sum_i G_{i,\alpha}=0, \quad \sum_i \bm c_iG_{i,\alpha}=\partial_t(\bar{u}_\alpha\bm u+P\bm E_\alpha),
\end{equation}
\begin{equation}\label{2.3.6}
  \sum_iF_{i,\alpha}=\bar{F}_\alpha,\quad \sum_i\bm c_iF_{i,\alpha}=0.
\end{equation}
Based on the moment conditions, we can get the following equations through summing Eqs.~\eqref{2.3.3b} and \eqref{2.3.3c} over $j$,
\begin{subequations}
\begin{equation}
\partial_{t_1}\bar{u}_\alpha+\nabla_1\cdot (\bar{u}_\alpha\bm u+P\bm E_\alpha)=F_\alpha^{(1)},
\label{2.3.7a}
\end{equation}
\begin{equation}
\partial_{t_2}\bar{u}_\alpha+\frac{\Delta t}{2}\partial_{t_1}^2\bar{u}_\alpha+\nabla_1\cdot \sum \bm c_jf_j^{(1)}+\frac{\Delta t}{2}\nabla_1\cdot \partial_{t_1}(\bar{u}_\alpha\bm u+P\bm E_\alpha)=F_\alpha^{(2)}+\frac{\Delta t}{2}\partial_{t_1}F_\alpha^{(1)}.
\label{2.3.7b}
\end{equation}
\label{2.3.7}
\end{subequations}
According to Eq.~\eqref{2.3.7a}, one can obtain
\begin{equation}
\partial_{t_1}^2\bar{u}_\alpha=\partial_{t_1}F_\alpha^{(1)}-\partial_{t_1}\nabla_1\cdot (\bar{u}_\alpha\bm u+P\bm E_\alpha).
\label{2.3.8}
\end{equation}
If the derivatives of time and space are interchangeable, Eq.~\eqref{2.3.7b} can be
rewritten as
\begin{equation}
\partial_{t_2}\bar{u}_\alpha=F_\alpha^{(2)}-\nabla_1\cdot\sum \bm c_jf_j^{(1)}.
\label{2.3.9}
\end{equation}
With the aid of Eq.~\eqref{2.2.4}, one can get the following equation through
multiplying the Eq.~\eqref{2.3.3b} by $\bm c_j$ and summing it over $j$,

\begin{equation}
\partial_{t_1}(\bar{u}_\alpha\bm u+P\bm E_\alpha)+\nabla_1\cdot c_s^2\bar{u}_\alpha\bm I=-\tilde{\bm S}\sum \bm c_jf_j^{(1)}+\bm M_{1,G}^{(1)},
\label{2.3.10}
\end{equation}
where  $\bm M_{1,G}^{(1)}=\sum \bm
c_jG_j^{(1)}$, and $\tilde{\bm S}$ is a D-dimensional diagonal matrix, which ia related to the diffusion coefficient $\nu$.
The relationship are $tilde{\bm S}=s\bm I$ and $\frac{c_s^2}{s}=\nu$.
From Eq.~\eqref{2.3.10}, one can obtain
\begin{equation}
\sum \bm c_jf_j^{(1)}=-\tilde{\bm S}^{-1}\left[\partial_{t_1}(\bar{u}_\alpha\bm u+P\bm E_\alpha)+\nabla_1\cdot c_s^2\bar{u}_\alpha\bm I-\bm M_{1,G}^{(1)}\right],
\label{2.3.11}
\end{equation}
and
\begin{equation}
\partial_{t_2}\bar{u}_\alpha = F_\alpha^{(2)}+\nabla_1\cdot c_s^2\tilde{\bm S}^{-1}\nabla_1\cdot u_\alpha\bm I+\nabla_1\cdot \tilde{\bm S}^{-1}\left[\partial_{t_1}(\bar{u}_\alpha\bm u+P\bm E_\alpha)-M_{1,G}^{(1)} \right].\\
\label{2.3.12}
\end{equation}
Because
\begin{equation}
\bm M_{1,G}=\partial_{t}(\bar{u}_\alpha\bm u+P\bm E_\alpha),
\label{2.3.13}
\end{equation}
 if $\nu\bm I= c_s^2\tilde{\bm S}^{-1}$ or $\nu=c_s^2/s$, with the help of Eq. \eqref{2.3.12}, one can deduce
\begin{equation}\label{2.3.14}
\partial_{t_2}\bar{u}_\alpha = F_\alpha^{(2)}+\nabla_1\cdot \nu  \nabla_1\cdot \bm u_\alpha\bm I.
\end{equation}
Through combining the results at $\varepsilon$ and $\varepsilon^2$
scales, i.e., Eqs. \eqref{2.3.7a} and \eqref{2.3.14}, we
can recover the convection-diffusion-system based Navier-Stokes-equations~\eqref{2.1.2} correctly.

Next, we will recover the convection-diffusion-system based NSEs~\eqref{2.1.2} from the second evolution equation
\eqref{2.2.8} with the CE analysis. With the Taylor expansion, Eq.
\eqref{2.2.8} can be expressed as
\begin{equation}
D_jf_{j,\alpha}+\frac{h}{2}D^2_jf_{j,\alpha}=-\tilde{\Lambda}_{jk}f_{k,\alpha}^{ne}+\frac{h}{2}(-\tilde{\Lambda}_{jk}D_kf_{k,\alpha}^{ne})+\bar{G}_{j,\alpha}+F_{j,\alpha}+\frac{h}{2}(\partial_tF_{j,\alpha}+\partial_t\bar{G}_{j,\alpha}).
\label{2.3.15}
\end{equation}
Then we can expand Eq.~\eqref{2.3.15} at different orders of $\varepsilon$,
\begin{subequations}
\begin{equation}
O(\varepsilon^0): -\tilde{\Lambda}_{jk}(f_{k,\alpha}^{(0)}-f_{k,\alpha}^{eq})=0,\Leftrightarrow f_{j,\alpha}^{(0)}=f_{j,\alpha}^{eq},
\label{2.3.16a}
\end{equation}
\begin{equation}
O(\varepsilon^1): D_{1j}f^{(0)}_{j,\alpha}=-\tilde{\Lambda}_{jk}f^{(1)}_{k,\alpha}+\bar{G}^{(1)}_{j,\alpha}+F^{(1)}_{j,\alpha},
\label{2.3.16b}
\end{equation}
\begin{equation}
\begin{split}
O(\varepsilon^2): \partial_{ t_2}f_{j,\alpha}^{(0)}+D_{1j}f_{j,\alpha}^{(1)}+\frac{h}{2}D_{1j}^2f_{j,\alpha}^{(0)}=&-\tilde{\Lambda}_{jk}f^{(2)}_{k,\alpha}+\frac{h}{2}(-\tilde{\Lambda}_{jk}D_{1k}f_{k,\alpha}^{(1)})+\bar{G}^{(2)}_{j,\alpha}+F^{(2)}_{j,\alpha}\\
&+\frac{h}{2}(\partial_{t_1}F_{j,\alpha}^{(1)}+\partial_{t_1}\bar{G}_{j,\alpha}^{(1)}).
\label{2.3.16c}
\end{split}
\end{equation}
\label{2.3.16}
\end{subequations}
With the help of Eq.~\eqref{2.3.16b}, Eq.~\eqref{2.3.16c} can be rewritten as
\begin{equation}
\partial_{ t_2}f_{j,\alpha}^{(0)}+D_{1j}f_{j,\alpha}^{(1)}+\frac{h}{2}(\nabla_1\cdot \bm c_jF_{j,\alpha}^{(1)}+\nabla_1\cdot \bm c_j\bar{G}_{j,\alpha}^{(1)})=-\tilde{\Lambda}_{jk}f^{(2)}_{k,\alpha}+\bar{G}^{(2)}_{j,\alpha}+F^{(2)}_{j,\alpha}.
\label{2.3.17}
\end{equation}
Summing Eqs.~\eqref{2.3.17} and \eqref{2.3.16b} over $j$, one can obtain
\begin{subequations}
\begin{equation}
\partial_{t_1}\bar{u}_\alpha+\nabla_1\cdot (\bar{u}_\alpha\bm u+P\bm E_\alpha)=F_\alpha^{(1)},
\label{2.3.18a}
\end{equation}
\begin{equation}
\partial_{t_2}\bar{u}_\alpha+\nabla_1\cdot \sum \bm c_jf_{j,\alpha}^{(1)}=F_\alpha^{(2)}-\frac{h}{2}\nabla_1\cdot M_{1\bar{G}}^{(1)},
\label{2.3.18b}
\end{equation}
\label{2.3.18}
\end{subequations}
where $M_{1\bar{G}}^{(1)}$ is the first-order moment condition of source term $\bar{G}$.
Multiplying the Eq.~\eqref{2.3.16b} by $\bm c_j$ and summing it over $j$, we
can derive
\begin{equation}
\sum \bm c_jf_j^{(1)}=-\tilde{\bm S}^{-1}[\partial_{t_1}(\bar{u}_\alpha\bm u+P\bm E_\alpha)+\nabla_1\cdot \bar{u}_\alpha c_s^2\bm I-\bm M_{1,\bar{G}}^{(1)}],
\label{2.3.19}
\end{equation}
where Eq.~\eqref{2.2.4} has been used. Substituting above equation and $M_{1,\bar{G}}^{(1)}=(\bm I+\frac{h}{2}\tilde{\bm S})^{-1}\partial_{t_1}(\bar{u}_\alpha\bm u+P\bm E_\alpha)$ into
Eq.~\eqref{2.3.18b}, we have
\begin{equation}
\partial_{t_2}\bar{u}_\alpha = F_\alpha^{(2)}+\nabla_1\cdot \nu  \nabla_1\cdot \bm u_\alpha\bm I,
\label{2.3.20}
\end{equation}
where $\nu \bm I=\tilde{\bm S}^{-1}c_s^2$ and $\nu=c_s^2/s$.
According to the
results at $\varepsilon$ and $\varepsilon^2$ scales, i.e.,
Eqs.~\eqref{2.3.18a} and \eqref{2.3.20}, Eq.~\eqref{2.1.2} can be recovered
exactly.

Then we will discuss the calculation the pressure $P$. According to Eq. \eqref{2.3.8}, one can obtain
\begin{eqnarray}\label{2.3.21}
  \sum_i\bm c_{i,\alpha}f_{i,\alpha}^{ne}=&\sum_i\bm c_{i,\alpha}f_{i,\alpha}-\sum_i\bm c_{i,\alpha}f_{i,\alpha}^{eq}\nonumber\\
  =& \sum_i\bm c_{i,\alpha}f_{i,\alpha}-(\bar{u}_\alpha u_\alpha +P)\nonumber\\
  =& -\frac{c_s^2}{s}\nabla_\alpha \bar{u}_\alpha,
\end{eqnarray}
if we sum Eq. \eqref{2.3.21} over $\alpha$ and adopt the continuity equation in Eq. \eqref{2.1.1}, one can obtain
\begin{equation}\label{2.3.22}
  \sum_{\alpha=1}^d\sum_i\bm c_{i,\alpha}f_{i,\alpha}-(|\bm u|^2+dP)=-\frac{c_s^2}{s}R,
\end{equation}
From Eq. \eqref{2.3.22}, we can get
\begin{equation}\label{2.3.23}
  P=\frac{1}{d}(\sum_{\alpha=1}^d\sum_i\bm c_{i,\alpha}f_{i,\alpha}+\frac{c_s^2}{s_1}R-|\bm u|^2).
\end{equation}

\textbf{Remark 1:} In the above work, if Eq. \eqref{2.3.22} is ensured, the continuity equation \eqref{2.1.1} must be satisfied. It indicated that continuity equation \eqref{2.1.1}
can not be considered and the evolution equations \eqref{2.2.1} and \eqref{2.2.8} with $\alpha=0$ can be ignored.

\textbf{Remark 2:} Because the source terms $G_{j,\alpha}$ and $\bar{G}_{j,\alpha}$ are on the order of $O(Ma^2)$, and the coefficient of the $G_{j,\alpha}$ and $\bar{G}_{j,\alpha}$ the is $\Delta t$ and $h$ in the two evolution equations. Thus the source terms $\Delta t G$ and $h G$ are on the order of $O(\delta tMa^2)$, which can be neglected in the computation. In Eq. \eqref{2.2.8},
it can be found that the terms $\frac{h^2}{2}\partial_t\bar{F}_{j,\alpha}$ and $\frac{h^2}{2}\partial_t\bar{G}_{j,\alpha}$ are on the order of $O(\delta t^2)$, thus, they can be omitted.
In this case, the evolution equations \eqref{2.2.1} and \eqref{2.2.8} can be rewritten as
\begin{equation}\label{2.3.24}
\begin{split}
  \hat{f}_{j,\alpha}(\bm x,t+\Delta t)=& \hat{f}^+_{j,\alpha}(\bm x,t)-\Delta t\bm c_j\cdot\nabla f_{j,\alpha}(\bm x,t+\frac{1}{2}\Delta t)\\
  & +\Delta tF_{j,\alpha}(\bm x,t)+\frac{\Delta t^2}{2}\partial_tF_{j,\alpha}(\bm x,t),
  \end{split}
\end{equation}
and
\begin{equation}
 \bar{f}_{j,\alpha}(\bm x,t+h)=\bar{f}^+_{j,\alpha}(\bm x-\bm c_jh,t)+hF_{j,\alpha}.
 \label{2.3.25}
\end{equation}

\subsection{The computation of some physical quantities}

For MDF-FDLBM, some quantities can be directly computed by the first-order moment of the non-equilibrium distribution function.
According to Eq. \eqref{2.3.21}, the velocity gradient can be obtained
\begin{equation}\label{2.4.1}
  \nabla_\beta u_\alpha=-\frac{s_1}{c_s^2}\sum_i \bm c_{i,\beta}f^{ne}_{i,\alpha}, \quad (\alpha\neq 0).
\end{equation}
The velocity divergence $\nabla\cdot \bm u$, strain rate tensor $\bm S$, shear tress $\sigma$ and the antisymmetric velocity gradient tensor $\bm \Omega$ can be derived
\begin{equation}\label{2.4.2}
  \nabla\cdot \bm u=\sum^d_{\alpha=1}\nabla_\alpha u_\alpha=-\frac{s_1}{c_s^2}\sum^d_{\alpha=1}\sum_i\bm c_{i,\alpha}f^{ne}_{i,\alpha},
\end{equation}
\begin{equation}\label{2.4.3}
  S_{\alpha\beta}=\frac{1}{2}(\nabla_\beta u_\alpha)(\nabla_\alpha u_\beta)=-\frac{s_1}{2c_s^2}\sum_i(\bm c_{i,\beta}f^{ne}_{i,\alpha}+\bm c_{i,\alpha}f^{ne}_{i,\beta}), \quad (\alpha\neq0,\beta\neq0),
\end{equation}
\begin{equation}\label{2.4.4}
  \sigma_{\alpha\beta}=2\rho_0\nu S_{\alpha\beta}=-\rho_0\sum_i(\bm c_{i,\beta}f^{ne}_{i,\alpha}+\bm c_{i,\alpha}f^{ne}_{i,\beta}),  \quad (\alpha\neq0,\beta\neq0),
\end{equation}
\begin{equation}\label{2.4.5}
  \Omega_{\alpha\beta}=\frac{1}{2}(\nabla_\beta u_\alpha-\nabla_\alpha u_\beta)=-\frac{s_1}{2c_s^2}\sum_i(\bm c_{i,\beta}f^{ne}_{i,\alpha}-\bm c_{i,\alpha}f^{ne}_{i,\beta}),  \quad (\alpha\neq0,\beta\neq0).
\end{equation}

\section{\label{sec:level3}Stability analysis of the MDF-FDLBM }
In this section, we will discuss the stability of the MDF-FDLBM.
To compare the stability of the MDF-FDLBM and IFDLBM \cite{IFDLBM}, the linear stability analysis is also implemented to the IFDLBM.
And the stability region of MDF-FDLBM and IFDLBM are compared under different conditions.

\subsection{ The stability analysis of the MDF-FDLBM}

As a common tool to evaluate the numerical stability of LBMs, the Von Neumann method will be used for the MDF-FDLBM.
For simplicity, the source terms will be neglected.
There are $d+1$ evolution equations in MDF-FDLBM.
However, as it mentioned in \textbf{Remark 2}, only the evolution equations $(\alpha=1,2,...,d)$ affects the stability.
Because the evolution equation of $\alpha=0$ do not influence the calculation of $P$ and $\bm u$.
Therefore, only the evolution equations of $\alpha=1,2,...,d$ need to be analyzed.
For brevity, we consider the D2Q5 model for example,
\begin{subequations}
\begin{equation}
\textbf{c}_{j}=\left(
\begin{matrix}
   0& 1 &0 &-1 &0 \\
   0& 0& 1 &0 &-1 \\
  \end{matrix}
  \right)c,\\
\label{3.1.0}
\end{equation}
\begin{equation}
w_{j=0-5}=1/5.
\end{equation}
\end{subequations}
Starting from the following two evolution equations,
\begin{eqnarray}\label{3.1.1}
  f_j(\bm x,t+\Delta t)+0.5\bm\omega_{jk}\left[f_k-f^{eq}_k\right](\bm x,t+\Delta t)&=& f_j(\bm x,t)-0.5\bm\omega_{jk}\left[f_k(\bm x,t)-f^{eq}_k(\bm x,t)\right]\nonumber\\
  &&-\Delta t\bm c_j\cdot\nabla f_j(\bm x,t+\frac{\Delta t}{2} ),
\end{eqnarray}
\begin{eqnarray}\label{3.1.2}
  g_j(\bm x,t+\Delta t)+0.5\bm\omega_{jk}\left[g_k-g^{eq}_k\right](\bm x,t+\Delta t)&=& g_j(\bm x,t)-0.5\bm\omega_{jk}\left[g_k(\bm x,t)-g^{eq}_k(\bm x,t)\right]\nonumber\\
  &&-\Delta t\bm c_j\cdot\nabla g_j(\bm x,t+\frac{\Delta t}{2} ),
\end{eqnarray}
where the $\bm\omega_{jk}=\Delta t \tilde{\bm\Lambda}_{jk}$. It is related to the relaxation time matrix $\bm S=(s_0,s_1,s_2,s_3,s_4)$.
Then we introduce the following linear expansion,
\begin{equation}\label{3.1.3}
  f_j(\bm x,t)=\overline{f_j^{eq}(\bm x,t)}+f_j'(\bm x,t),\quad g_j(\bm x,t)=\overline{g_j^{eq}(\bm x,t)}+g_j'(\bm x,t),
\end{equation}
where $g_j'(\bm x,t)$ and $f_j'(\bm x,t)$ are the fluctuating quantities. The global equilibrium distribution $\overline{g_j^{eq}(\bm x,t)}$ and $\overline{f_j^{eq}(\bm x,t)}$ are the constants and they will not change over time and space.
Substituting Eq. \eqref{3.1.3} into Eqs. \eqref{3.1.2} and \eqref{3.1.1}, then adding the generation equations, one can get
\begin{equation}
\begin{split}
\left[(1+\frac{\bm\omega_{jk}}{2})\delta_{jk}+\frac{\bm\omega_{jk}}{2}\Gamma1_{jk}\right]f'_j(\bm x,t+\Delta t)+\left[(1+\frac{\bm\omega_{jk}}{2})\delta_{jk}+\frac{\bm\omega_{jk}}{2}\Gamma2_{jk}\right]g'_j(\bm x,t+\Delta t) =& \\ \left[(1-\frac{\bm\omega_{jk}}{2})\delta_{jk}+\frac{\bm\omega_{jk}}{2}\Gamma1_{jk}\right]f'_k(\bm x,t)-\delta_{jk}\Delta t \bm c_k\cdot\nabla f'_j\left(\bm x,t+\frac{\Delta t}{2}\right)&\\
 +\left[(1-\frac{\bm\omega_{jk}}{2})\delta_{jk}+\frac{\bm\omega_{jk}}{2}\Gamma2_{jk}\right]g'_k(\bm x,t)-\delta_{jk}\Delta t \bm c_k\cdot\nabla g'_j\left(\bm x,t+\frac{\Delta t}{2}\right)&,
\label{3.1.4}
\end{split}
\end{equation}
where the $\bm\Gamma1_{jk}$ and $\bm\Gamma2_{jk}$ can be calculated by
\begin{equation}\label{3.1.5}
  \bm\Gamma1_{jk}=\frac{\partial f_j^{eq}(\bm{x},t)}{\partial f_k(\bm{x},t)}+\frac{\partial g_j^{eq}(\bm{x},t)}{\partial f_k(\bm{x},t)},
  \bm\Gamma2_{jk}=\frac{\partial f_j^{eq}(\bm{x},t)}{\partial g_k(\bm{x},t)}+\frac{\partial g_j^{eq}(\bm{x},t)}{\partial g_k(\bm{x},t)}.
\end{equation}
The $\bm\Gamma1_{jk}$ and $\bm\Gamma2_{jk}$ contain the derivative of the equilibrium distribution function, which is difficult to calculate directly.
The chain rule can be used to convert the derivative. For the D2Q5 model, $\bm u=(u_1,u_2)$. Then we can deduce
\begin{equation}\label{3.1.6}
\begin{split}
  \frac{\partial f_j^{eq}}{\partial f_k} & = \frac{\partial f_j^{eq}}{\partial u_1}\frac{u_1}{\partial f_k} =w_j\left(1+\frac{c_j\cdot \bm u+\frac{1}{2} c_{j,1} c_{j,1}}{c_s^2}\right),\\
  \frac{\partial f_j^{eq}}{\partial g_k}&=\frac{\partial f_j^{eq}}{\partial u_2}\frac{u_2}{\partial g_k} =\frac{w_j}{c_s^2}\left(c_{j\beta}u_\alpha-c_{j\alpha}u_\beta+\frac{1}{2}c_{j\alpha}c_{j\beta}\right),\\
  \frac{\partial g_j^{eq}}{\partial g_k}&=\frac{\partial g_j^{eq}}{\partial u_1}\frac{u_1}{\partial f_k} =w_j\left(1+\frac{c_j\cdot \bm u+\frac{1}{2}\bm c_{j\beta}\bm c_{j\beta}}{c_s^2}\right),\\
  \frac{\partial g_j^{eq}}{\partial f_k} & =\frac{\partial g_j^{eq}}{\partial u_2}\frac{u_2}{\partial g_k} =\frac{w_j}{c_s^2}\left(c_{j\alpha}u_\beta-c_{j\beta}u_\alpha+\frac{1}{2}c_{j\alpha}c_{j\beta}\right).
  \end{split}
\end{equation}
Let $S'_k(t)=[F'_k(t+\frac{\Delta t}{2}),G'_k(t+\frac{\Delta t}{2}),F'_k(t),G'_k(t)]^T$, where $F'_k(\bm \kappa,t)=\int f'_k(\bm x,t) \exp (-i\bm \kappa\cdot \bm x)d\bm x$ and $G'_k(\bm \kappa,t)=\int g'_k(\bm x,t) \exp (-i\bm \kappa\cdot \bm x)d\bm x$.
Applying the Fourier transform to Eq. \eqref{3.1.4}, the following equation is obtained
\begin{equation}\label{3.1.7}
  S'_k(\bm \kappa,t+\frac{1}{2}\Delta t)=\tilde{\bm G}S'_k(\bm \kappa,t),
\end{equation}
where the $\tilde{\bm G}$ is the growth matrix. Its definition is
\begin{equation}\label{3.1.8}
\begin{split}
  \tilde{\bm G}=&diag[\bm I+0.5\bm \omega(\bm I-\bm\Gamma1),\bm I+0.5\bm \omega(\bm I-\bm\Gamma2),\bm I,\bm I]^{-1} \\
  &diag[\bm I-r\bm T,\bm I-r\bm T,I-0.5\bm \omega(\bm I+\bm\Gamma1),I-0.5\bm \omega(\bm I+\bm\Gamma2)],
  \end{split}
\end{equation}
where $\bm \kappa=(\kappa_x,\kappa_y)$ is the wave number, $r=\Delta t/\Delta x$ and $\bm T=diag(T_0,T_1,...T_q)$. For the mixed
difference scheme,
\begin{equation}
\begin{split}
 T_j=&i(1-\eta)(\sin\vartheta_{jx}+\sin\vartheta_{jy})+\frac{\eta}{2}[6-4\exp(-\vartheta_{jx})-4\exp(-\vartheta_{jy})\\
 &+\exp(-2i\vartheta_{jx})+\exp(-2i\vartheta_{jy})].
\label{3.1.9}
\end{split}
\end{equation}
It should be noted that the growth matrix $\tilde{\bm G}$ is a $2q\times 2q$ diagonal matrix. For the D2Q5 model, the size of $\tilde{\bm G}$ is $10\times 10$.
Now, the stability of Eq. \eqref{3.1.1} is equal to the stability of the linear
system Eq. \eqref{3.1.7}. According to the von Neumann stability condition, if the spectral radius
of growth matrix $\tilde{\bm G}$ is less than 1, the linear system Eq. \eqref{3.1.7} is stable.
However, The growth matrix contains many variables, so it is necessary to determine some variables to analyze the numerical stability of the MDF-FDLBM.

\begin{figure}[htbp]
\centering
\subfigure[$s'=0.5s$]{\includegraphics[scale=0.5]{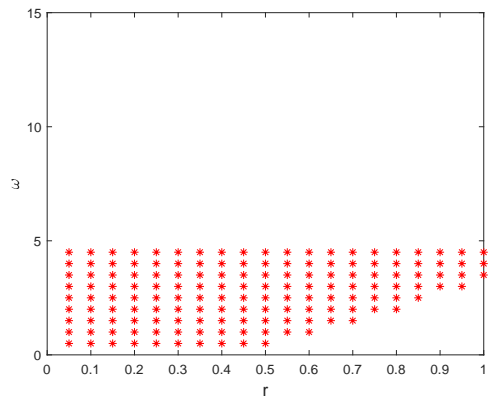}}
\subfigure[$s'=1.0s$]{\includegraphics[scale=0.5]{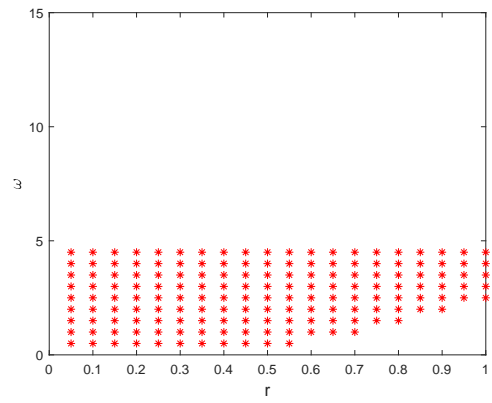}}

\subfigure[$s'=1.5s$]{\includegraphics[scale=0.5]{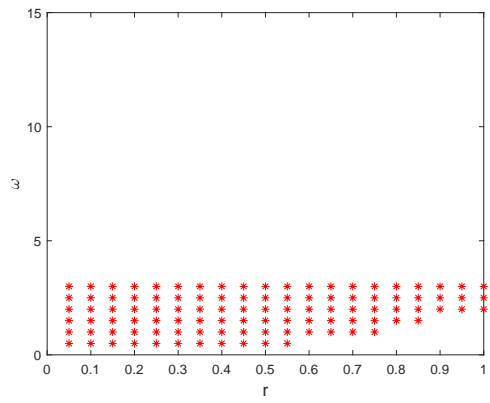}}
\subfigure[$s'=4.0s$]{\includegraphics[scale=0.5]{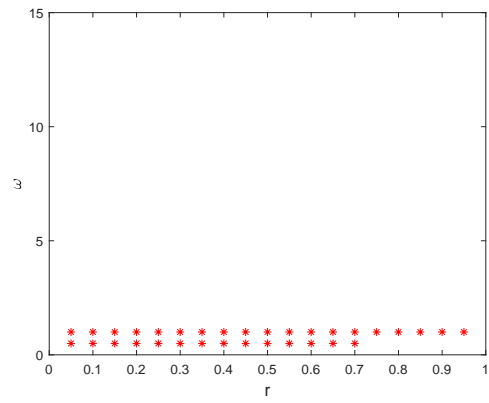}}

 \caption{The stability region of MDF-FDLBM with different $s'$ } \label{fig:stable-MDF-FDLBM-s}
\end{figure}

\begin{figure}[htbp]
\centering
\subfigure[$r=0.25$]{\includegraphics[scale=0.5]{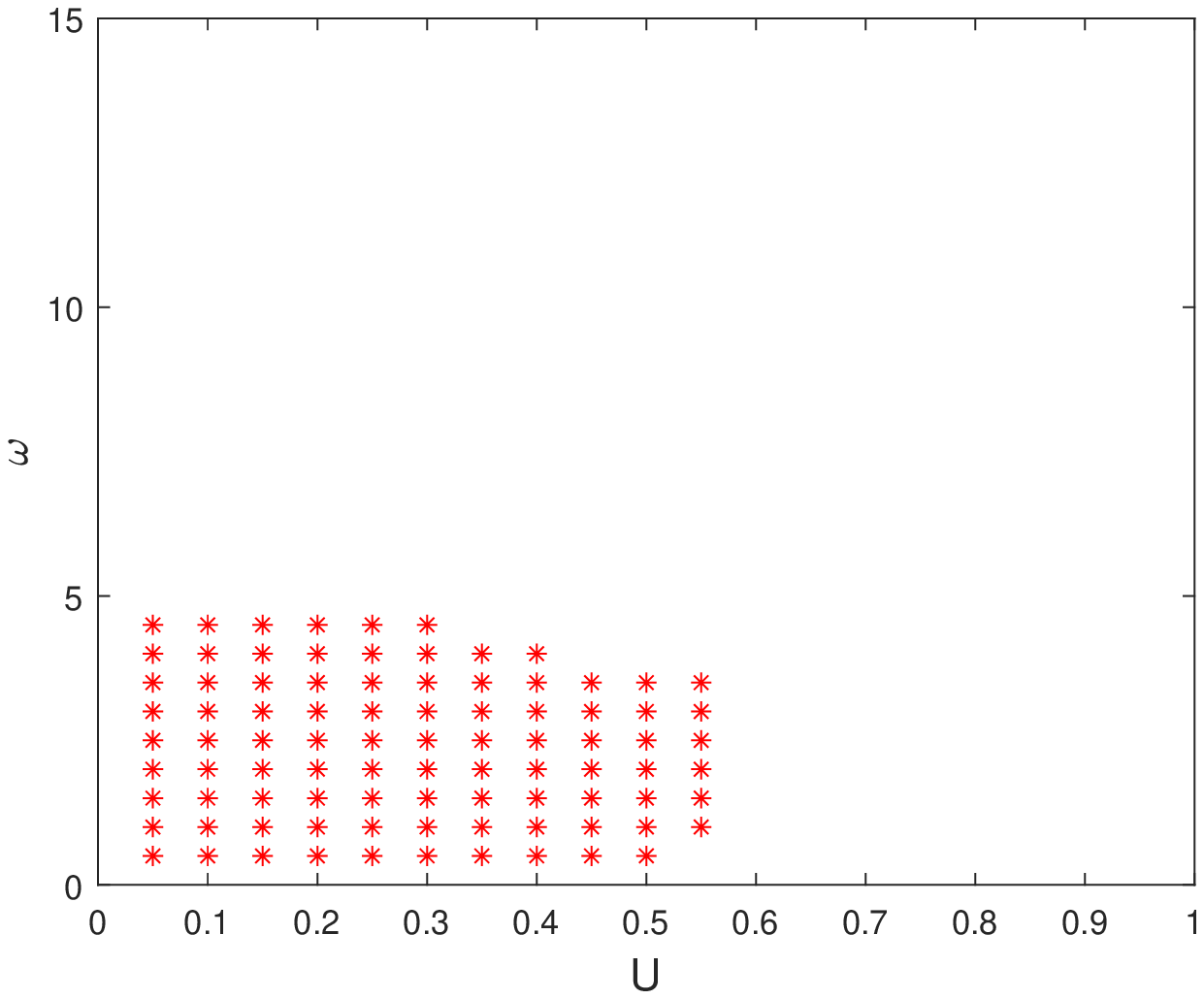}}
\subfigure[$r=0.5$]{\includegraphics[scale=0.5]{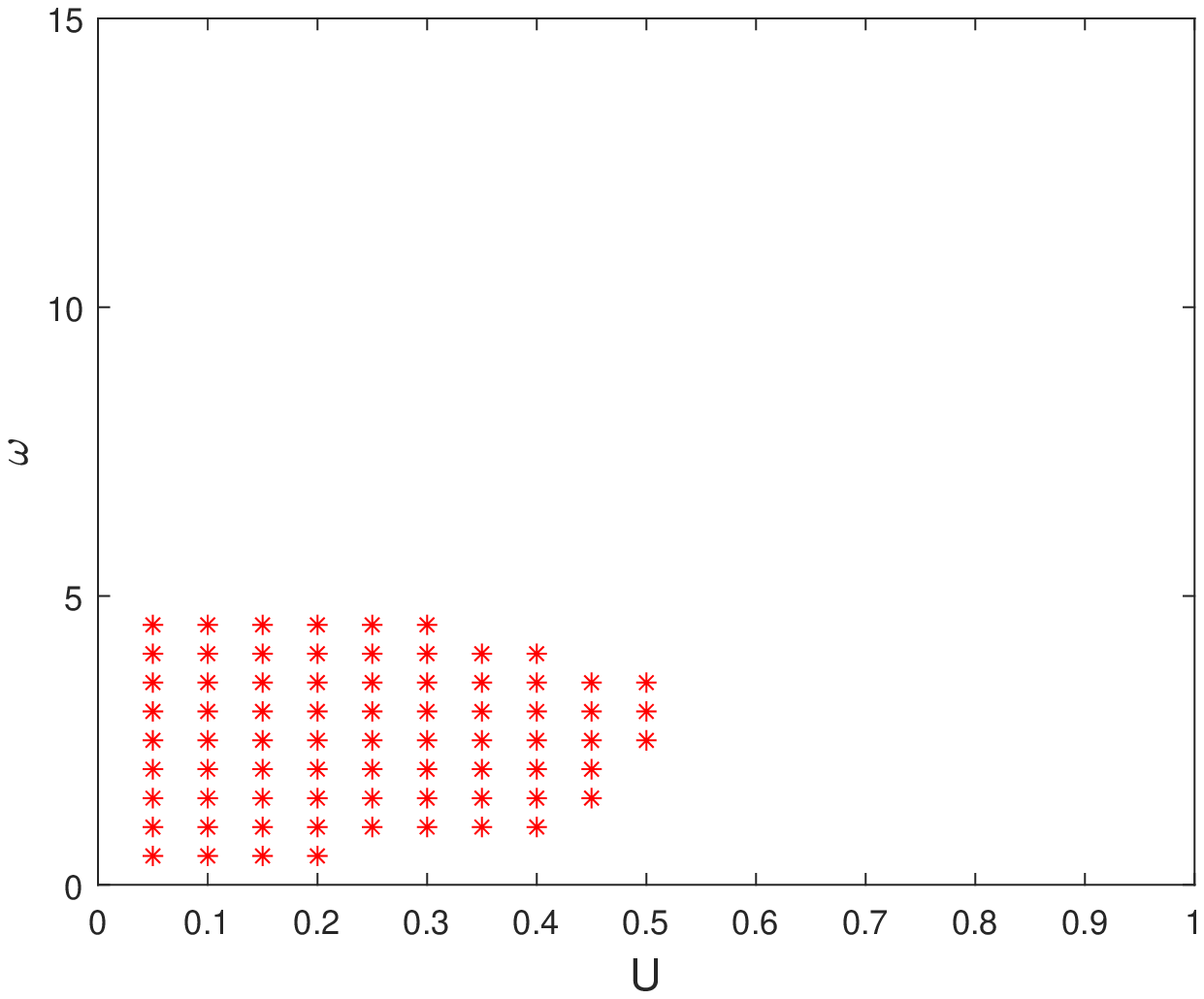}}

\subfigure[$r=0.75$]{\includegraphics[scale=0.5]{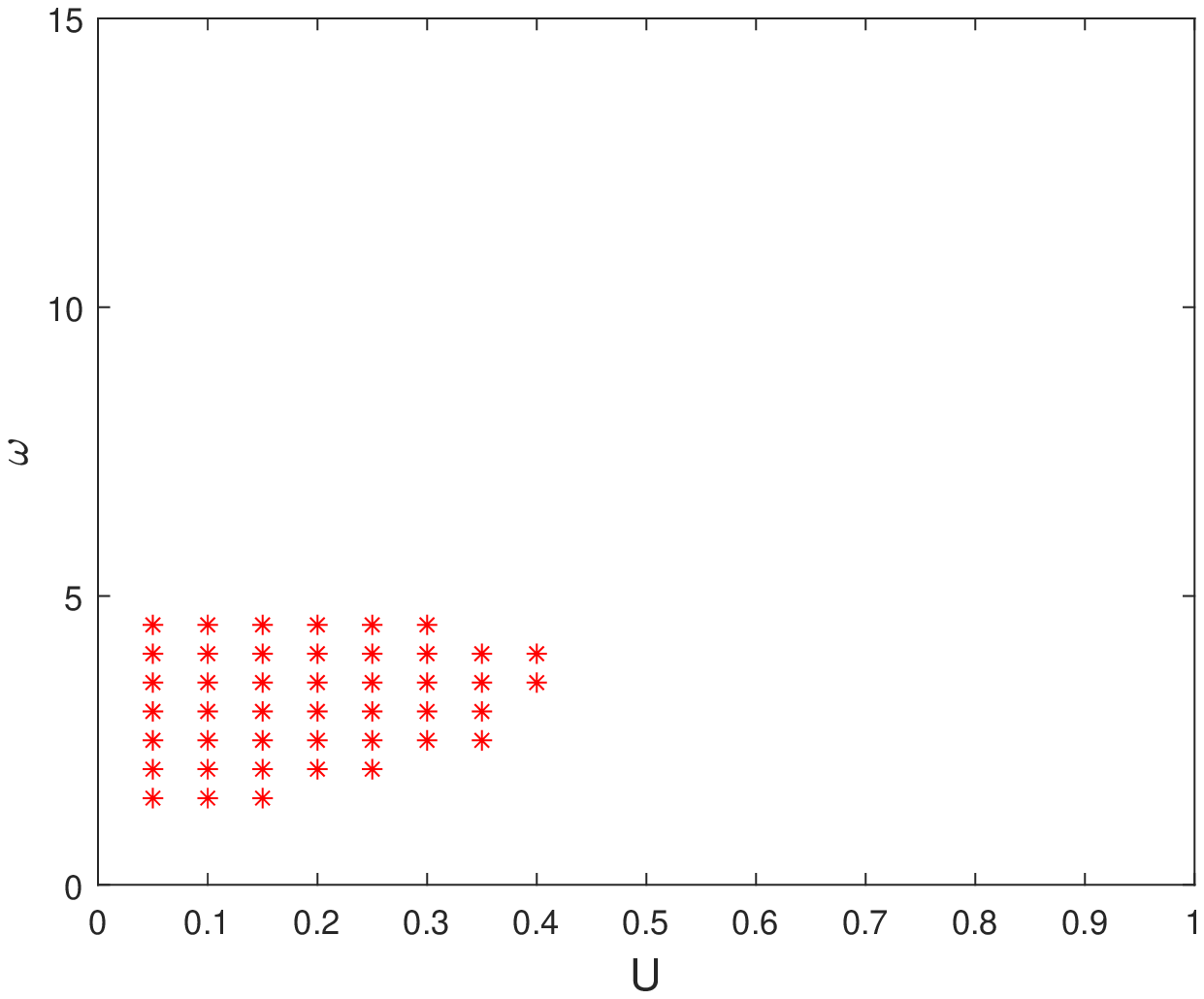}}
\subfigure[$r=0.9$]{\includegraphics[scale=0.5]{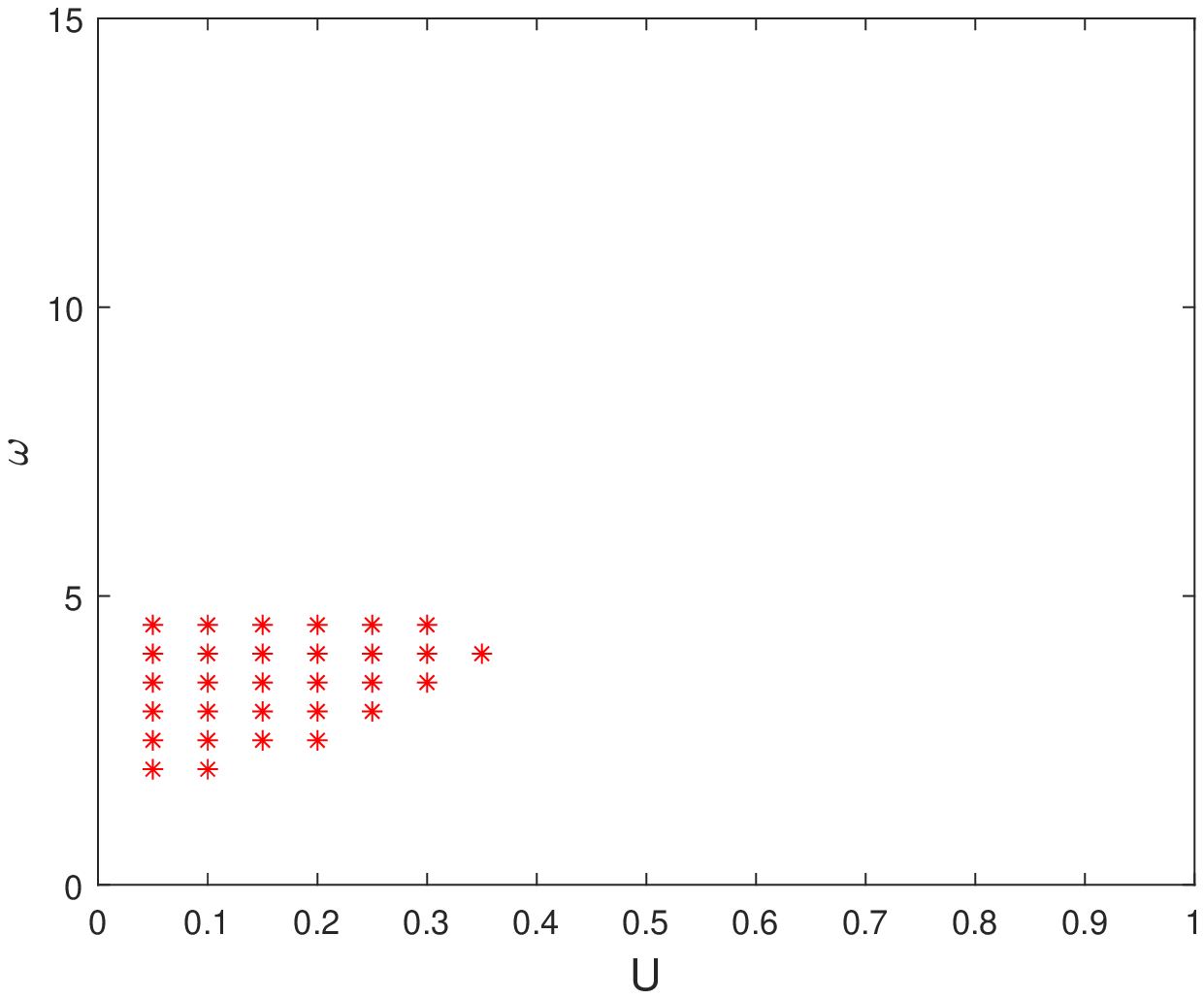}}

 \caption{The stability region of MDF-FDLBM with different CFL condition number } \label{fig:stable-MDF-FDLBM-r}
\end{figure}

Now, we consider the stability analysis of the linear system Eq. \eqref{3.1.7}. First of all, we set $\bm u=(0.5,0.5)^T$ and $c=1.0$.
For the multi-relaxation model, the relaxation parameters are set as $s_0=s_3=s_4=s'$ and $s_1=s_2=s$ for MDF-FDLBM. Then the stability regions are presented in Figs. \ref{fig:stable-MDF-FDLBM-s} and
\ref{fig:stable-MDF-FDLBM-r}.
It can be found that the largest area of the stability region appears when $s'=s$.
When $s'$ increases or decreases, the area of stability region of MDF-FDLBM will decreases.
Besides, the velocity $c$ is taken as $1$, so the CFL condition number is equal to $r$.
We also analysis the relationship between the stability region and the CFL condition number with $s'=1.5s$.
As shown in Fig. \ref{fig:stable-MDF-FDLBM-r}, the stability region areas decrease with the increase of $r$.

\subsection{ The stability analysis of the IFDLBM}

The von Neumann method is conducted to the IFDLBM. For simplicity, the source term is neglected and $D2Q9$ model is adopted.\\
D2Q9:
\begin{subequations}
\begin{equation}
\textbf{c}_{j}=\left(
\begin{matrix}
   0& 1 &0 &-1 &0 &1 &-1 &-1 &1\\
   0& 0& 1 &0 &-1  &1 &1 &-1 &-1\\
  \end{matrix}
  \right)c,\\
\label{3.2.0}
\end{equation}
\begin{equation}
w_0=4/9,w_{j=1-4}=1/9,w_{j=5-9}=1/36.
\end{equation}
\end{subequations}
According to the Ref. \cite{IFDLBM}, the evolution equation of IFDLBM without source term can be written as
\begin{eqnarray}\label{3.2.1}
  f_j(\bm x,t+\Delta t)+0.5\bm\omega_{jk}\left[f_k-f^{eq}_k\right](\bm x,t+\Delta t)&=& f_j(\bm x,t)-0.5\bm\omega_{jk}\left[f_k(\bm x,t)-f^{eq}_k(\bm x,t)\right]\nonumber\\
  &&-\Delta t\bm c_j\cdot\nabla f_j(\bm x,t+\frac{\Delta t}{2} ).
\end{eqnarray}
Then the linear expansion $f_j(\bm x,t)=\overline{f_j^{eq}(\bm x,t)}+f_j'(\bm x,t)$ is conducted to Eq. \eqref{3.2.1}, we can obtain
\begin{equation}
\begin{split}
 \left[(1+\frac{\bm\omega_{jk}}{2})\delta_{jk}+\frac{\bm\omega_{jk}}{2}\Gamma_{jk}\right]f'_j(\bm x,t+\Delta t) = &\left[(1-\frac{\bm\omega_{jk}}{2})\delta_{jk}+\frac{\bm\omega_{jk}}{2}\Gamma_{jk}\right]f'_k(\bm x,t)\\
 &-\delta_{jk}\Delta t \bm c_k\cdot\nabla f'_j\left(\bm x,t+\frac{\Delta t}{2}\right),
\label{3.2.2}
\end{split}
\end{equation}
where the term $\bm\Gamma_{jk}=\partial f_j^{eq}(\bm{x},t)/\partial f_k(\bm{x},t)$. It should be noticed that the equilibrium distribution function of IFDLBM is different from that of MDF-FDLBM.
This is also the main reason for the different stability regions of the two models.
For the IFDLBM, the $\bm\Gamma_{jk}$ can be calculated by
\begin{subequations}
\begin{equation}
 \frac{\partial f_j^{eq}}{\partial f_k} =
 \begin{cases}
 w_j\frac{\bm c_j\cdot\bm c_j}{c_s^2}+\frac{w_j(\bm c_j\bm c_j-c_s^2\bm I):(\bm c_j\cdot\nabla\bm u\bm u)}{2c_s^4}, \quad &if \quad {k = 0},\\
 -\frac{\lambda_j}{\lambda_0}+\frac{\lambda_j\omega_0}{c_s^2\lambda_0}(\bm c_j\cdot \bm u)+w_j\frac{\bm c_j\cdot\bm c_j}{c_s^2}+\frac{w_j(\bm c_j\bm c_j-c_s^2\bm I):(\bm c_j\cdot\nabla\bm u\bm u)}{2c_s^4}, \quad &if \quad {k \neq 0}.
 \end{cases}
 \label{3.2.3}
\end{equation}
\end{subequations}
Let $Q'_k(t)=[F'_k(t+\frac{\Delta t}{2}),F'_k(t)]^T$. Applying the Fourier transform to Eq. \eqref{3.2.2}, the following equation is obtained
\begin{equation}
\left(
\begin{matrix}
   \bm I+0.5\bm \omega-0.5\bm \omega\bm\Gamma & 0 \\
   0& \bm I \\
  \end{matrix}
  \right)
  Q'_k(\bm \kappa,t+\frac{1}{2}\Delta t)=\left(
  \begin{matrix}
   -r\bm T & I-0.5\bm \omega+0.5\bm \omega\bm\Gamma \\
   I& 0 \\
  \end{matrix}\right)
Q'_k(\bm \kappa,t)
  ,\\
\label{3.2.4}
\end{equation}
and we have
\begin{equation}\label{3.2.5}
  S'_k(\bm \kappa,t+\frac{1}{2}\Delta t)=\tilde{\bm G}S'_k(\bm \kappa,t),
\end{equation}
From Eq.~\eqref{3.2.4}, the growth matrix $\hat{\bm G}$ can be determined as
\begin{equation}
\bm \hat{\bm G}  =\left(
\begin{matrix}
   \bm I+0.5\bm \omega-0.5\bm \omega\bm \Gamma & 0 \\
   0& I \\
  \end{matrix}
  \right)^{-1}
  \left(
\begin{matrix}
   -r\bm T & \bm I-0.5\bm \omega+0.5\bm \omega\bm \Gamma \\
   \bm I& 0 \\
  \end{matrix}\right)\\.
\label{3.2.6}
\end{equation}
It should be noted that, for IFDLBM with D2Q9 model, the growth matrix $\hat{\bm G}$ is an $18\times 18$ matrix and it is no longer a diagonal matrix.
If the spectral radius of growth matrix $\hat{\bm G}$ is less than 1, the linear system Eq. \eqref{3.2.4} is stable. However, too many free parameters make it difficult to analyze matrix $\hat{\bm G}$.
Therefore, we set $\bm u=(0.5,0.5)^T$ and $c=1.0$. Then it is clear that the spectral radius of the matrix $\hat{\bm G}$ is the function of parameters $\bm \omega$ and $r$. As we know $c=1.0$, so $r$ is equals to the CFL condition number.
For the multi-relaxation model, the relaxation parameters are taken as $s_0=s_1=s_2=s_4=s_6=s_7=s_8=s'$ and $s_3=s_5=s$ for IFDLBM.

\begin{figure}[htbp]
\centering
\subfigure[$s'=0.5s$]{\includegraphics[scale=0.5]{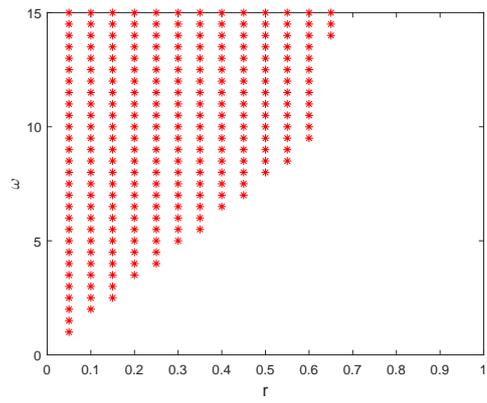}}
\subfigure[$s'=1.0s$]{\includegraphics[scale=0.5]{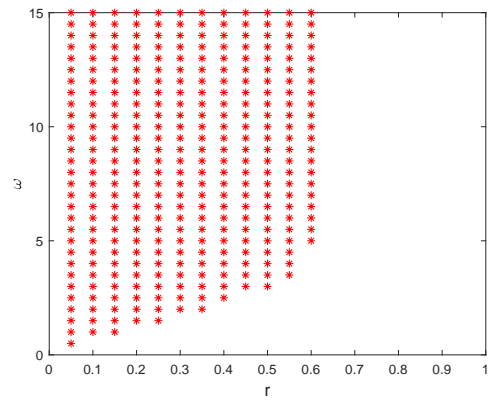}}

\subfigure[$s'=1.5s$]{\includegraphics[scale=0.5]{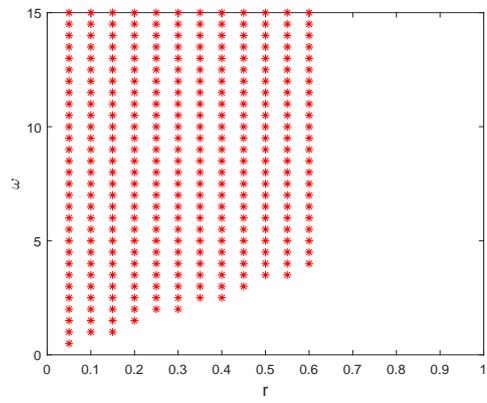}}
\subfigure[$s'=2.0s$]{\includegraphics[scale=0.5]{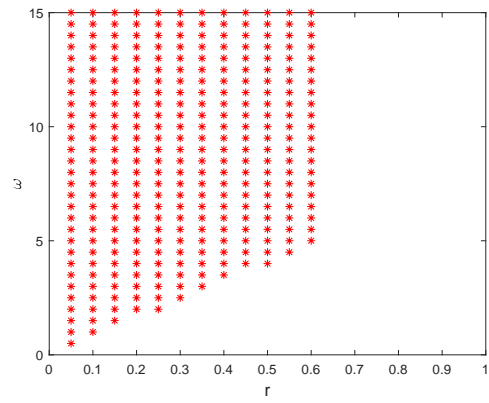}}

 \caption{The stability region of IFDLBM with different $s'$ } \label{fig:stable-IFDLBM-s}
\end{figure}

\begin{figure}[htbp]
\centering
\subfigure[$r=0.1$]{\includegraphics[scale=0.5]{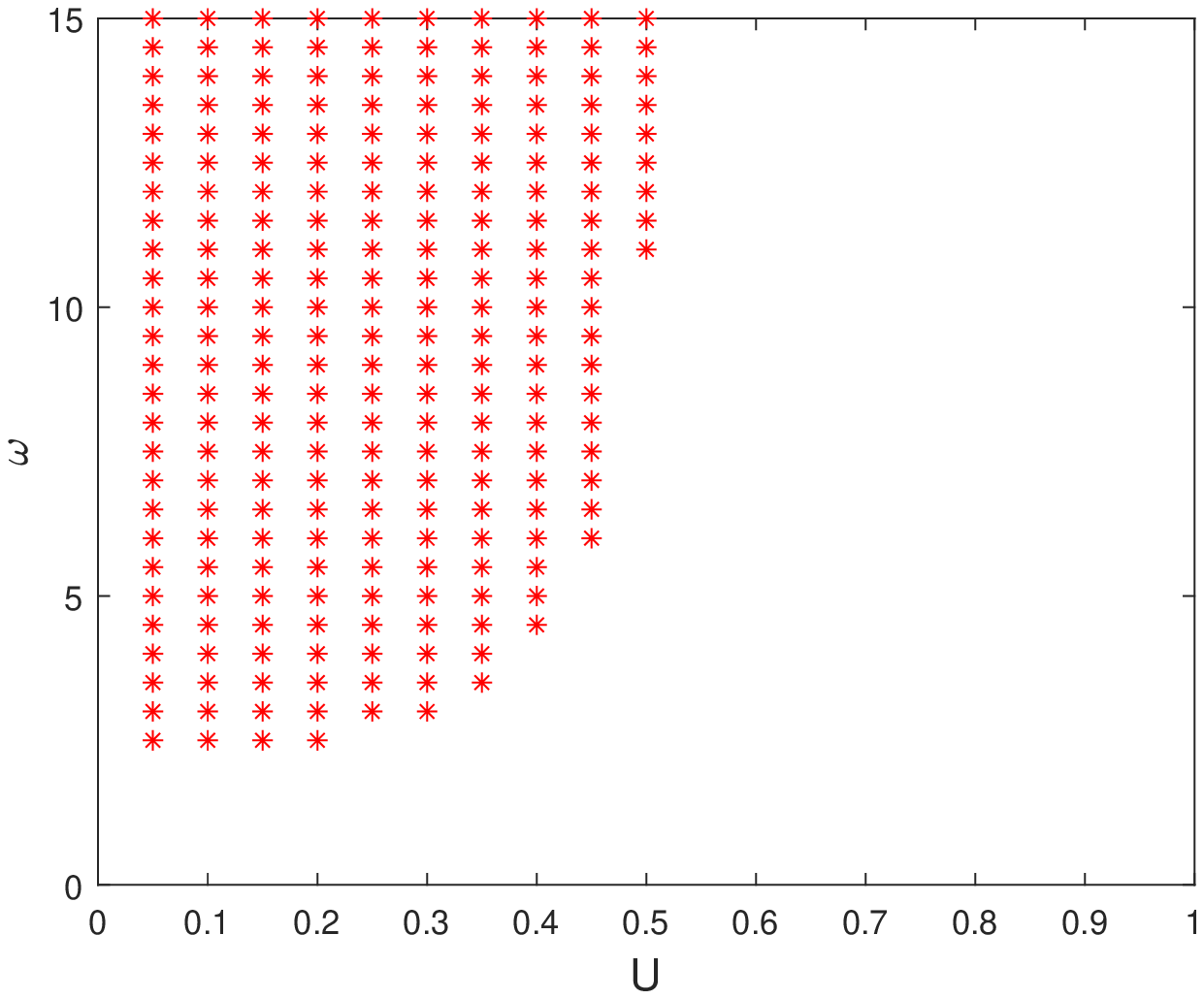}}
\subfigure[$r=0.25$]{\includegraphics[scale=0.5]{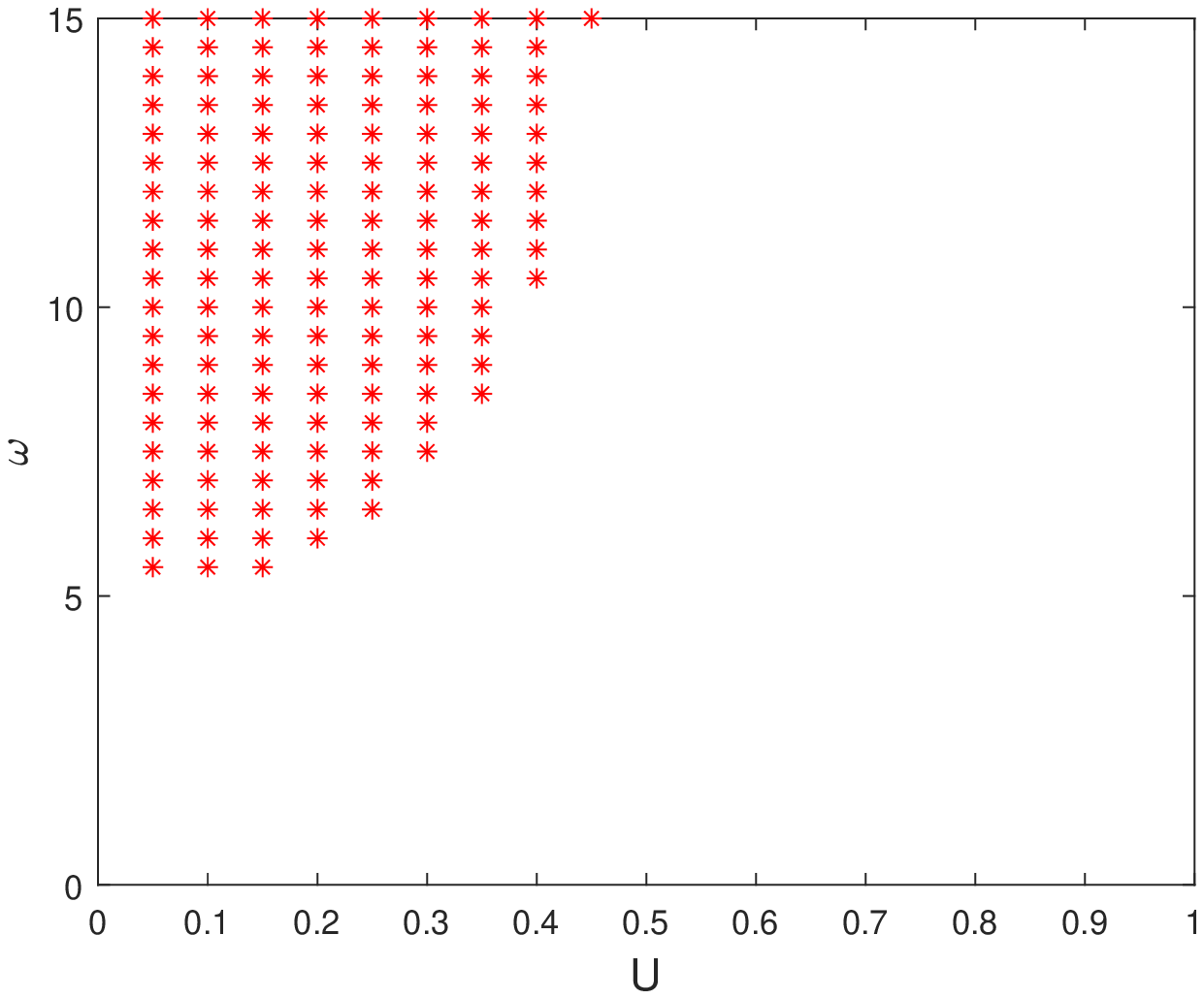}}

\subfigure[$r=0.5$]{\includegraphics[scale=0.5]{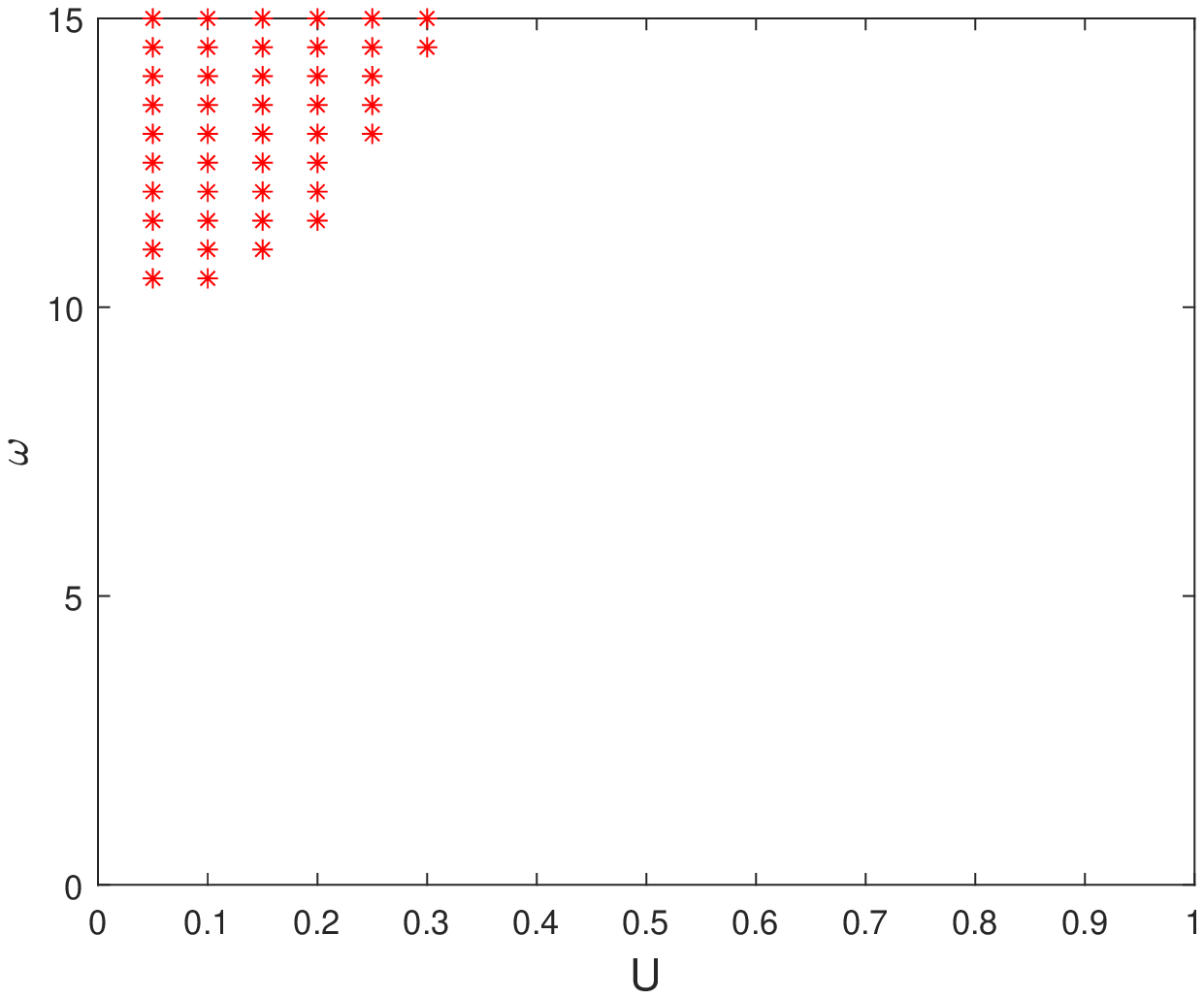}}
\subfigure[$r=0.7$]{\includegraphics[scale=0.5]{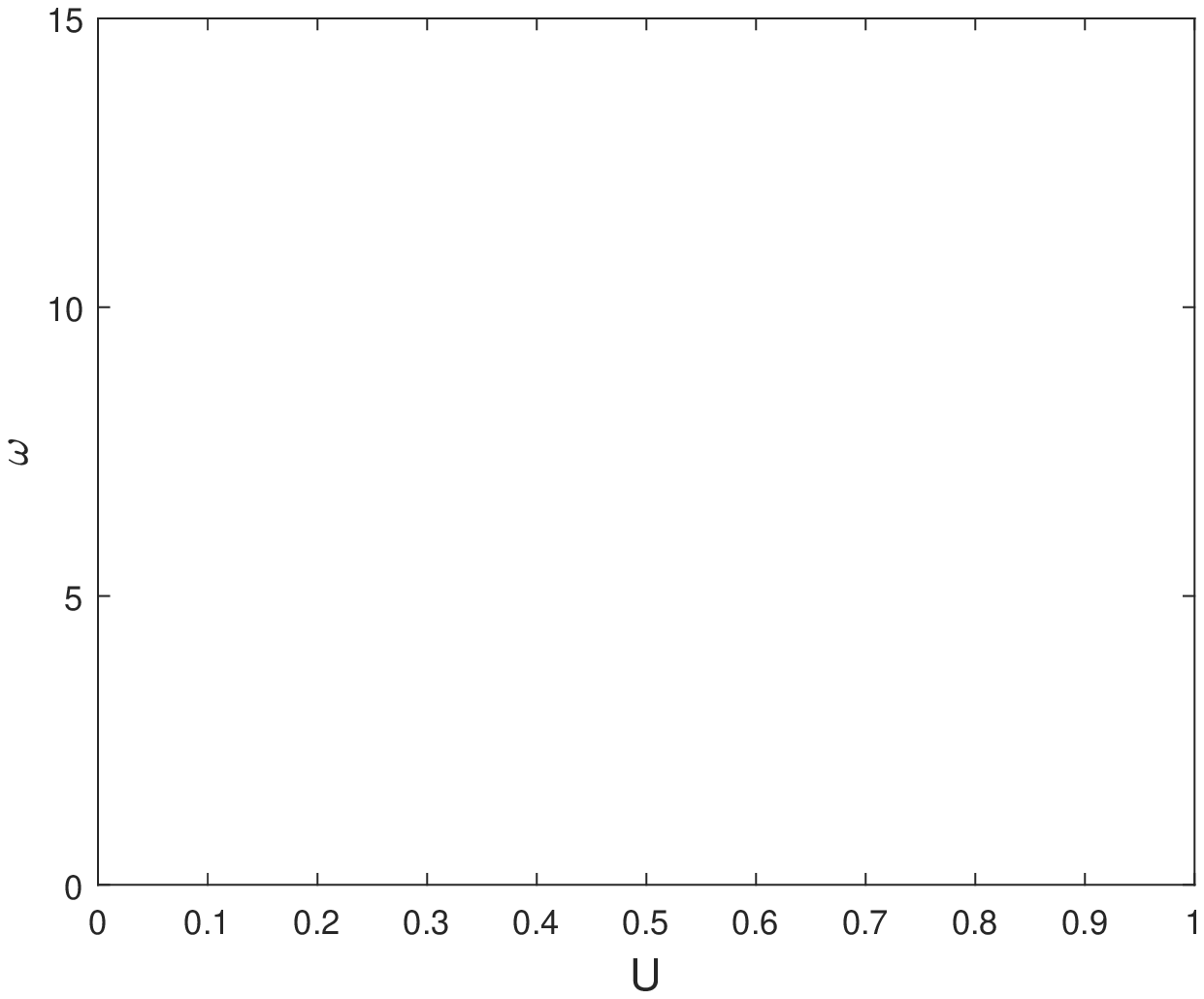}}

 \caption{The stability region of IFDLBM with different CFL condition number } \label{fig:stable-IFDLBM-r}
\end{figure}

Fig. \ref{fig:stable-IFDLBM-s} shows the stability domains of IFDLBM with different $s'$. When $s' =1.5s$, the area of the stable region is the largest, and when $s'$ is less than $s$, the area of the stable region will reduce significantly.
Compared with the stability region of MDF-FDLBM, we can find the IFDLBM will be more stable with high $\omega$. Because $\omega=\Delta t s$ and $c_s^2/s=\nu$, it can be deduced that the IFDLBM will be more stable under the condition of low $\nu$.
Besides, we also fixed $s'=1.5s$ and presented the stability domains of IFDLBM with different $r$ in Fig. \ref{fig:stable-IFDLBM-r}. It is clear that the stability region areas decrease with the increase of $r$.
This phenomenon is similar to the results of some stability analyses.
Compared with Fig. \ref{fig:stable-MDF-FDLBM-r}, we can find that MDF-FDLBM is more stable when the large CFL condition number is taken.

\section{\label{sec:level4}Numerical example }
In this section, four different numerical examples are used
to test the present MDF-FDLBM, including the four-roll mill problem, the periodic flow,
the poiseuille flow and the lid driven flow. Unless otherwise
specified, the distribution function is initialized by the equilibrium
distribution function, i.e., $\hat{f}_j=f^{eq}_j$. We use the D2Q5 lattice model and $s'=1.5s$ for all MDF-FDLBM in this section. To test the
accuracy of the MDF-FDLBM, the following global relative error (GRE) is
adopted
\begin{equation}
GRE=\frac{\sum_i|\phi(\bm x_i,t)-\phi^*(\bm x_i,t)|}{\sum_i|\phi^*(\bm x_i,t)|},
\label{4.0}
\end{equation}
where $\phi(\bm x,t)$ and $\phi^*(\bm x,t)$ denote numerical and analytical
solutions, respectively.

\subsection{ The two-dimensional four-roll mill problem}

\begin{figure}[htbp]
\centering
\subfigure{\includegraphics[scale=0.5]{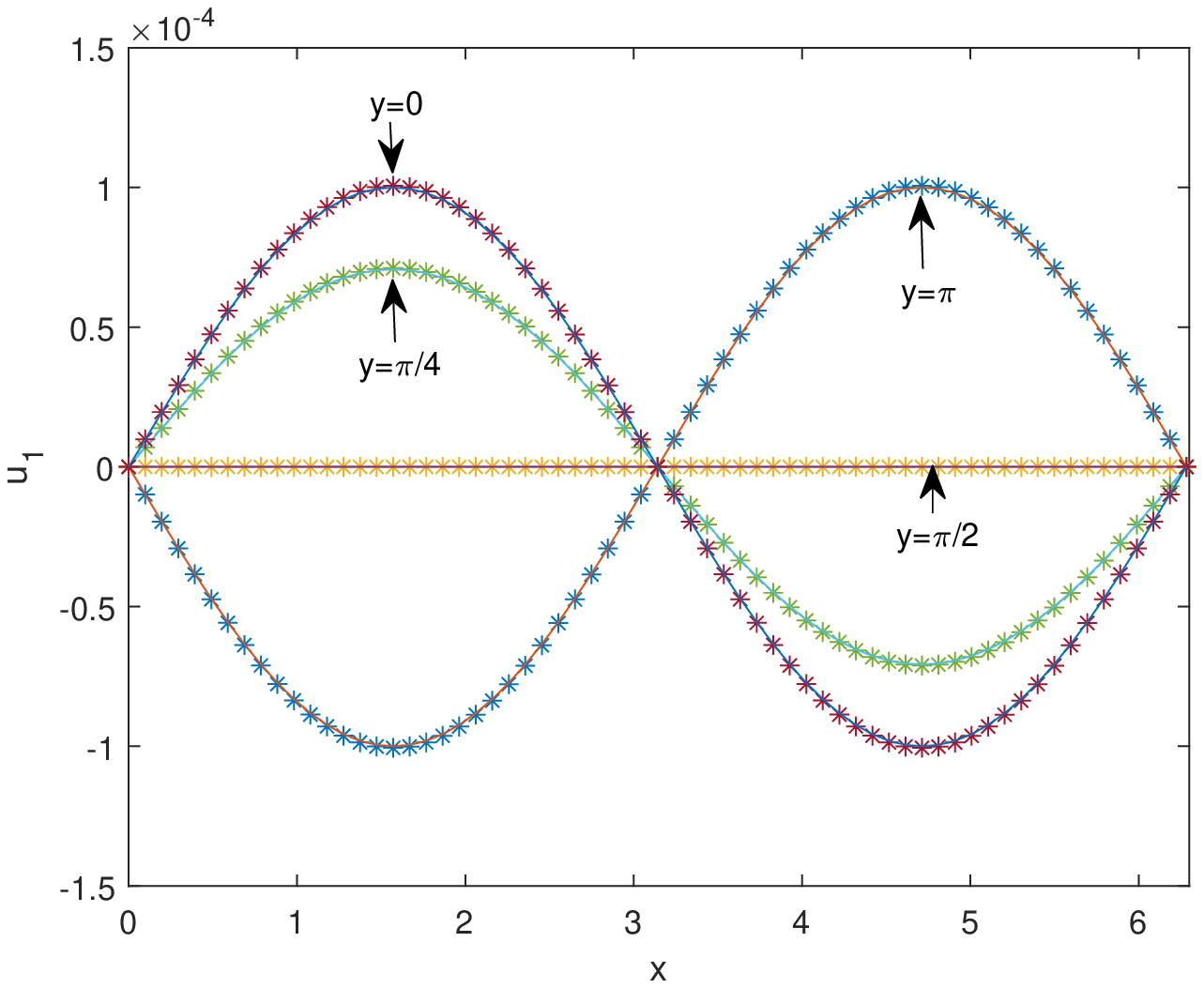}}
\subfigure{\includegraphics[scale=0.5]{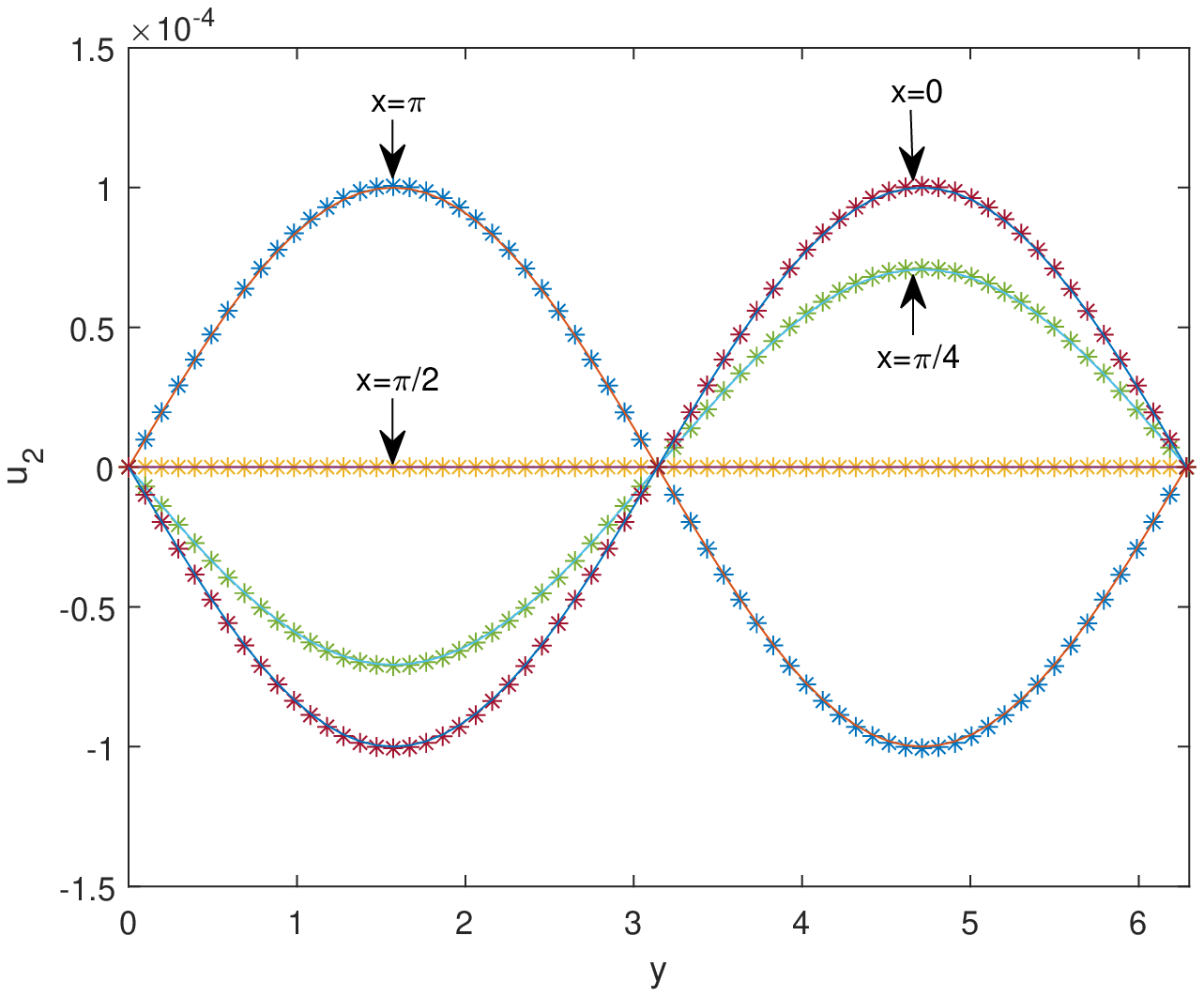}}

 \caption{The MDF-FDLBM numerical and analytical solution of velocity at different positions [symbol:numercial solution, solid line: analytical solution]. } \label{fig:four-uv}
\end{figure}

\begin{figure}[htbp]
\centering
\subfigure{\includegraphics[scale=0.5]{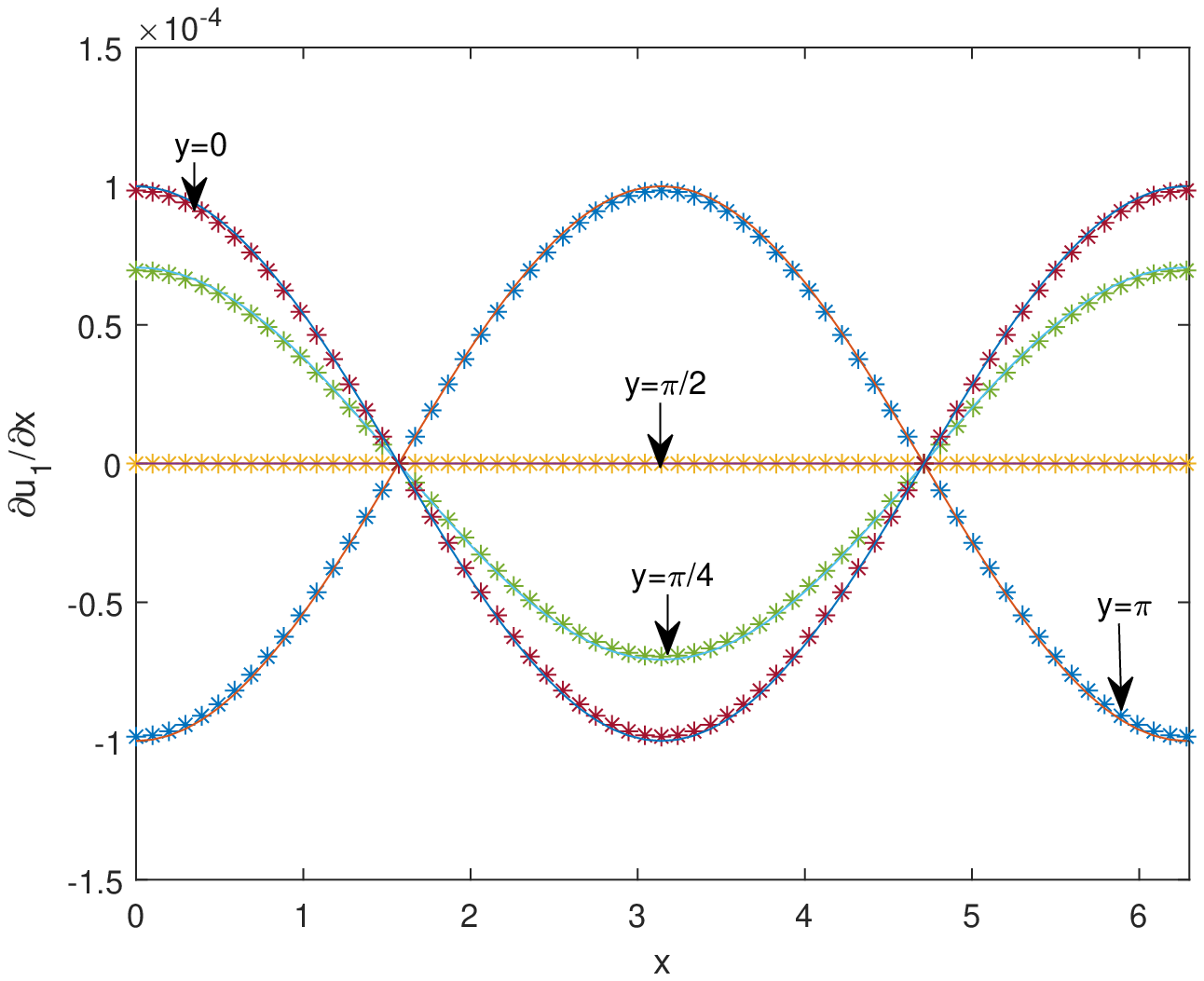}}
\subfigure{\includegraphics[scale=0.5]{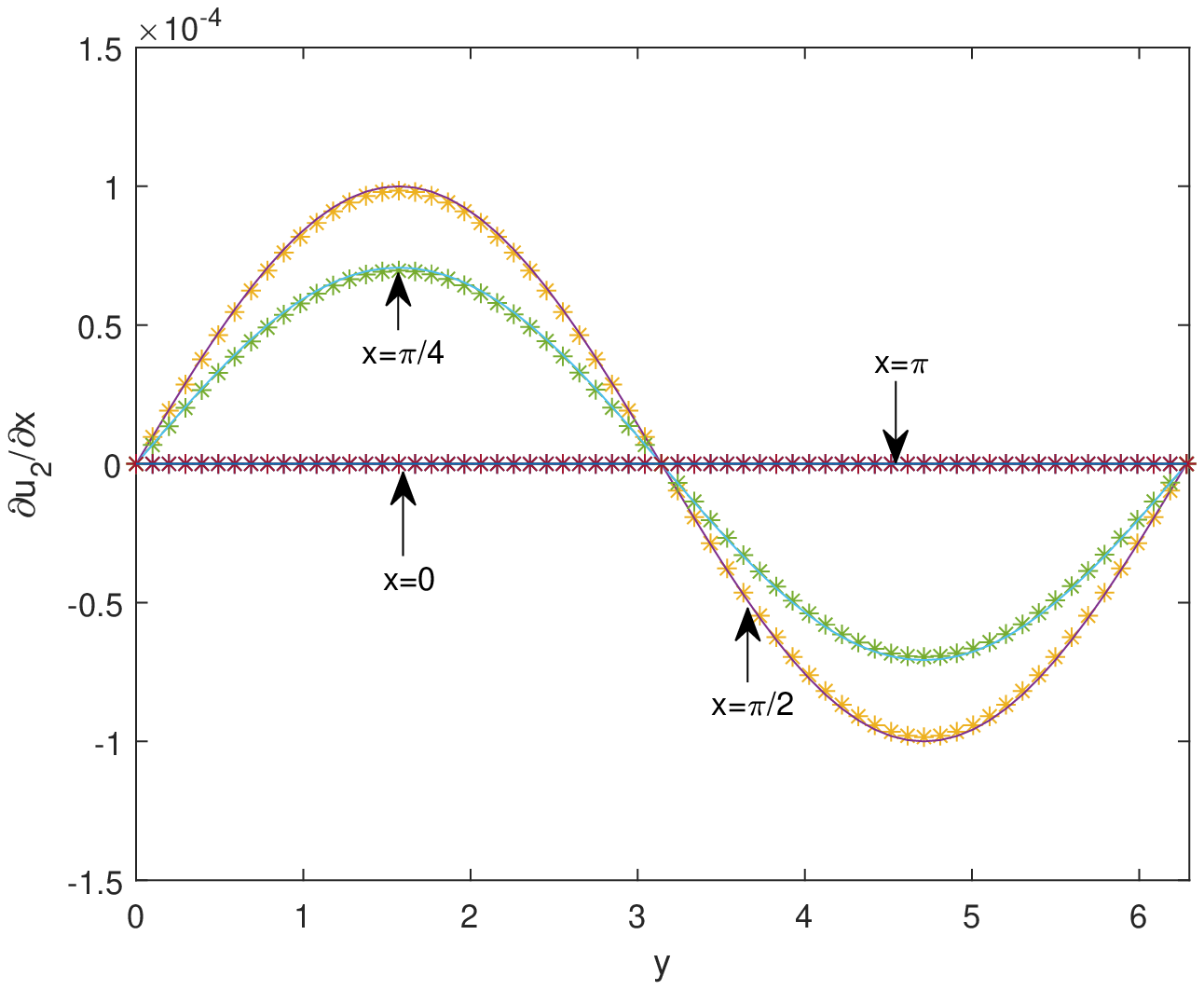}}

\subfigure{\includegraphics[scale=0.5]{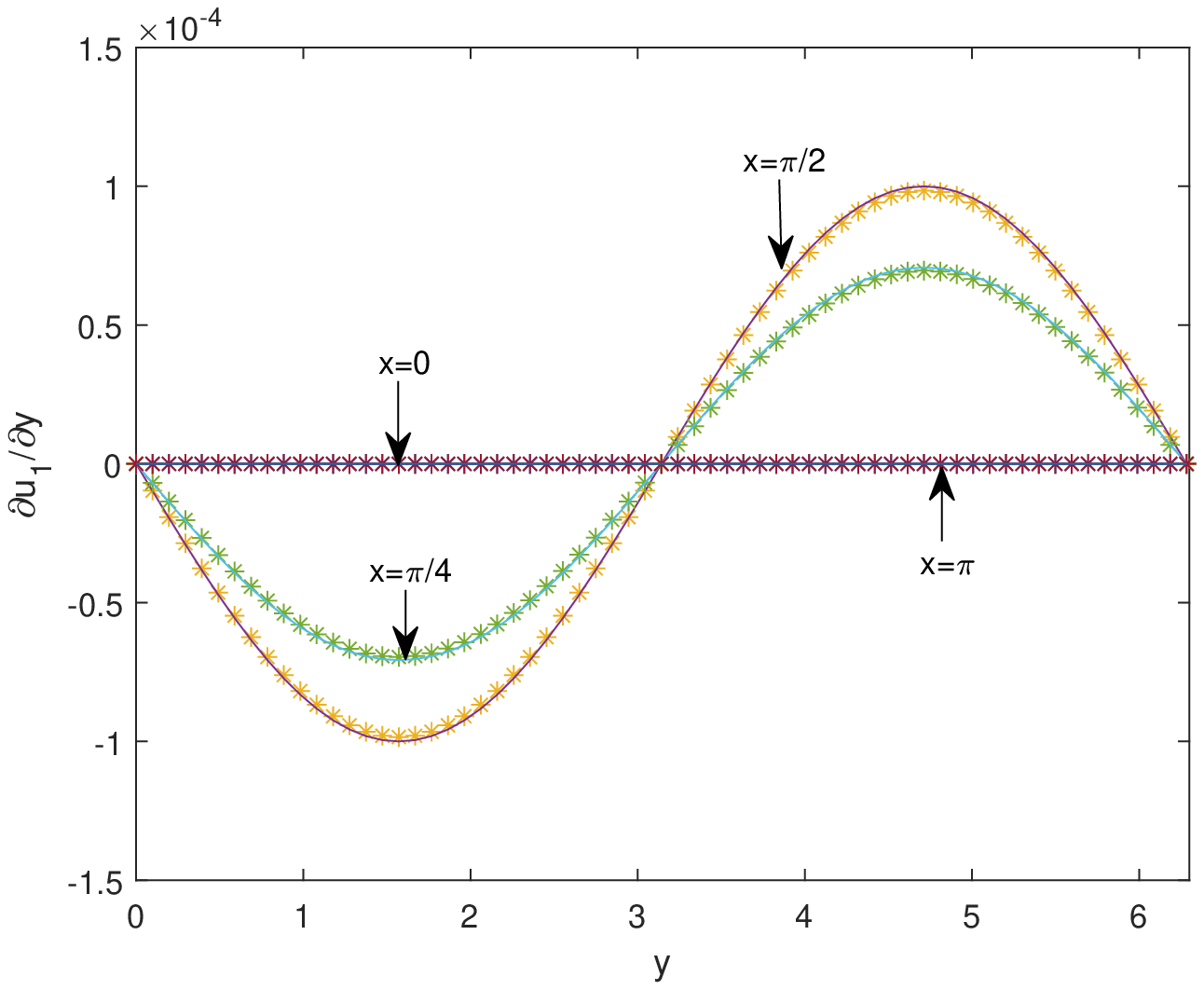}}
\subfigure{\includegraphics[scale=0.5]{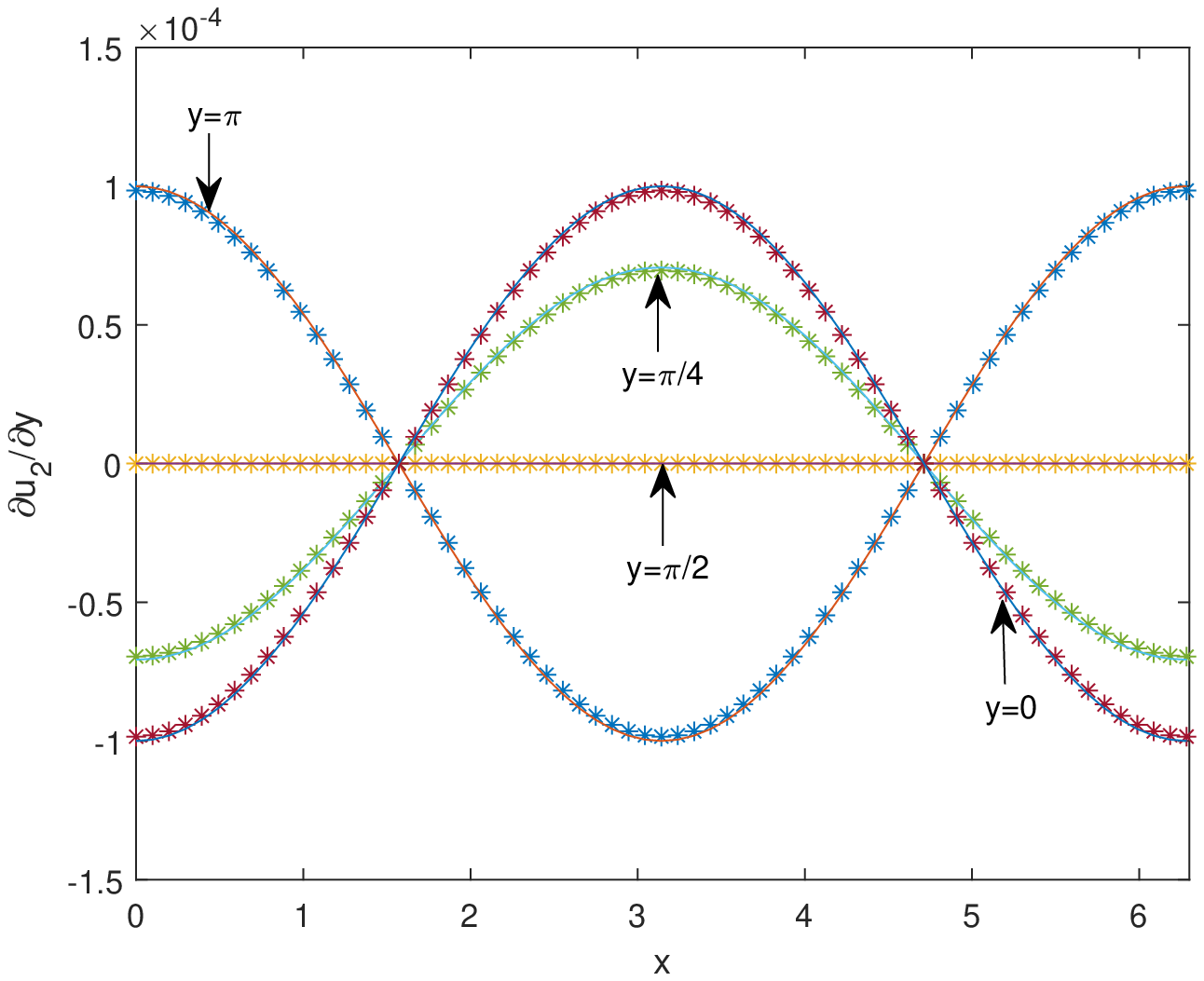}}

 \caption{The MDF-FDLBM numerical and analytical solution of velocity gradient at different positions [symbol:numercial solution, solid line: analytical solution]. } \label{fig:four-u1v1}
\end{figure}

\begin{figure}[htbp]
\centering
\subfigure{\includegraphics[scale=0.5]{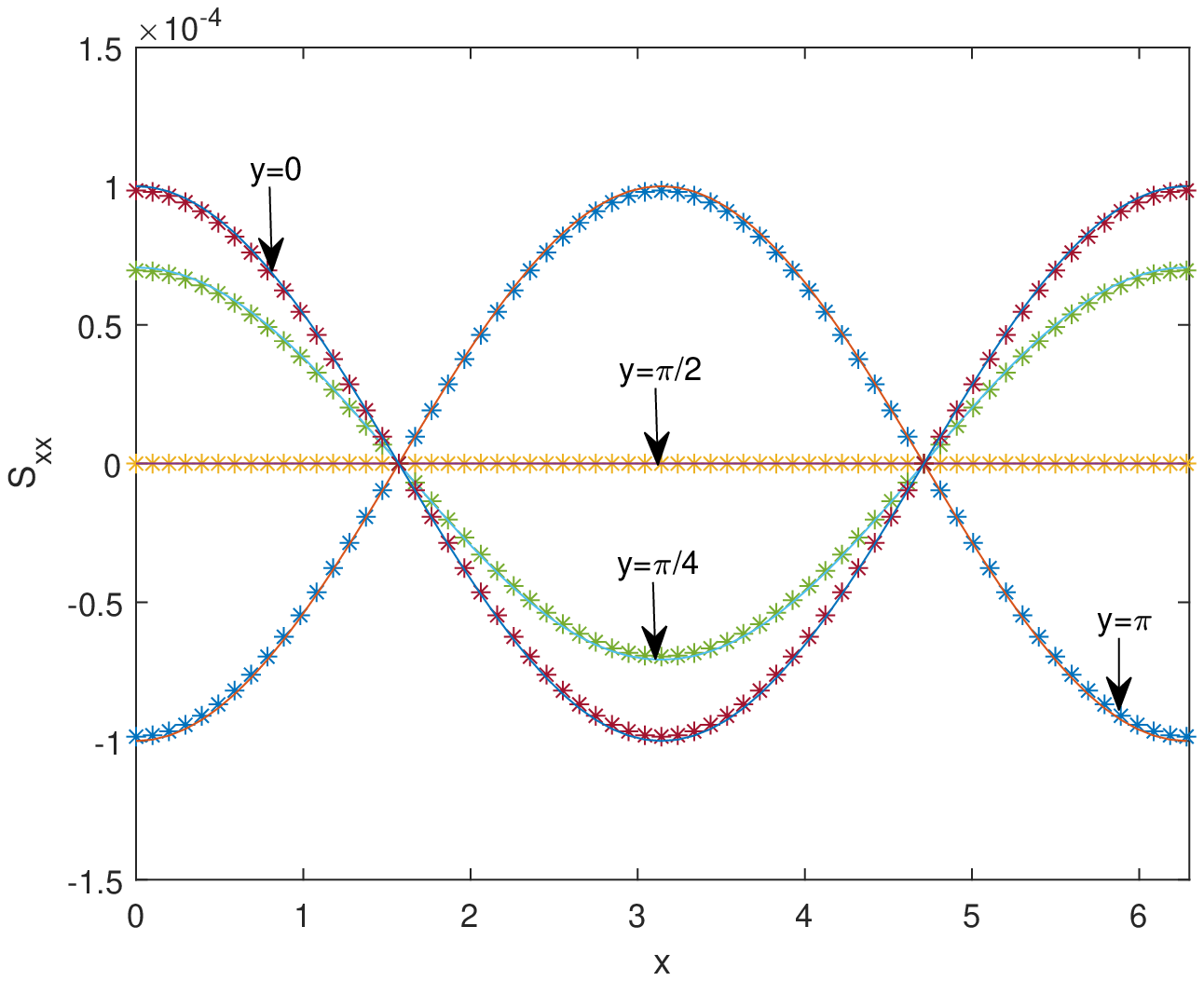}}
\subfigure{\includegraphics[scale=0.5]{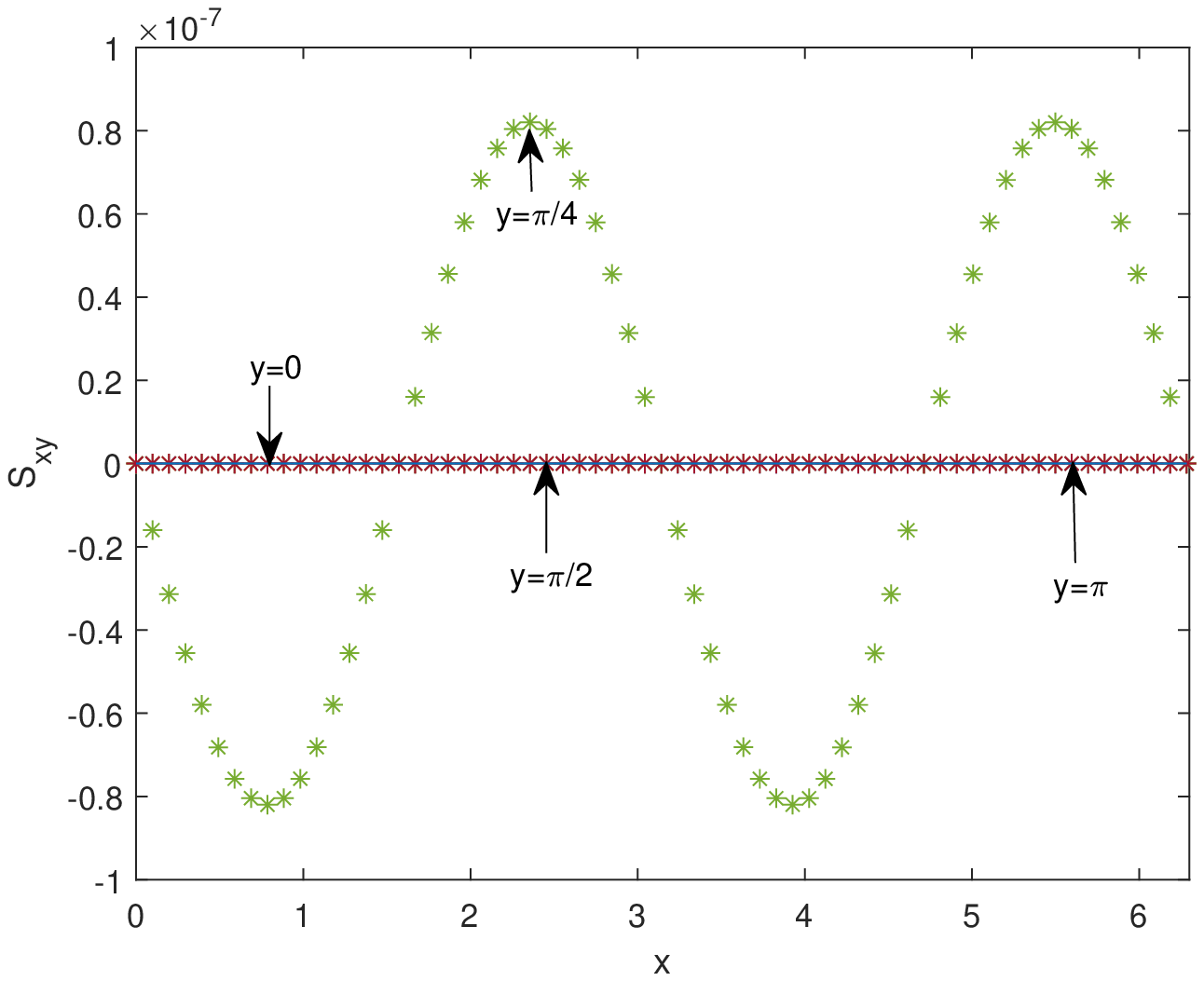}}

 \caption{The MDF-FDLBM numerical and analytical solution of strain rate tensor at different positions [symbol:numercial solution, solid line: analytical solution]. } \label{fig:four-Sxy}
\end{figure}

\begin{figure}[htbp]
\centering
\subfigure{\includegraphics[scale=0.5]{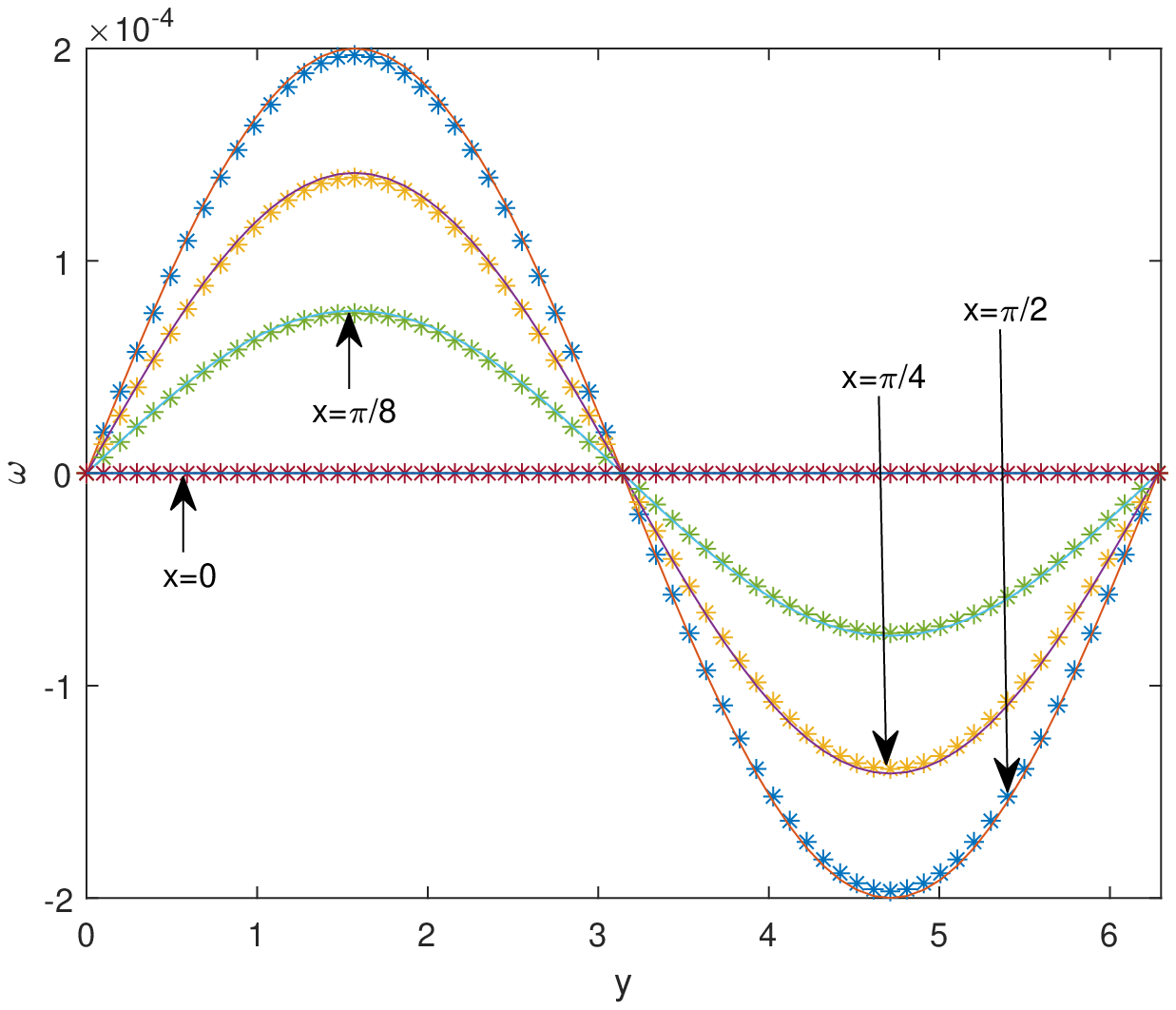}}

 \caption{The MDF-FDLBM numerical and analytical solution of vorticity at different positions [symbol:numercial solution, solid line: analytical solution]. } \label{fig:four-omega}
\end{figure}

To test the validity of the MDF-FDLBM, we use the MDF-FDLBM to simulate the two-dimensional four-roll mill problem.
The physical domain is fixed in $[0,2\pi]\times[0,2\pi]$, and the periodic boundary condition is used for four boundaries.
The force term can be given by
\begin{equation}\label{4.1.1}
\begin{split}
  &F_1=U_0^2\sin(x)\cos(x)+2\nu U_0\sin(x)\cos(y),\\
  &F_2=U_0^2\sin(y)\cos(y)+2\nu U_0\sin(y)\cos(x).
\end{split}
\end{equation}
The analytical solution of velocity is defined as,
\begin{equation}\label{4.1.2}
  u_1=U_0\sin(x)\cos(y),\quad u_2=-U_0\sin(y)\cos(x).\\
\end{equation}
According to the analytical solutions of velocity, we can get the analytical solutions of velocity gradient, velocity divergence, stain rate tensor,
\begin{equation}\label{4.1.3}
  \begin{split}
  &\frac{\partial u_1}{\partial x}=-\frac{\partial u_2}{\partial y}=U_0\cos(x)\cos(y),\quad \frac{\partial u_1}{\partial y}=-\frac{\partial u_2}{\partial x}=-U_0\sin(x)\sin(y),\\
  &\nabla\cdot \bm u=\frac{\partial u_1}{\partial x}+\frac{\partial u_2}{\partial y}=0,\\
  &S_{xx}=-S_{yy}=U_0\cos(x)\cos(y),\quad S_{xy}=S_{yx}=0.\\
  &\omega=\frac{\partial u_2}{\partial x}-\frac{\partial u_1}{\partial y}=2U_0\sin(x)\sin(y).
  \end{split}
\end{equation}
In our simulation, the lattice grids is set as $64\times 64$, and we take $U_0=0.0001$, $\nu=0.01$, $CFL=0.5$. Figs. \ref{fig:four-uv}, \ref{fig:four-u1v1}, \ref{fig:four-Sxy} and \ref{fig:four-omega} show the numerical solutions of velocity, velocity gradient, stain rate tensor and vorticity of MDF-FDLBM at different locations, respectively.
It can be found that the numerical results of these physical quantities agree well with the analytical solutions.
The results show that MDF-FDLBM is effective to solve the convection-diffusion system based NSEs, and the calculation schemes of these physical quantities are correct.

\begin{figure}[htbp]
\centering
\subfigure{\includegraphics[scale=0.5]{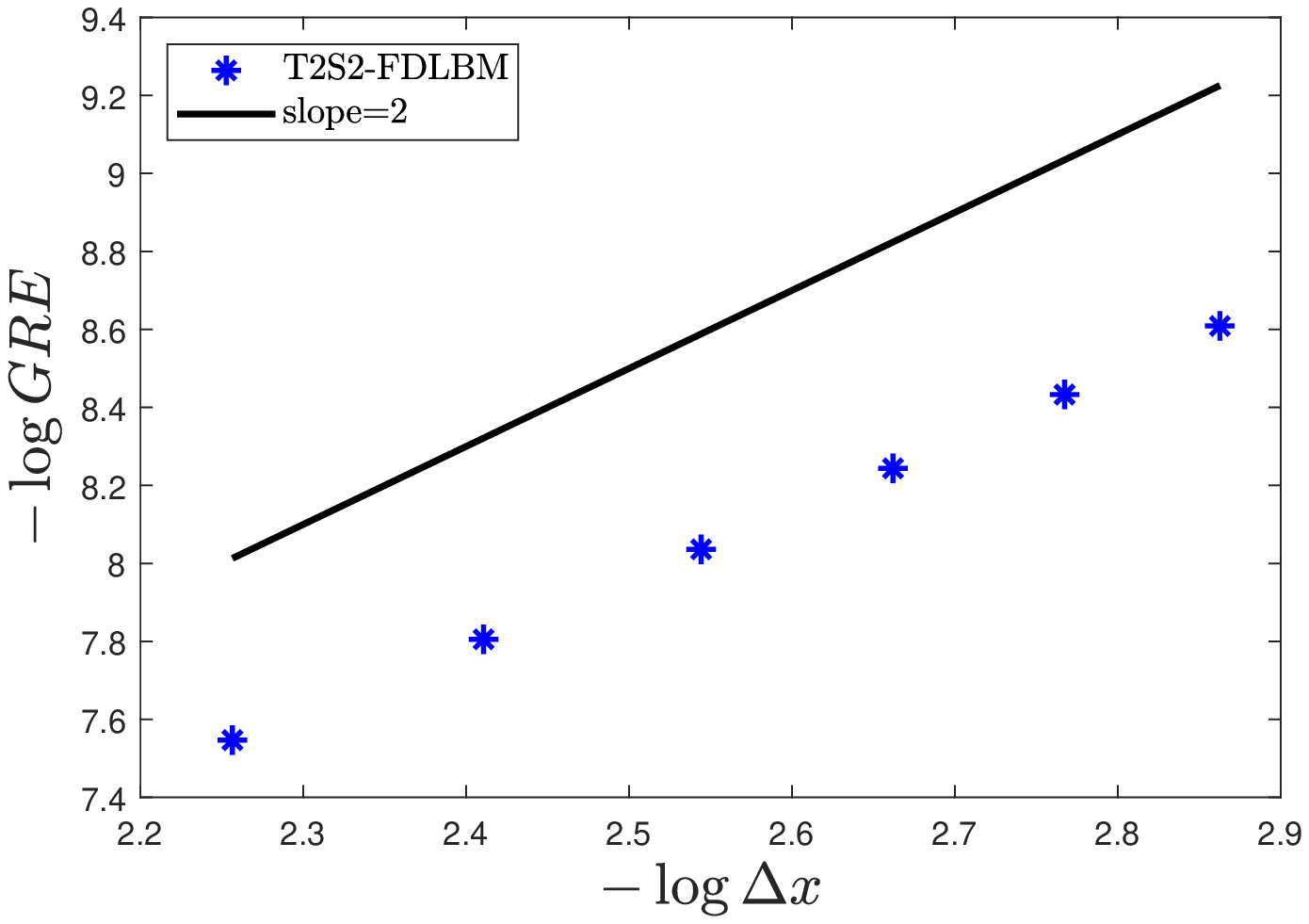}}
\subfigure{\includegraphics[scale=0.5]{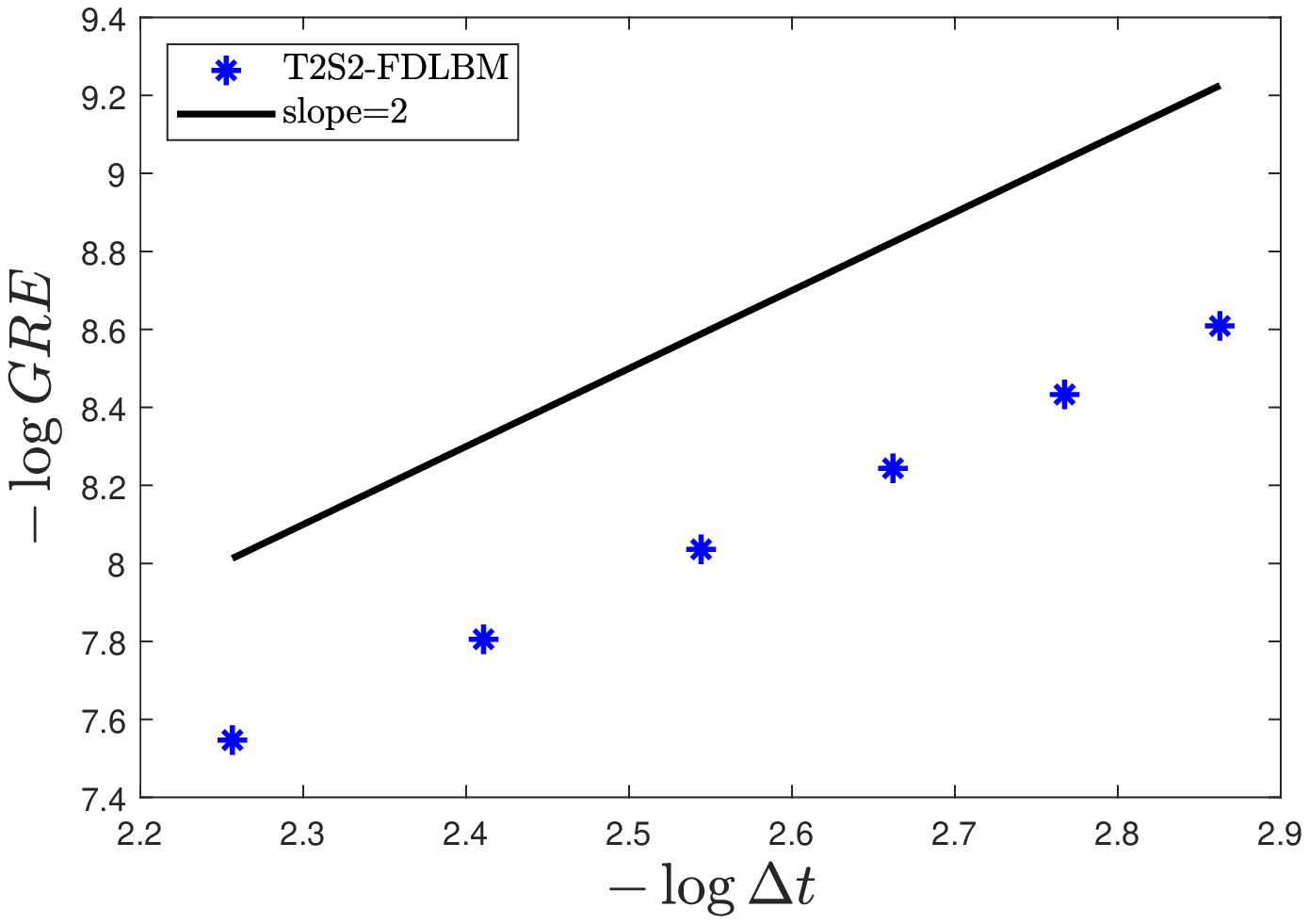}}

 \caption{The GRE of MDF-FDLBM in different lattice space and time step } \label{fig:four-rate}
\end{figure}

In addition, we need to measure the convergence rate of the MDF-FDLBM. The use of periodic boundary conditions in the four-roll mill problem ensures that the convergence rate of the MDF-FDLBM is not affected by boundaries.
In theory, the MDF-FDLBM has a second-order convergence rate in space and time, so the error is $O(\delta t^2+\delta x^2)$.
Some numerical simulations are carried out at different lattice grids ($N\times N=60\times 60,70\times 70,80\times 80, 90\times 90,100\times 100,1100\times 110$) with $dt=dx$.
The results are shown in Fig. \ref{fig:four-rate}. From the figure, we can find that the MDF-FDLBM has second-order accuracy in space and time, which is consistent with theoretical accuracy.

\subsection{ The periodic flow}

\begin{figure}[htbp]
\centering
\subfigure{\includegraphics[scale=0.5]{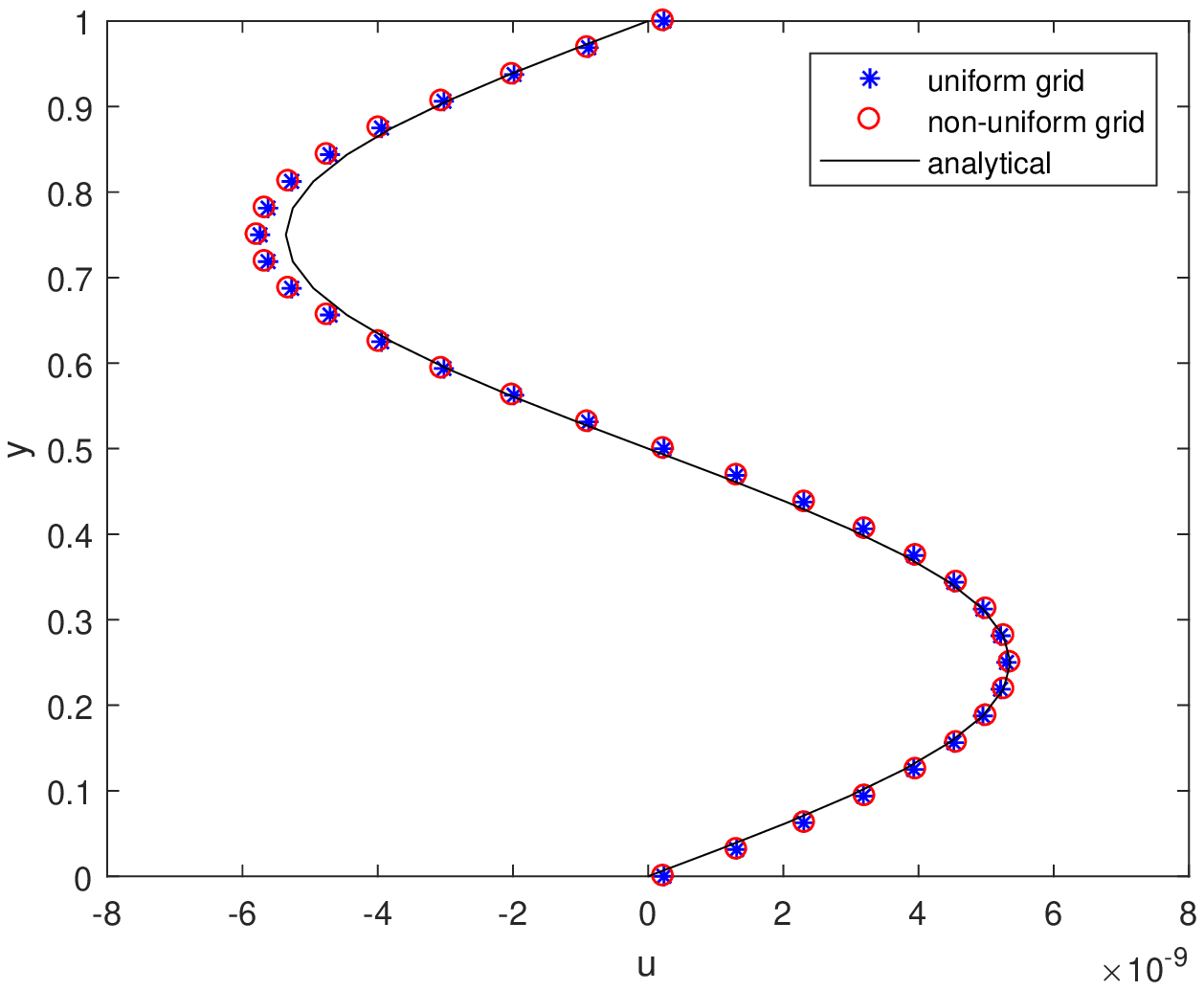}}
\subfigure{\includegraphics[scale=0.5]{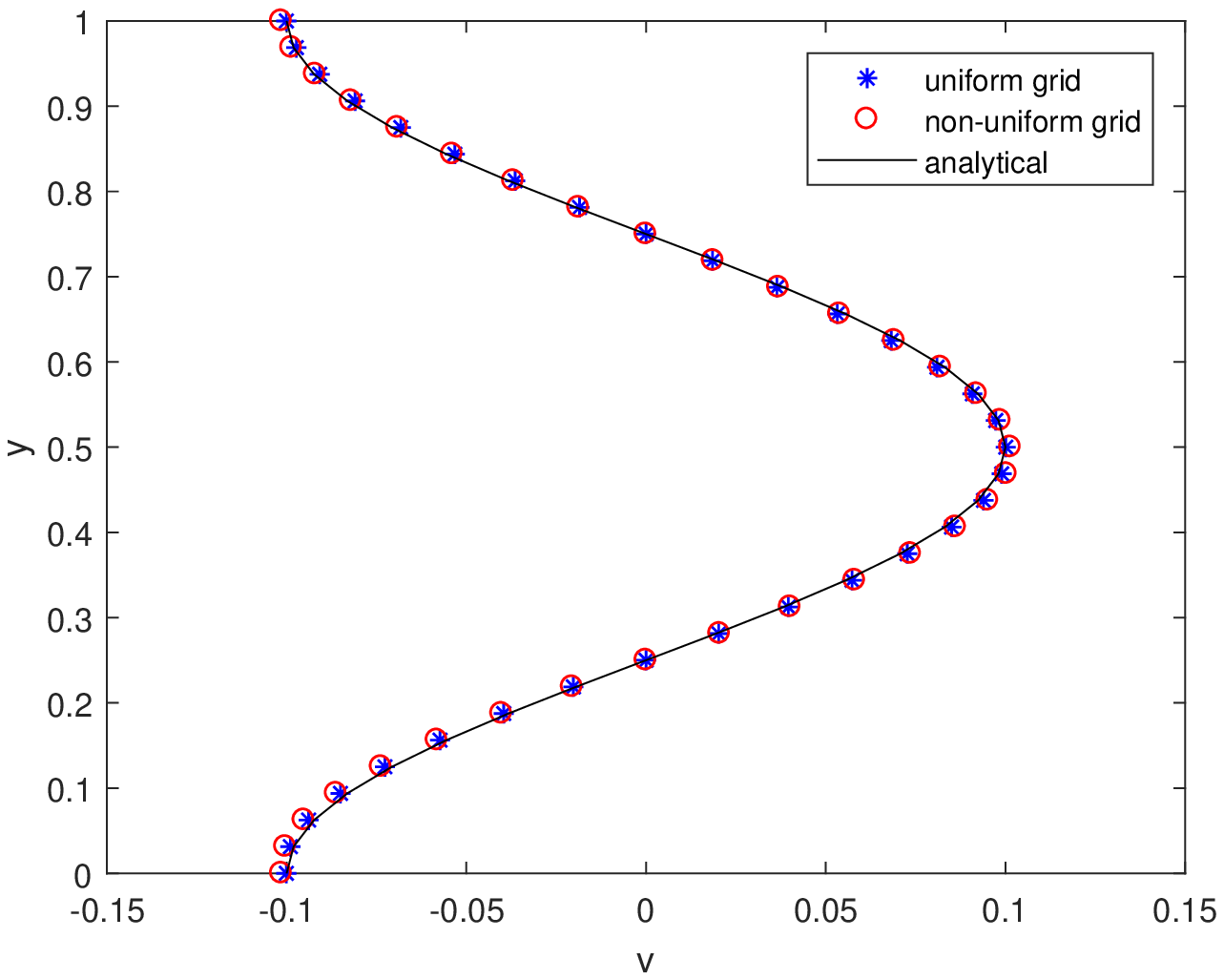}}
\subfigure{\includegraphics[scale=0.5]{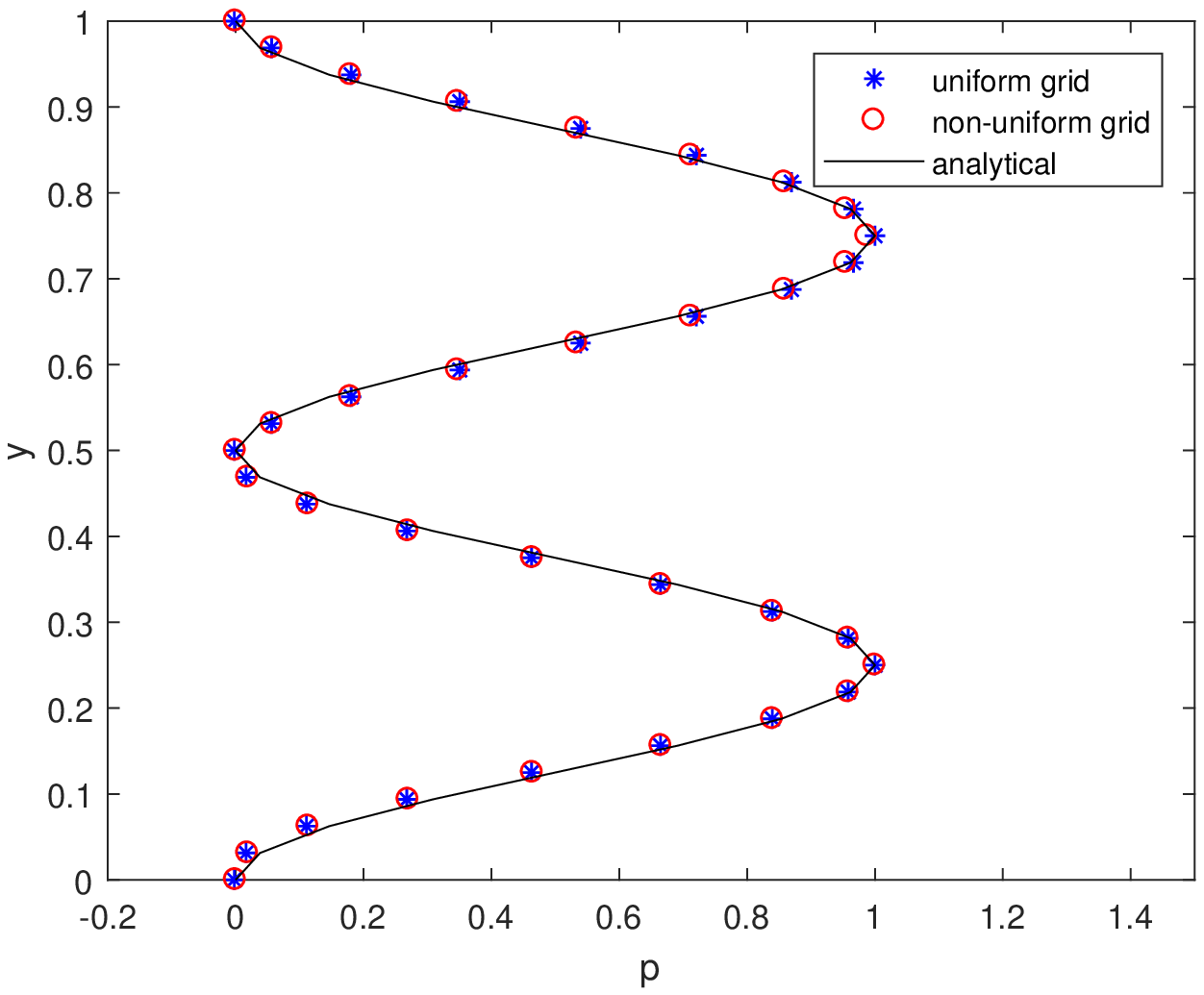}}
 \caption{The MDF-FDLBM numerical and analytical solutions of the velocity and pressure. } \label{fig:periodic-uv}
\end{figure}

\begin{figure}[htbp]
\centering
\subfigure[$\partial u/\partial x$]{\includegraphics[scale=0.5]{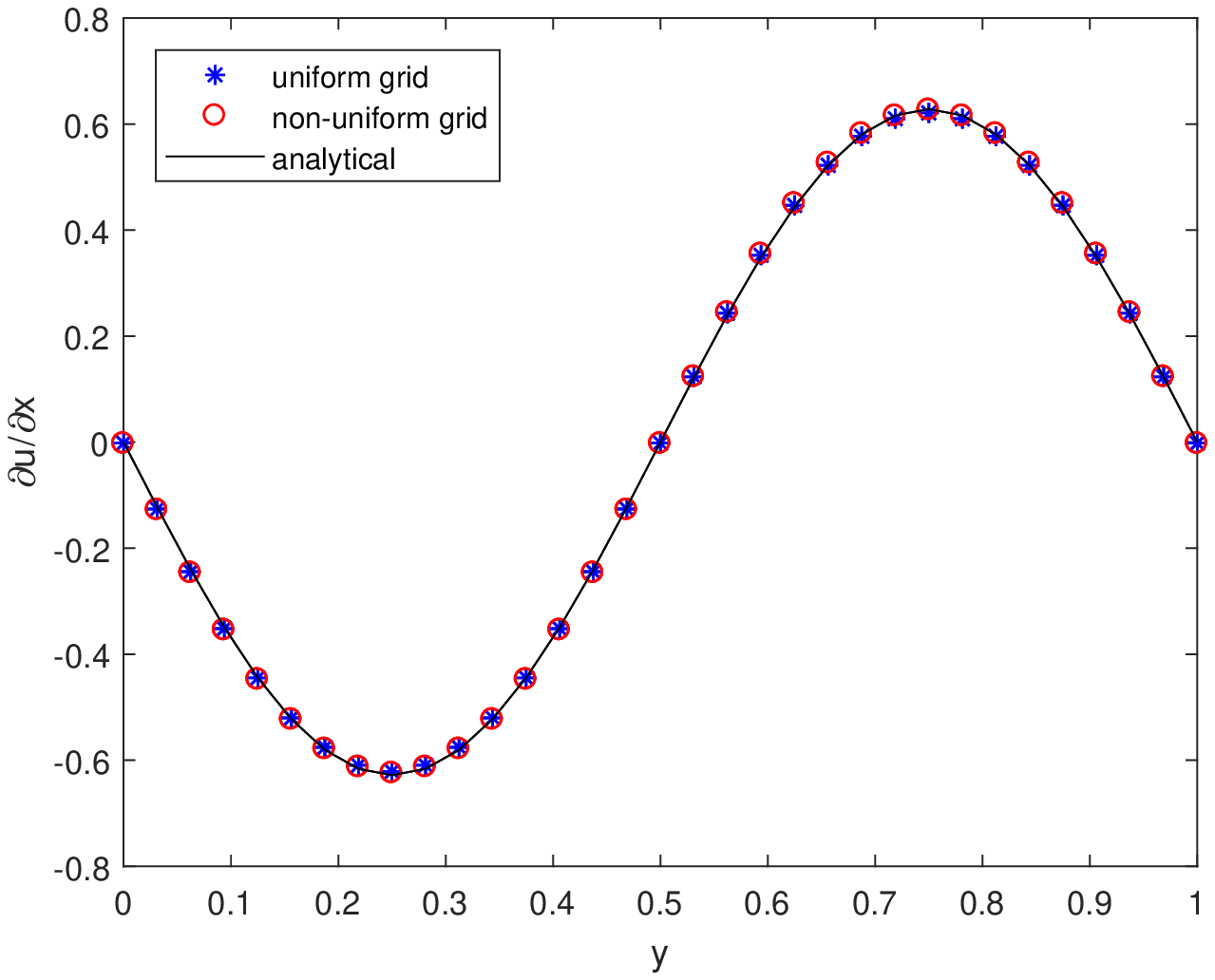}}
\subfigure[$\partial v/\partial y$]{\includegraphics[scale=0.5]{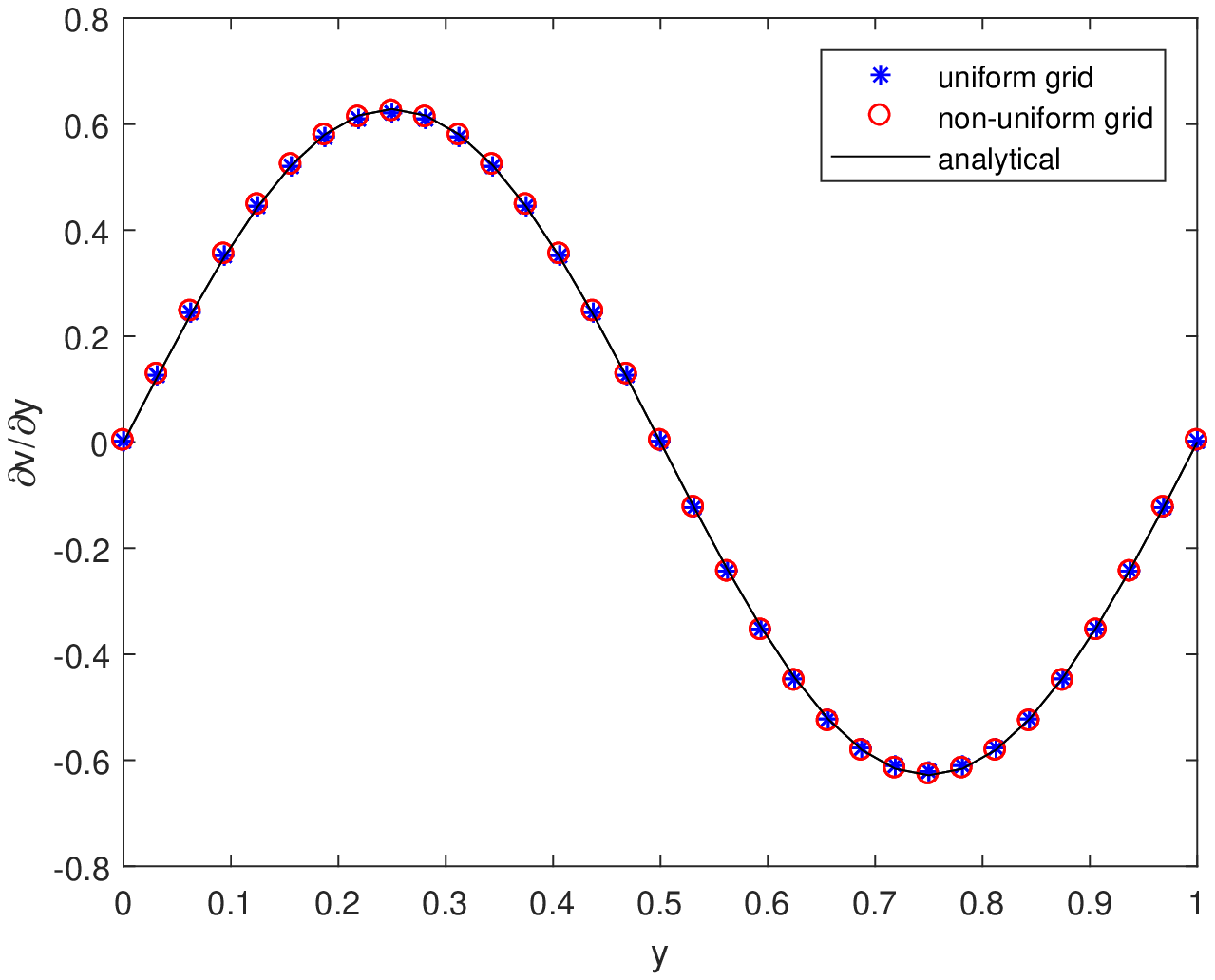}}
\subfigure[$\nabla\cdot \bm u$]{\includegraphics[scale=0.5]{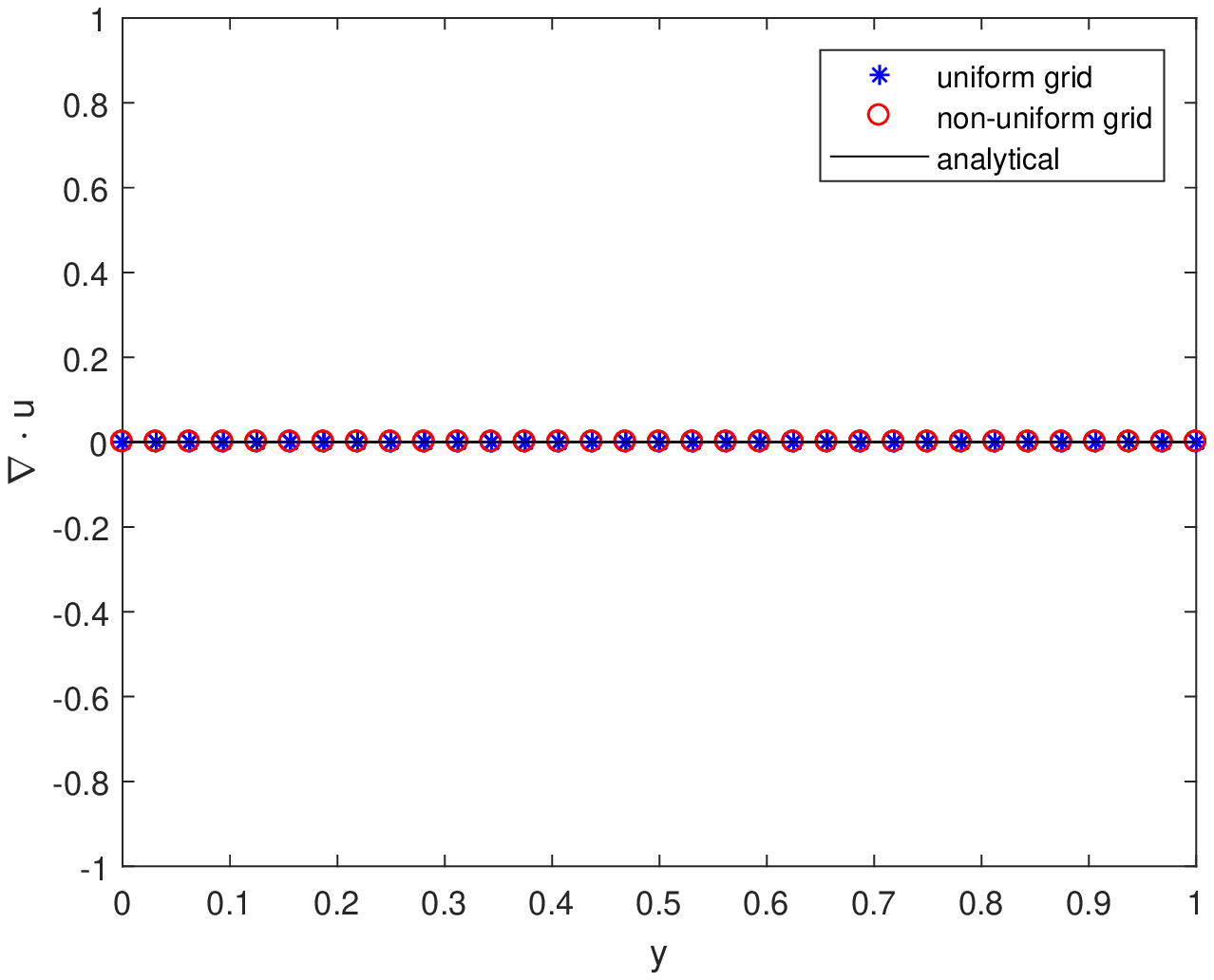}}
\subfigure[$S_{xx}$]{\includegraphics[scale=0.5]{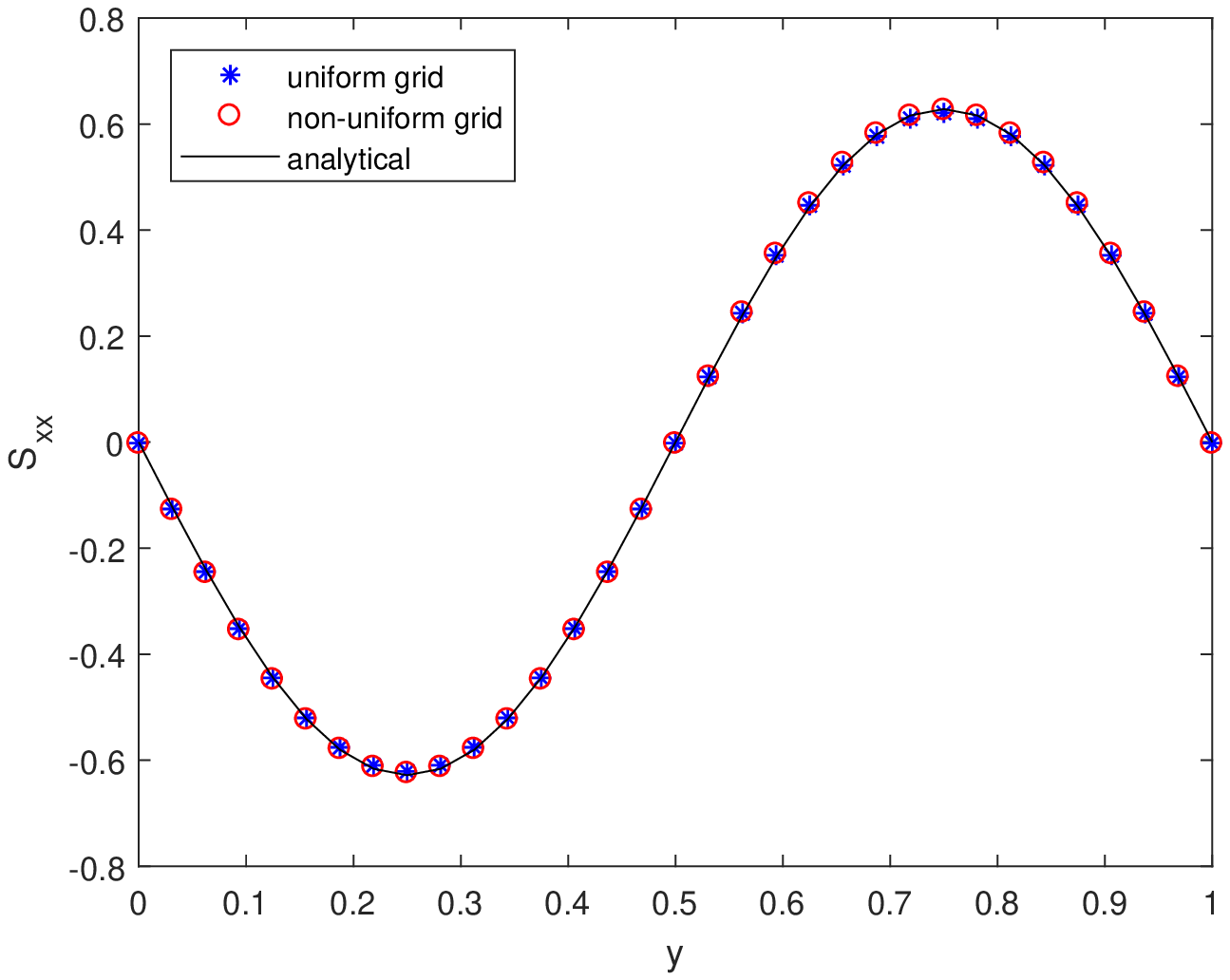}}
\subfigure[$S_{yy}$]{\includegraphics[scale=0.5]{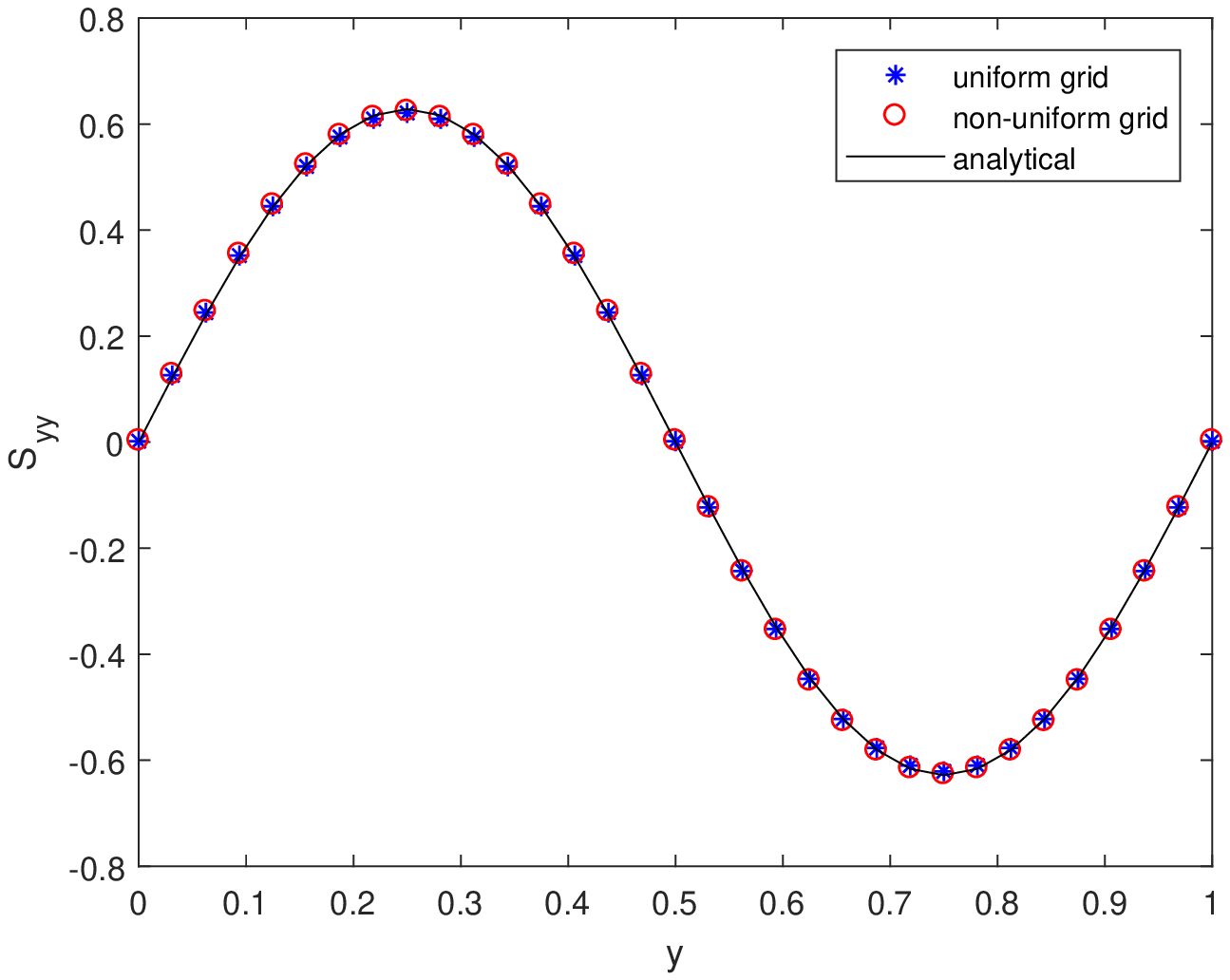}}
\subfigure[$\Omega_{xy}$]{\includegraphics[scale=0.5]{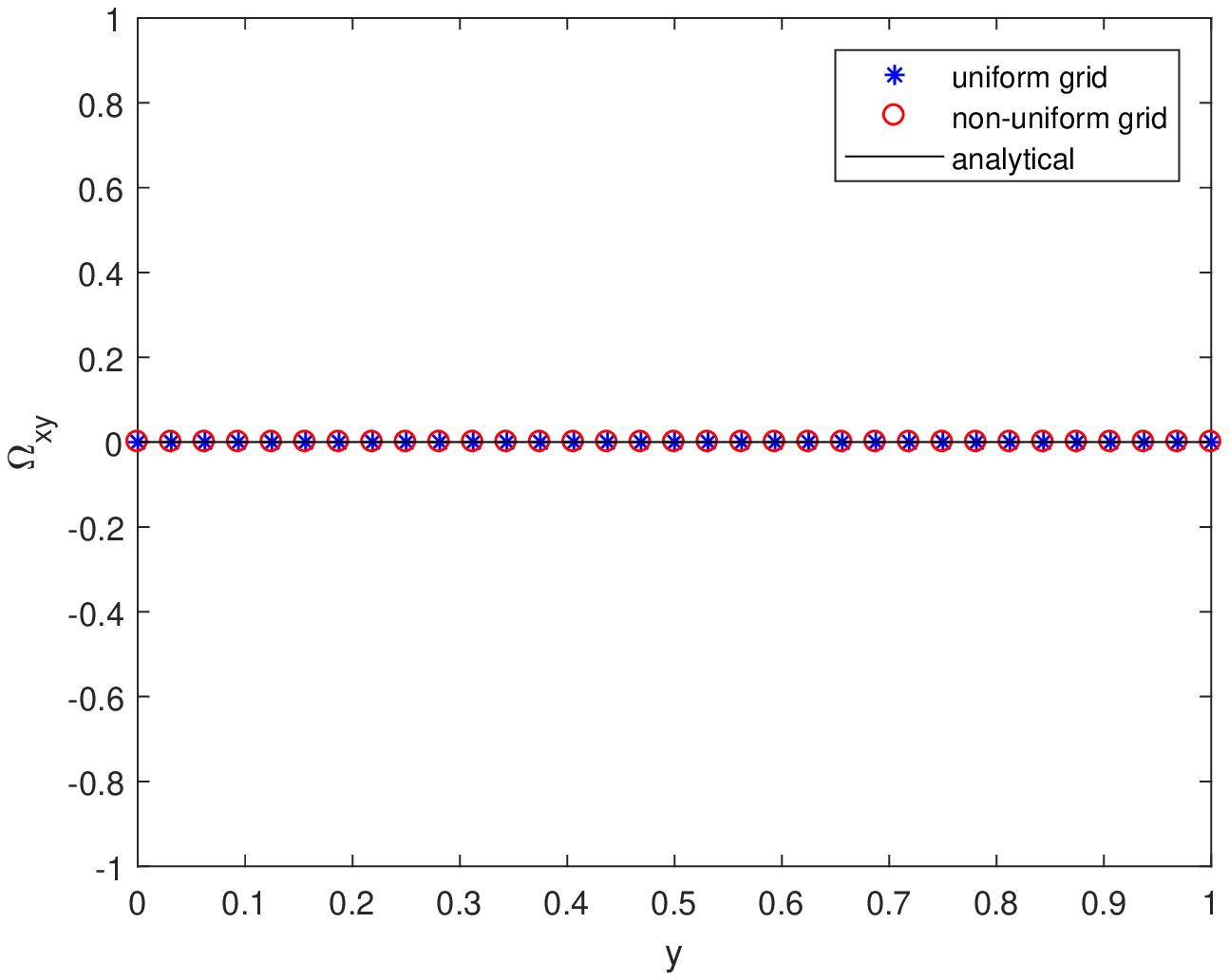}}
 \caption{The numerical solution and analytical solutions of some physical quantity. } \label{fig:periodic-uxy}
\end{figure}

Now, we are going to use the periodic flow to verify the performance of the MDF-FDLBM. It is a steady state problem and the external force term can be expressed by
\begin{equation}\label{4.2.1}
\begin{split}
  &F_x(x,y)=8\pi^2\nu u_0 \sin 2\pi x\sin 2\pi y,\\
  &F_y(x,y)=8\pi^2\nu u_0 \cos 2\pi x\cos 2\pi y,
  \end{split}
\end{equation}
where the kinetic viscosity $\nu$ is set to be 0.01 and $u_0=0.1$. Under the proper initial and boundary conditions, the analytical solutions can be obtained,
\begin{equation}\label{4.2.2}
\begin{split}
  &p(x,y)=\frac{1}{4}\rho_0u_0^2(\cos 4\pi x-\cos 4\pi y),\\
  &u(x,y)=u_0\sin 2\pi x\sin 2\pi y,\\
  &v(x,y)=u_0\cos 2\pi x\cos 2\pi y,
  \end{split}
\end{equation}
where $\rho_0=1.0$, $\bm u=(u,v)$ is the velocity and $p$ is the presure.
We implemented a simulation with a fixed computational region of $[0,1]\times [0,1]$, utilizing periodic boundary conditions on the four boundaries.
The CFL condition number is set to $0.5$ and the weight coefficient of mixed difference scheme $\eta=0.8$.
The uniform and non-uniform grids are used to simulate this problem, respectively.
The results of velocity and pressure with grid $32\times 32$ and $64\times 32$ are shown in Fig. \ref{fig:periodic-uv},
which demonstrates good agreement with the analytical solution.
Some physical quantities are also calculated through the non-equilibrium distribution function.
And the numerical results are presented in the Fig. \ref{fig:periodic-uxy}.
The numerical results of those physical quantities are in good agreement with the analytical solution.
This suggests that computational schemes utilizing the non-equilibrium distribution function are effective,
as is the MDF-FDLBM combined with non-uniform grid when solving a convection-diffusion system based on NSEs.

Besides, the computational efficiency of the MDF-FDLBM is also a key characteristic that we focus on.
Therefore, we recorded the computing time and GREs of IFDLBM and MDF-FDLBM respectively.
In order to effectively compare the computational efficiency of the two models,
the evolution of both models are stopped at $10000$ steps, and the periodic flow tends to be stable at this time.
The computational time and GREs of the two models are shown in the Tab. \ref{table:periodic-Ma}.
From the Tab. \ref{table:periodic-Ma}, it can be found the GREs of MDF-FDLBM are smaller than that of IFDLBM.
At the same time, the computational time of MDF-FDLBM is smaller than that of IFDLBM.
Compared to IFDLBM, MDF-FDLBM saves more than $35\% $ in computation time.
There are two reasons for this phenomenon.
Firstly, D2Q9 model is used in IFDLBM, while the MDF-FDLBM adopts the D2Q5 model.
There are $9\times9=81$ cycles are involved in the IFDLBM evolution process.
For the MDF-FDLBM, two evolution equations are calculated in simulation, and there are $2\times 5\times 5=50$ cycles in one evolutionary step.
Secondly, the MDF-FDLBM adopts a linear equilibrium distribution function, while the equilibrium distribution function in IFDLBM contains quadratic terms. Obviously, the calculation of equilibrium distribution function in MDF-FDLBM is simpler than that in IFDLBM.
Therefore, the reduction of the cycles number results in the reduction of the computation time.
For this problem, the computational efficiency of MDF-FDLBM is higher than that of IFDLBM, and the MDF-FDLBM is more accurate than that of IFDLBM.

\renewcommand\arraystretch{0.9}
\begin{table}[!htbp]
  \vspace{1ex}
\caption{Example 1: A comparison of the GREs and CPU time between the IFDLBM and MDF-FDLBM.} \label{table:periodic-Ma}
\centering
 \vspace{1ex} \resizebox{\textwidth}{!}{
\begin{tabular}{p{5em}<{\centering}p{1em}p{10em}<{\centering}p{1em}p{7em}<{\centering}p{1em}p{7em}<{\centering}p{1em}p{7em}<{\centering}p{1em}p{7em}<{\centering}}
\hline\hline
  $\nu$&  & model & &$16\times 16$ & & $32\times 32$ & &$48\times 48$ & & $64\times 64$  \\
\hline
  \multirow{5}{5em}{$\nu=0.005$} &  & GRE of IFDLBM & &$3.1899\times 10^{-2}$ & & $6.3791\times 10^{-3}$ & &$4.9637\times 10^{-3}$ & & $4.2204\times 10^{-3}$    \\
  &  & time of IFDLBM & &$2.58s$ & & $9.53s$ & &$21.01s$ & & $37.24s$    \\
  &  & GRE of MDF-FDLBM & &$2.3754\times 10^{-2}$ & & $6.1662\times 10^{-4}$ & &$2.2326\times 10^{-4}$ & & $2.7564\times 10^{-4}$    \\
  &  & time of MDF-FDLBM & &$1.52s$ & & $5.93s$ & &$12.31s$ & & $22.93s$    \\
  &  & decrease percentage & &$41.09\%$ & & $37.78\%$ & &$41.41\%$ & & $38.43\%$    \\
\hline
\multirow{5}{5em}{$\nu=0.001$} &  & GRE of IFDLBM & &$1.0719\times 10^{-1}$ & & $1.0994\times 10^{-2}$ & &$2.0904\times 10^{-3}$ & & $3.5852\times 10^{-4}$    \\
  &  & time of IFDLBM & &$2.6s$ & & $9.53s$ & &$21.00s$ & & $37.02s$    \\
  &  & GRE of MDF-FDLBM & &$1.6495\times 10^{-2}$ & & $6.1275\times 10^{-3}$ & &$4.2685\times 10^{-4}$ & & $7.8567\times 10^{-5}$    \\
  &  & time of MDF-FDLBM & &$1.51s$ & & $5.94s$ & &$12.29s$ & & $23.37s$    \\
  &  & decrease percentage & &$41.92\%$ & & $37.67\%$ & &$41.48\%$ & & $36.87\%$    \\
\hline\hline

\end{tabular}}
\end{table}

\subsection{The two-dimensional Poiseuille flow}

The two-dimensional Poiseuille flow is taken account for testing the stability of the MDF-FDLBM. This problem is driven by a constant external force ($F_1=1.0\times 10^{-6}$) in a channel, and its analytical solution of velocity, velocity gradient, velocity divergence, strain rate tensor and vorticity can be expressed as
\begin{equation}\label{4.3.1}
\begin{split}
 & u_1=\frac{F_1H^2}{2\nu}\left[\frac{y}{H}-\left(\frac{y}{H}\right)^2\right],\quad u_2=0,\\
 &\frac{\partial u_1}{\partial x}=\frac{\partial u_2}{\partial x}=\frac{\partial u_2}{\partial y}=0,\quad \frac{\partial u_1}{\partial y}=\frac{F_1H}{2\nu}\left[1-2\frac{y}{H}\right],\\
 &\nabla\cdot u=\frac{\partial u_1}{\partial x}+\frac{\partial u_2}{\partial y}=0,\\
 &S_{xx}=S_{yy}0,\quad S_{xy}=S_{yx}=\frac{F_1H}{4\nu}\left[1-2\frac{y}{H}\right],\\
 &\omega=\frac{\partial u_2}{\partial x}-\frac{\partial u_1}{\partial y}=-\frac{F_1H}{2\nu}\left[1-2\frac{y}{H}\right].
\end{split}
\end{equation}

\begin{figure}[htbp]
\centering
\subfigure{\includegraphics[scale=0.5]{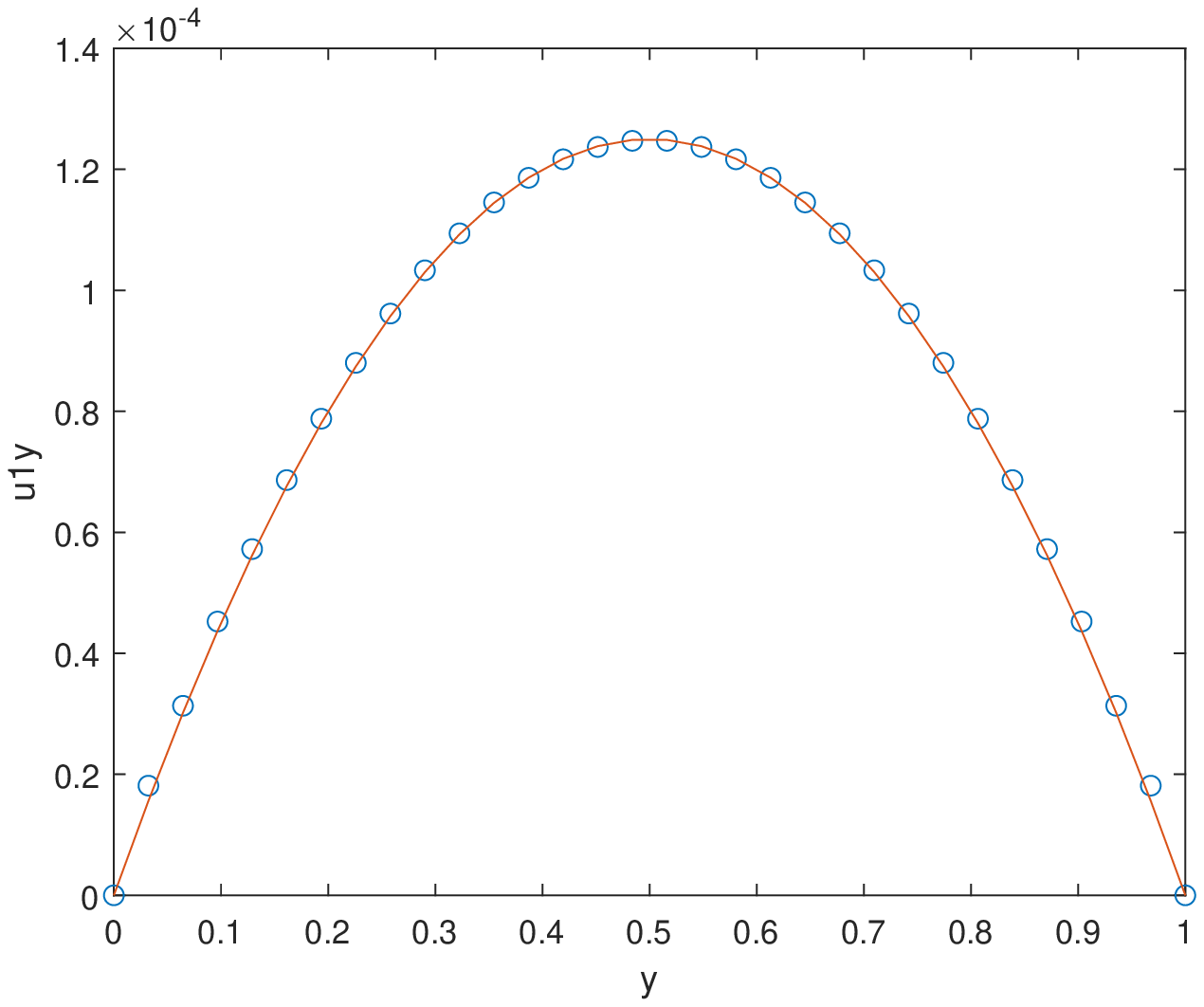}}
\subfigure{\includegraphics[scale=0.5]{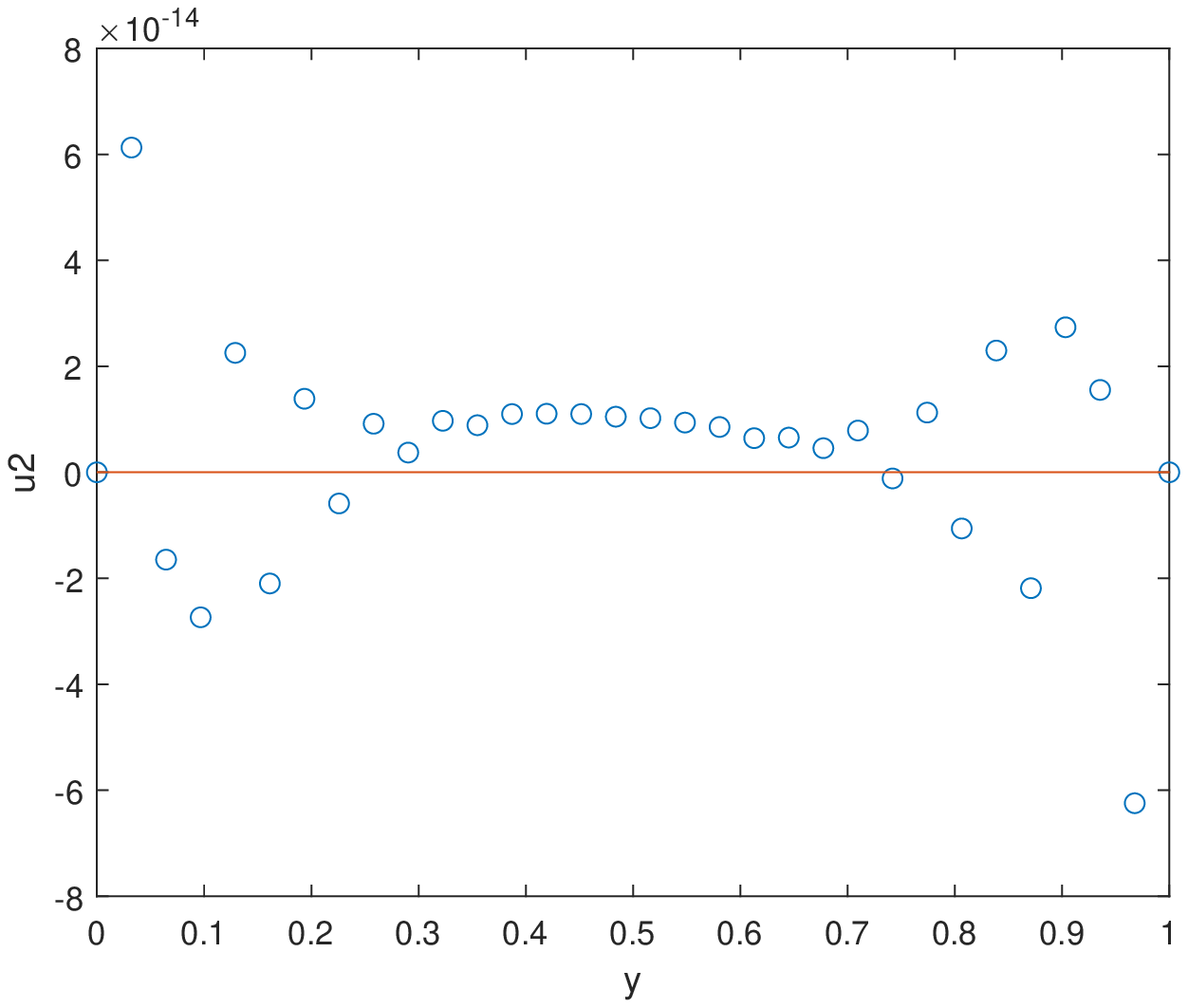}}

 \caption{The MDF-FDLBM numerical and analytical solution of velocity at different positions [symbol:numercial solution, solid line: analytical solution]. } \label{fig:poiseuille-uv}
\end{figure}

\begin{figure}[htbp]
\centering
\subfigure{\includegraphics[scale=0.5]{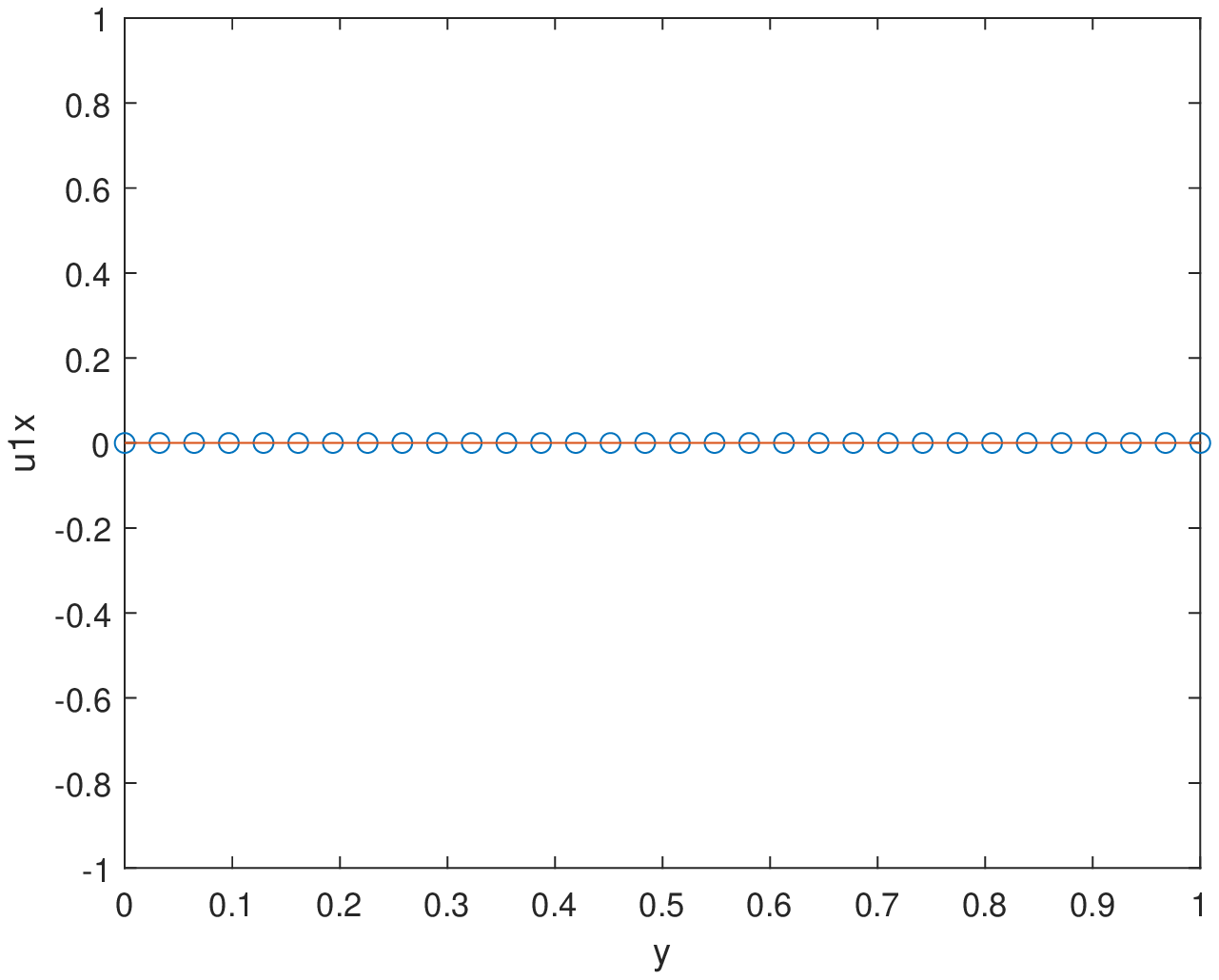}}
\subfigure{\includegraphics[scale=0.5]{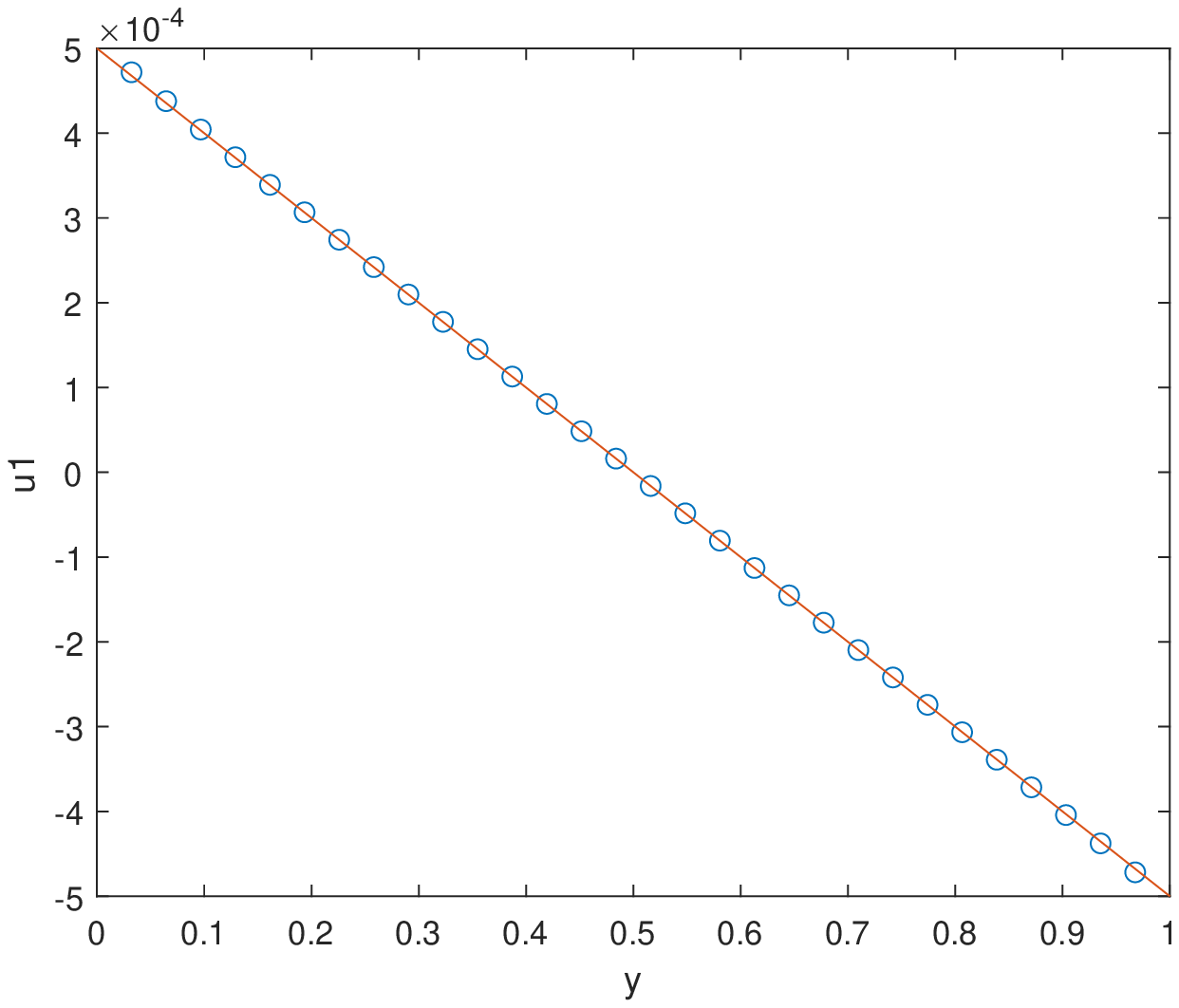}}

 \caption{The MDF-FDLBM numerical and analytical solution of velocity gradient at different positions [symbol:numercial solution, solid line: analytical solution]. } \label{fig:poiseuille-u1v1}
\end{figure}

\begin{figure}[htbp]
\centering
\subfigure{\includegraphics[scale=0.5]{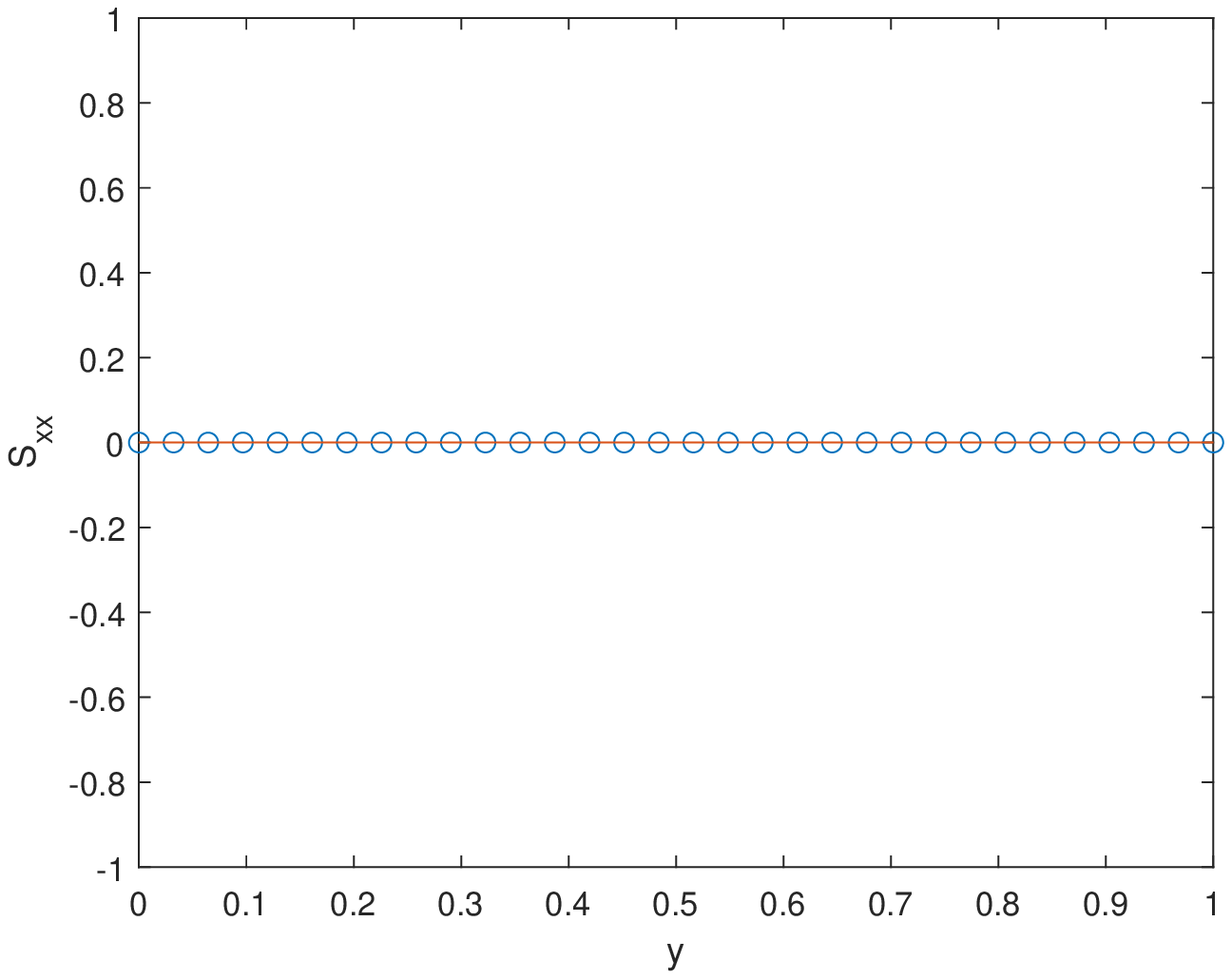}}
\subfigure{\includegraphics[scale=0.5]{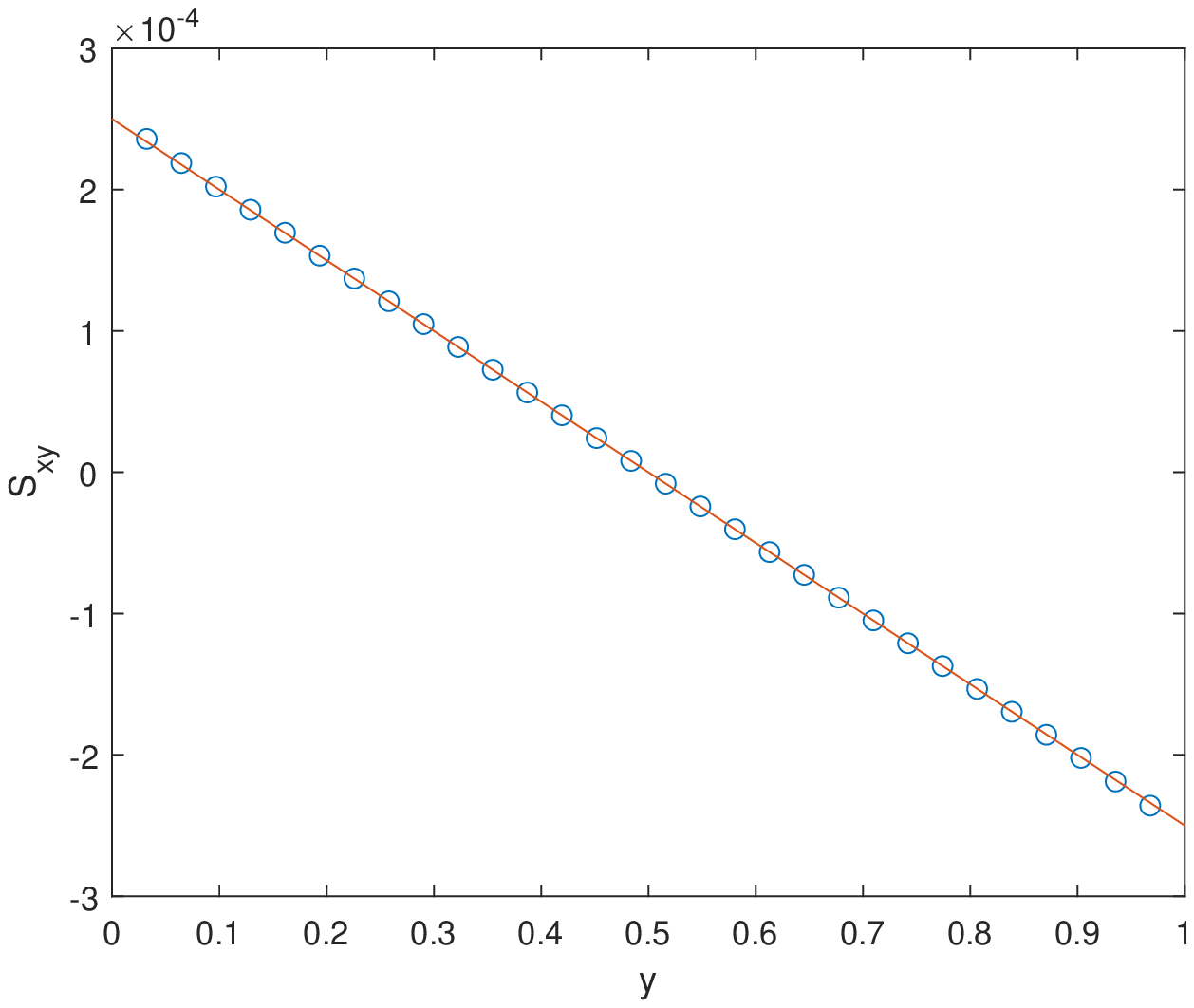}}

 \caption{The MDF-FDLBM numerical and analytical solution of strain rate tensor at different positions [symbol:numercial solution, solid line: analytical solution]. } \label{fig:poiseuille-Sxy}
\end{figure}

\begin{figure}[htbp]
\centering
\subfigure{\includegraphics[scale=0.5]{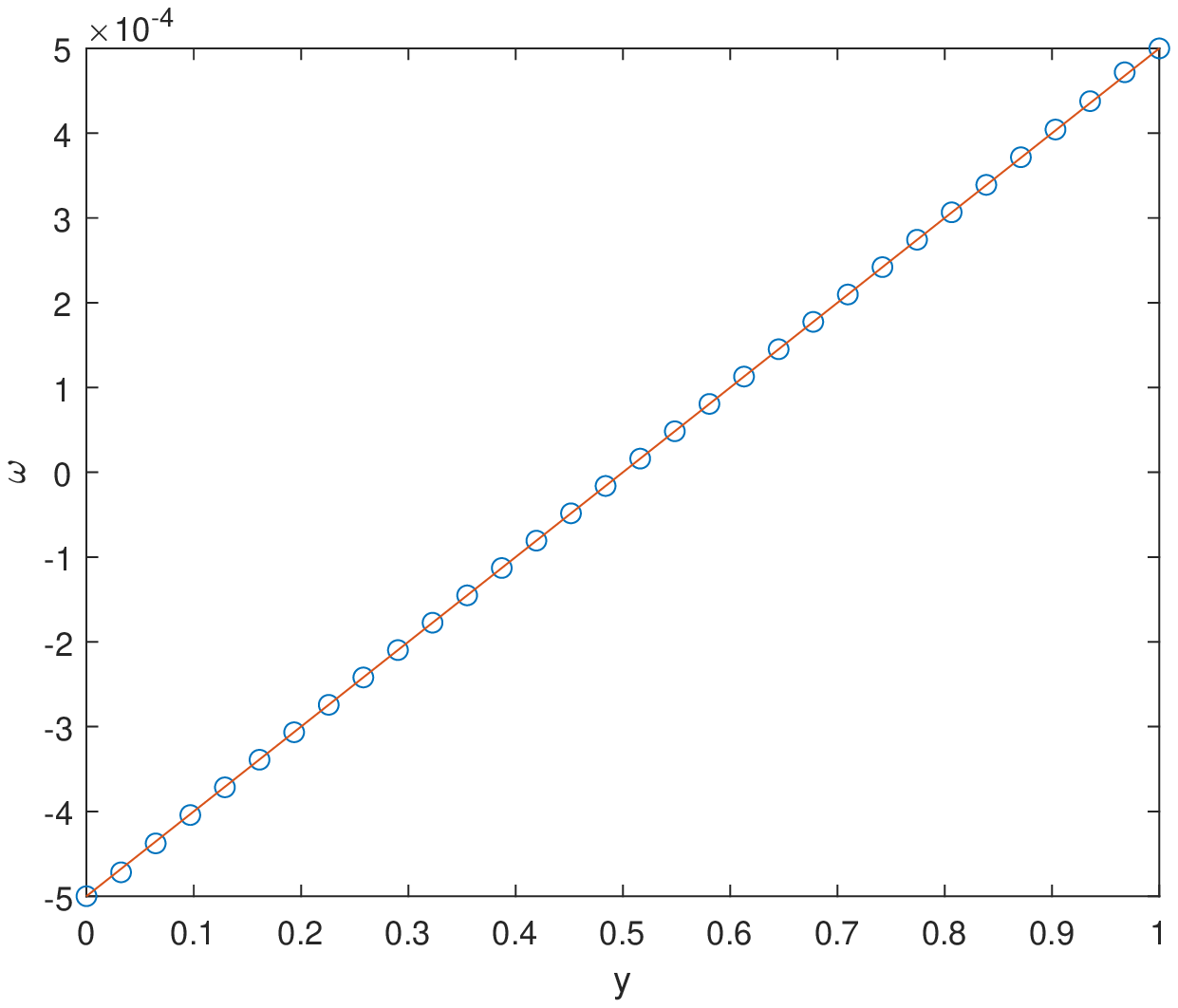}}
\subfigure{\includegraphics[scale=0.5]{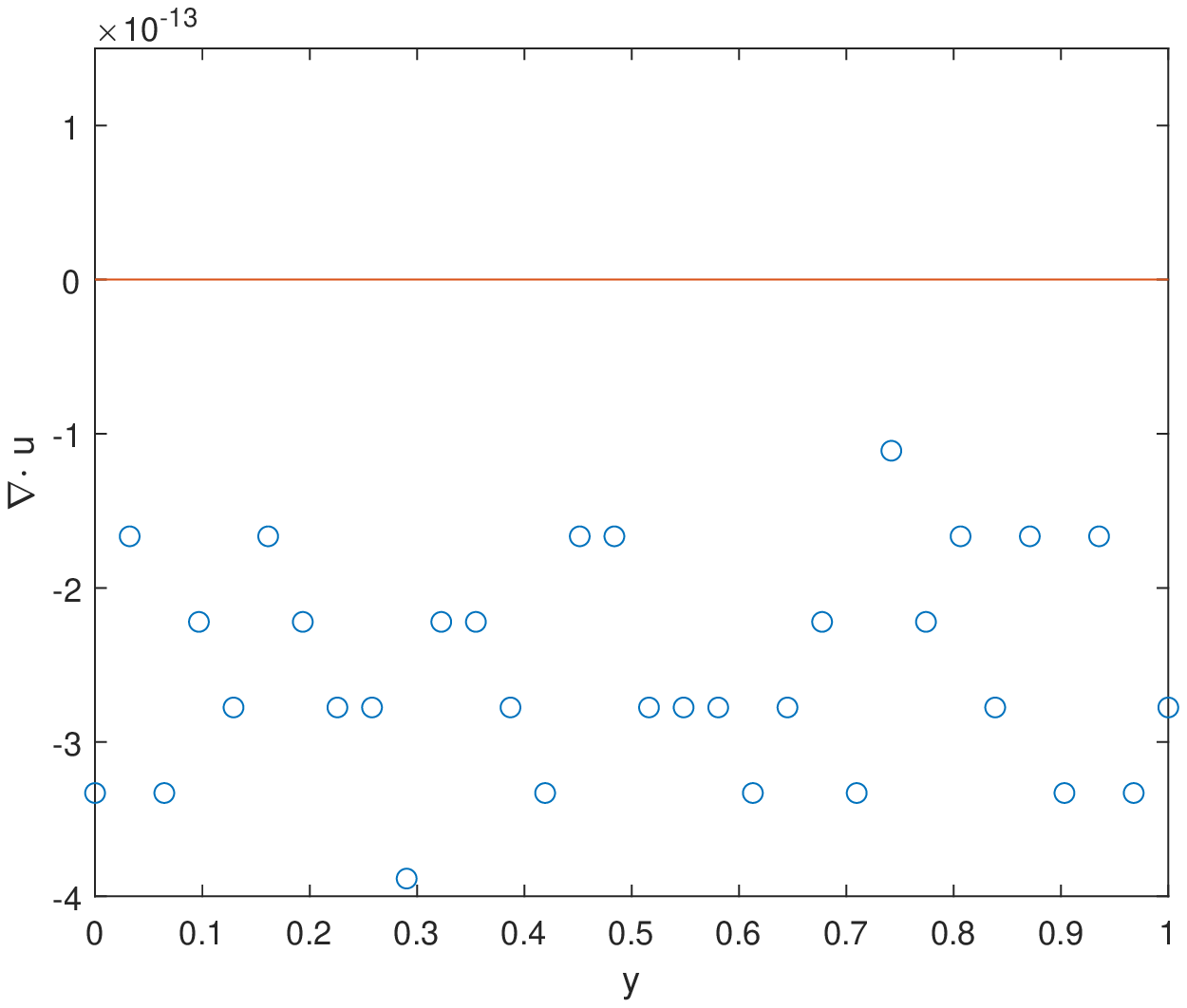}}

 \caption{The MDF-FDLBM numerical and analytical solution of vorticity at different positions [symbol:numercial solution, solid line: analytical solution]. } \label{fig:poiseuille-omega}
\end{figure}

In our simulation, the lattice size is set as $32\times 32$ for the computational domain $[0,1]\times[0,1]$. The periodic boundary condition is adopted in $x$ direction and the non-equilibrium extrapolation scheme is applied to treat the top and bottom walls. Figs. \ref{fig:poiseuille-uv},\ref{fig:poiseuille-u1v1},\ref{fig:poiseuille-Sxy},\ref{fig:poiseuille-omega} show the numerical results of MDF-FDLBM with $\nu=0.001$. It can be found that the results of velocity $\bm u$, velocity gradient ($\partial u_1/\partial x$ and $\partial u_1/\partial y$), velocity divergence $\nabla\cdot \bm u$, strain rate tensor ($S_{xx}$ and $S_{xy}$) and the vorticity $\omega$ agree well with the analytical solution.

We also conduct some comparation between the MDF-FDLBM, IFDLBM and MDF-LBM. The results are presented in Tab. \ref{table:poiseuille-FD-LB}. It can be noticed that the stability of MDF-FDLBM is better than the MDF-LBM while the errors of both models are similar. Besides, from Tab. \ref{table:poiseuille-FD-LB}, we can find that the MDF-FDLBM is more stable with large $\nu$, while the IFDLBM is more stable with small $\nu$. This numerical result is consistent with the conclusion obtained from the stability analysis. In addition, it is observed that the GREs of MDF-FDLBM are smaller than that of IFDLBM.

In order to further verify the conclusion of stability analysis, we use MDF-FDLBM and IFDLBM to simulate this example under different CFL condition number.
The external force term $F$ is set to $10^{-4}$, the grid size is set to $32\times 32$ and $c=1.0$.
The simulation results of different CFL condition number are shown in Tab. \ref{table:poiseuille-CFL}. It can be found that the stability of MDF-FDLBM is better than IFDLBM when the CFL condition number is large, and the error of MDF-FDLBM is still small with large CFL condition number.
In addition, the GRE of MDF-FDLBM is smaller than IFDLBM, indicating that MDF-FDKBM is more accurate than IFDLBM in this example.
Meanwhile, the calculation time of the two models are also recorded in Tab. \ref{table:poiseuille-CFL}. It can be seen that the calculation time of MDF-FDKBM is obviously less than that of IFDLBM. This phenomenon is the same as in Tab. \ref{table:periodic-Ma}. The reason for this phenomenon are the same. In this example, the reduction time of MDF-FDLBM is more than $56\%$. This also indicates that the computational efficiency of MDF-FFLBM is higher than that of IFDLBM.

\renewcommand\arraystretch{0.9}
\begin{table}[!htbp]
  \vspace{1ex}
\caption{Example 3: A comparison of the GREs between the MDF-FDLBM and IFDLBM.} \label{table:poiseuille-FD-LB}
\centering
 \vspace{1ex} \resizebox{\textwidth}{!}{
\begin{tabular}{p{5em}<{\centering}p{1em}p{8em}<{\centering}p{1em}p{6em}<{\centering}p{6em}<{\centering}p{6em}<{\centering}p{6em}<{\centering}p{6em}<{\centering}p{6em}<{\centering}}
\hline\hline
  &  & &  & $F=10^{-1}$   & $F=10^{-2}$  &  $F=10^{-3}$   & $F=10^{-4}$   & $F=10^{-5}$  & $F=10^{-6}$     \\
\hline
 \multirow{2}{5em}{$\nu=0.01$} & &MDF-FDLBM & & $3.4607\times 10^{-3}$ & $6.1900\times10^{-4}$ & $6.2142\times10^{-4}$ & $6.2145\times10^{-4}$ & $6.2145\times10^{-4}$  &  $6.2145\times10^{-4}$ \\                       & &IFDLBM & & $--$ & $2.7500\times10^{-2}$ & $2.7488\times10^{-2}$ & $2.7488\times10^{-2}$ & $2.7488\times10^{-2}$  &  $2.7488\times10^{-2}$ \\
 \hline
  \multirow{2}{5em}{$\nu=0.001$} & &MDF-FDLBM & & $--$ & $1.3363\times10^{-1}$ & $6.4287\times10^{-3}$ & $4.7497\times10^{-4}$ &  $4.7447\times10^{-4}$  &  $4.7446\times10^{-4}$\\
                          & &IFDLBM & & $--$ & $--$ & $1.5170\times10^{-2}$ & $1.5337\times10^{-2}$ & $1.5339\times10^{-2}$  &  $1.5339\times10^{-2}$ \\
 \hline
  \multirow{2}{5em}{$\nu=0.0001$} & &MDF-FDLBM & & $--$ & $--$ & $--$ & $--$&  $3.4607\times10^{-3}$  &  $2.9603\times10^{-4}$ \\
                          & &IFDLBM & & $--$ & $--$ & $--$ & $6.4776\times10^{-2}$ & $6.4639\times10^{-2}$  &  $6.4639\times10^{-2}$ \\
\hline\hline

\end{tabular}}
\end{table}

\renewcommand\arraystretch{0.9}
\begin{table}[!htbp]
  \vspace{1ex}
\caption{Example 3: A comparison of the GREs between the MDF-FDLBM and IFDLBM with different CFL condition number.} \label{table:poiseuille-CFL}
\centering
 \vspace{1ex} \resizebox{\textwidth}{!}{
\begin{tabular}{p{5em}<{\centering}p{1em}p{10em}<{\centering}p{1em}p{6em}<{\centering}p{6em}<{\centering}p{6em}<{\centering}p{6em}<{\centering}p{6em}<{\centering}}
\hline\hline
  &    & & & $CFL=0.1$   & $CFL=0.3$  &  $CFL=0.5$   & $CFL=0.7$   & $CFL=0.9$      \\
\hline
 \multirow{4}{5em}{$\nu=0.01$} & & GRE of MDF-FDLBM &  & $4.0605\times 10^{-3}$ & $4.0336\times10^{-3}$ & $3.9957\times10^{-3}$ & $4.2918\times10^{-3}$ & $6.8811\times10^{-3}$   \\
                        & & time of MDF-FDLBM &  & $116.47s$ & $38.84s$ & $23.35s$ & $16.79s$ & $13.01s$   \\
                        & & GRE of IFDLBM &  & $2.7488\times10^{-2}$ & $3.2652\times10^{-2}$ & $3.7857\times10^{-2}$ & $4.3119\times10^{-2}$ & $4.8433\times10^{-2}$   \\
                        & & time of IFDLBM & & $266.89s$ & $90.52s$ & $56.09s$ & $38.33s$ & $29.89s$   \\
                        & & time reduction ratio & & $56.36\%$ & $57.39\%$ & $58.37\%$ & $56.20\%$ & $56.47\%$   \\
 \hline
  \multirow{4}{5em}{$\nu=0.001$} & & GRE of MDF-FDLBM &  & $2.0447\times 10^{-3}$ & $1.7724\times10^{-3}$ & $1.5278\times10^{-3}$ & $1.4421\times10^{-3}$ & $1.6319\times10^{-3}$   \\
                        & & time of MDF-FDLBM &  & $115.91s$ & $38.65s$ & $23.37s$ & $16.79s$ & $13.01s$   \\
                        & & GRE of IFDLBM &  & $5.0468\times10^{-2}$ & $1.1061\times10^{-1}$ & $1.8049\times10^{-1}$ & $2.6347\times10^{-1}$ & $--$   \\
                        & & time of IFDLBM & & $280.08s$ & $102.99s$ & $61.61s$ & $43.99s$ & $--$   \\
                        & & time reduction ratio & & $58.62\%$ & $62.29\%$ & $62.07\%$ & $61.83\%$ & $--$   \\
 \hline
  \multirow{4}{5em}{$\nu=0.0005$} & & GRE of MDF-FDLBM &  & $4.1387\times 10^{-3}$ & $3.8526\times10^{-3}$ & $3.6170\times10^{-3}$ & $3.4875\times10^{-3}$ & $3.4832\times10^{-3}$   \\
                        & & time of MDF-FDLBM &  & $116.16s$ & $38.75s$ & $23.35s$ & $16.66s$ & $13.10s$   \\
                        & & GRE of IFDLBM &  & $7.4985\times10^{-2}$ & $2.0969\times10^{-1}$ & $3.9927\times10^{-1}$ & $7.6692\times10^{-1}$ & $--$   \\
                        & & time of IFDLBM & & $279.60s$ & $93.71s$ & $56.17s$ & $40.22s$ & $--$   \\
                        & & time reduction ratio & & $58.45\%$ & $58.65\%$ & $58.43\%$ & $58.58\%$ & $--$   \\
\hline\hline

\end{tabular}}
\end{table}

\subsection{The two-dimensional lid-driven cavity flow}

As a bench mark problem, the lid-driven cavity flow is simulated to test the capacity of the MDF-FDLBM, which is driven by a constant velocity $u_0=0.1$ of the top wall. The length of the square cavity is taken as 1. It is a complicated problem because the fluid mechanical phenomena is rich and there is no analytical solution available. The initial velocity and press are set to be $\bm u=0$ and $p=0$ and the four wall is treated by the non-equilibrium extrapolation scheme.

\begin{figure}[htbp]
\centering
\subfigure[$Re=400$]{\includegraphics[scale=0.35]{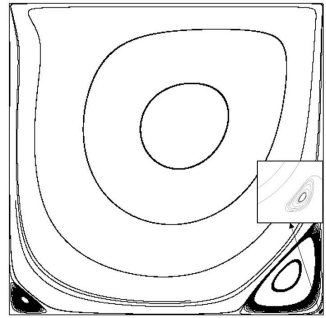}}
\subfigure[$Re=1000$]{\includegraphics[scale=0.35]{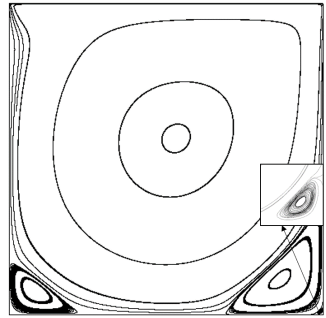}}
\subfigure[$Re=3200$]{\includegraphics[scale=0.365]{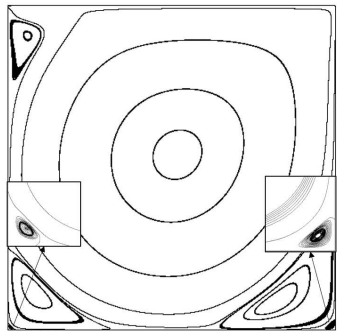}}
\subfigure[$Re=5000$]{\includegraphics[scale=0.35]{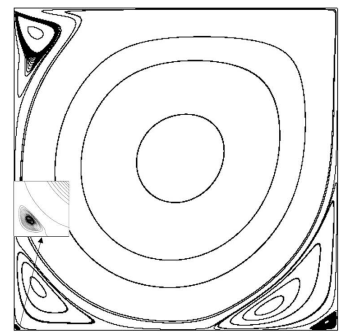}}
 \caption{The streamline of Lid driven flow under different $Re$ number } \label{fig:lid_stream}
\end{figure}

\begin{figure}[htbp]
\centering
\subfigure{\includegraphics[scale=0.5]{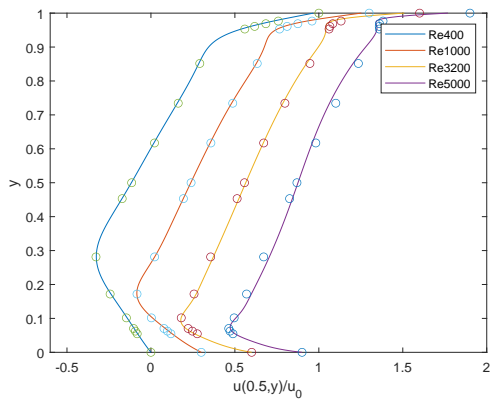}}
\subfigure{\includegraphics[scale=0.5]{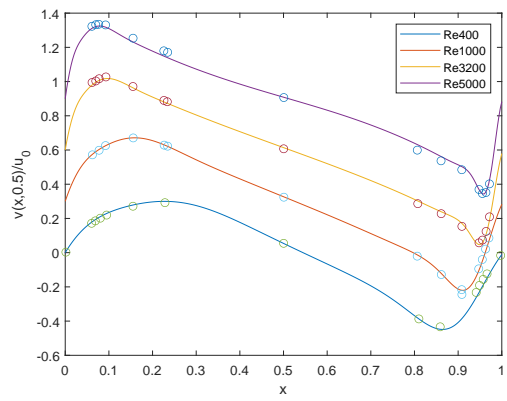}}

 \caption{Velocity profiles along the centerline at different Re, (o) is reference date \cite{ChenFDLBM2020}. [(a) $Re$=400, 1000, 3200, 5000 from left to right; (b) $Re$=400, 1000,
3200, 5000 from bottom to top.] } \label{fig:Lid-uv}
\end{figure}

In this part, we would conduct some simulations by the MDF-FDLBM at different Reynolds number ($Re=Lu_0/\nu$). In our simulation, the lattice size is chosen to be $128\times 128$ for $Re=400$ and $256\times 256$ for $Re=1000,3200,5000$. We take $c=1.0$ and $CFL=0.5$. The simulations results are presented in Fig. \ref{fig:lid_stream}. As we can seen, there are four vortices appear in the cavity when $Re \leq 1000$, the first one is a primary vortex at the center of the cavity, the others are the secondary vortices at the lower left and lower right corners and a third level vortex at the lower right corner. When Re number is up to $3200$ or $5000$, we can observe a secondary vortex appears in the upper left corner. And a new third level vortex appears at the lower left corner. Besides, the center of the primary vortex approaches the center of the cavity as $Re$ increases. These phenomena are consistent with previous work \cite{ChenFDLBM2020}. Compared with the simulation results of MDF-LBM \cite{Chai2022}, MDF-FDLBM can capture more flow details even for $N_x\times N_y=256\times 256$.
The profiles of velocity along vertical and horizontal lines are displayed in Fig. \ref{fig:Lid-uv}. It is obviously that the results are in good agreement with the available results. In Table 10, the locations of the vortices are also consistent with the previous work.

\renewcommand\arraystretch{0.9}
\begin{table}[!htbp]
  \vspace{1ex}
\caption{Example 3: A comparison of the GREs between the MDF-FDLBM and MDF-LBM.} \label{table:Lid-lication}
\centering
 \vspace{1ex} \resizebox{\textwidth}{!}{
\begin{tabular}{p{8em}<{\centering}p{1em}p{8em}<{\centering}p{2em}<{\centering}p{1em}p{6em}<{\centering}p{6em}<{\centering}p{6em}<{\centering}p{6em}<{\centering}}
\hline\hline
  &  & & & & $Re=400$   & $Re=1000$  &  $Re=3200$   & $Re=5000$        \\
\hline
 \multirow{4}{8em}{primary vortex} & & \multirow{2}{8em}{MDF-FDLBM} &$x$ & & $0.5568$ & $0.5330$ & $0.5176$ & $0.5144$ \\
                          & & &$y$ & & $0.6081$ & $0.5679$ & $0.5403$ & $0.5344$  \\
                          & &\multirow{2}{5em}{Ref\cite{UGlid1982}} &$x$ & & $0.5547$ & $0.5313$ & $0.5165$ & $0.5176$  \\
                          & & &$y$ & & $0.6055$ & $0.5626$ & $0.5469$ & $0.5373$  \\
 \hline
  \multirow{4}{8em}{primary vortex} & & \multirow{2}{8em}{MDF-FDLBM} &$x$ & & $0.0500$ & $0.0830$ & $0.0808$ & $0.0720$   \\
                          & & &$y$ & & $0.0486$ & $0.0781$ & $0.1186$ & $0.1355$ \\
                          & &\multirow{2}{5em}{Ref\cite{UGlid1982}} &$x$ & & $0.0508$ & $0.0859$ & $0.0859$ & $0.0784$  \\
                          & & &$y$ & & $0.0469$ & $0.0781$ & $0.1094$ & $0.1373$  \\
 \hline
  \multirow{4}{8em}{primary vortex} & & \multirow{2}{8em}{MDF-FDLBM} &$x$ & & $0.8879$ & $0.8662$ & $0.8178$ & $0.7937$ \\
                          & & &$y$ & & $0.1227$ & $0.1130$ & $0.0850$ & $0.0827$ \\
                          & &\multirow{2}{5em}{Ref\cite{UGlid1982}} &$x$ & & $0.8906$ & $0.8906$ & $0.8125$ & $0.8078$  \\
                          & & &$y$ & & $0.1250$ & $0.1250$ & $0.0859$ & $0.0745$  \\

\hline\hline
\end{tabular}}
\end{table}

	\section{Conclusions}\label{Conclusion}
	In this paper, the incompressible NSEs is transformed in to a convection-diffusion system. A MDF-FDLBM with the MRT model is proposed for the convection-diffusion system based NSEs.
Through the CE analysis, the incompressible NSEs can be recovered correctly from the MDF-FDLBM.
Additionally, some local scheme for the velocity gradient, velocity divergence, strain rate tensor, shear stress, and vorticity are proposed.
We also analyze the stability of the MDF-FDLBM and IFDLBM. It is found that
the IFDLBM will be more stable than that of MDF-FDLBM with smaller $\nu$,
and the MDF-FDLBM will be more stable than that of IFDLBM with large CFL condition number.

Some classic fluid problems are conducted to test the performance of the MDF-FDLBM, including
the two-dimensional four-roll mill problem, the periodic flow, the two-dimensional Poiseuille flow
and the two-dimensional lid-driven cavity flow. The results make great agreements with the
analytical solution or previous works. This shows that MDF-FDLBM and the local scheme of these physical quantities are valid.
Through the numerical testing, the MDF-FDLBM has a second-order convergence rate in space and time.
Besides, the stability of MDF-FDLBM and IFDLBM was compared by different examples.
The numerical results show that IFDLBM is more stable under small kinematic viscosity, while MDF-FDLBM is more stable under large CFL condition number. This is consistent with the conclusion of theoretical analysis.
Meanwhile, compared with IFDLBM, it can be found that the simulation results of MDF-FDLBM are more accurate, and the calculation efficiency of MDF-FDLBM is higher.

In light of the aforementioned advantages, it is reasonable to suggest that the MDF-FDLBM can be extended for investigating thermal flows and multiphase fluid systems subject to the incompressible NSEs and CDE, which will be considered in the future work.
	\section*{Acknowledgments}
	This work is supported by the National Natural Science Foundation of China
(Grants No. 51836003 and No. 12072127), the Graduates Innovation Fund,
Huazhong University of Science and Technology (No. 2020yjsCXCY034), and the
Nature Science Foundation of Hubei province (Grant No. 2020CFB384).

	\bibliographystyle{elsarticle-num}
	\bibliography{ref}
\end{document}